\documentclass[aip,graphicx,pof,amsmath,amssym,reprint,groupadress]{revtex4-1}
\usepackage{graphicx,color,bm,dcolumn,latexsym}
\usepackage{amsmath,amsbsy,amssymb,amsgen,amsfonts}
\usepackage{epstopdf} 
\usepackage{relsize}
\usepackage{enumerate}
\usepackage[breaklinks=true,bookmarks=true,colorlinks=true,linkcolor=blue,citecolor=blue,allcolors=blue]{hyperref}
\usepackage[toc,title]{appendix}
\definecolor{AgatheCommentColor}{rgb}{0.4 , 0.5 , 0.7}

\newcommand{\ben} {\begin{equation}}
\newcommand{\een} {\end{equation}}
\newcommand{\be} [1] {\begin{equation} \label{#1}}
\newcommand{\ee} {\end{equation}}
\newcommand{\bse} [1] {\begin{subequations} \label{#1}}
\newcommand{\ese} {\end{subequations}}
\newcommand{\ban} {\begin{eqnarray*} }
\newcommand{\ean} {\end{eqnarray*} }
\newcommand{\bea} {\begin{eqnarray}}
\newcommand{\eea} {\end{eqnarray}}

\def\solid{\protect\rule[1pt]{10.pt}{1pt}}
\def\solidthick{\protect\rule[2pt]{10.pt}{1pt}}
\def\solidshort{\protect\rule[2pt]{3.pt}{1pt}}
\def\dashed{\solidshort$\,$\solidshort$\,$\solidshort}
\def\chndot{\solidshort$\,\cdot\,$\solidshort}

\newcommand{\opencircle}
           {$\mathlarger{\mathlarger{\mathlarger{\circ}}}$}

\newcommand{\crossed}{$\mathlarger{\mathlarger{\times}}$}
\definecolor{agreen}{rgb}{0.4083 , 0.667 , 0.3958} 
\definecolor{a2green}{rgb}{0.2042 , 0.3333 , 0.1979}%
\definecolor{ablue}{rgb}{0.4 , 0.5 , 0.7}
\definecolor{a2blue}{rgb}{0.02 , 0.25 , 0.35}%
\definecolor{apink}{rgb}{0.8333 , 0.3875 , 0.5667} 
\definecolor{abrown}{rgb}{0.8333 , 0.6250 , 0.500} 
\definecolor{ared}{rgb}{0.75, 0.2042 , 0.2042} 
\definecolor{a2red}{rgb}{0.3750, 0.1021 , 0.1024} 
\definecolor{aturquoise}{rgb}{0.1250 , 0.4167 , 0.5208} 
\definecolor{apurple}{rgb}{0.7083 , 0.3542 , 0.7000} 
\definecolor{agrey}{rgb}{0.3917 , 0.3583 , 0.3792} 
\definecolor{ablack}{rgb}{0.0 , 0.0 , 0.0} 
\newcommand{\revision}[2]{#2}
\begin{document}
\title{%
  Forcing homogeneous turbulence in DNS of 
  particulate flow with interface resolution and %
  gravity 
} 
\affiliation{Institute for Hydromechanics, Karlsruhe Institute of
  Technology, 76131 Karlsruhe, Germany}
\author{Agathe Chouippe}
\email[]{agathe.chouippe@kit.edu}
\author{Markus Uhlmann}
\email[]{markus.uhlmann@kit.edu}
\date{7 November 2015 -- accepted for publication in {\it Phys.\ Fluids}}
\begin{abstract}
  We consider the case of finite-size spherical particles which are
  settling under gravity in a homogeneous turbulent background
  flow. Turbulence is forced with the aid of the random forcing
  method of Eswaran and Pope [Comput.\ Fluids, 16(3):257-278, 1988],
  while the solid particles are represented with an immersed-boundary
  method. 
  The forcing scheme is used to generate isotropic turbulence in
  vertically elongated boxes in order to warrant better decorrelation
  of the Lagrangian signals in the direction of gravity. 
  Since only a limited number of Fourier modes are forced, it is
  possible to evaluate the forcing field directly in physical space,
  thereby avoiding full-size transforms. 
  The budget of box-averaged kinetic energy is derived from the forced 
  momentum equations. 
  Medium-sized simulations for dilute suspensions at low Taylor-scale
  Reynolds number $Re_\lambda=65$, small density ratio
  $\rho_p/\rho_f=1.5$ and for two Galileo numbers $Ga=0$ and $120$ are
  carried out over long time intervals in order to exclude the
  possibility of slow divergence.  
  It is shown that the results at zero gravity are fully consistent
  with previous experimental measurements and available numerical
  reference data. 
  Specific features of the finite-gravity case are discussed with
  respect to a reduction of the average settling velocity, the
  acceleration statistics and the Lagrangian auto-correlations. 
\end{abstract}

\pacs{}%
\maketitle %
\section{Introduction}
\label{sec-intro}
Particulate flow is a rich scientific topic which bears a multitude of
open questions, some of which have broad practical consequences. 
One example is the settling speed of heavy solid particles which are
suspended in a fluid undergoing turbulent motion. 
Predicting its average value and higher statistical moments is
certainly of great interest in fields such as meteorology and 
chemical engineering. 
However, our current knowledge of the underlying dynamical processes
in particulate flow systems is still incomplete, to a large
extent due to a lack of detailed data covering the entire
multi-parameter space. 
As a consequence the predictive capabilities of existing
engineering-type models are still relatively limited. 

Interface-resolved direct numerical simulation (DNS) of idealized
particulate flows is starting to become a viable option for generating
high-fidelity data-sets complementing data obtained through laboratory
measurements.    
Most numerical studies in this spirit have either focused
upon plane channel
flow\cite{pan:97,kajishima:02,uhlmann:08a,villalba:12,picano:15}  
or upon homogeneous-isotropic
turbulence\cite{tencate:04,homann:10,YeoClimentMaxey:10,doychev:10a,lucci:10}.  
In homogeneous flows without mean deformation, a turbulent background
field is typically either obtained through a suitably chosen initial
condition with a subsequent energy decay\cite{doychev:10a,lucci:10},
or the turbulent fluctuations are generated 
by explicitly adding an energy input to the Navier-Stokes
equations\cite{tencate:04,homann:10,YeoClimentMaxey:10}.   
The latter approach has the obvious advantage of allowing for a
statistically-stationary state to be reached and maintained as long as
desired, thereby facilitating the simulation of processes occurring on
a long time-scale, such as particle cluster
formation\cite{uhlmann:14a}. 

Most turbulence forcing methods are formulated in Fourier space, and the
energy input is restricted to the largest scales, i.e.\ to low
wavenumbers\cite{jimenez:93,sullivan:94,ishihara:07,yeung:12}. 
One notable exception is the forcing scheme proposed by
Lundgren\cite{lundgren:03} (cf.\ also Ref.\citenum{rosales:05}), in
which an additional forcing term directly proportional to the velocity
vector itself is added to the momentum equation. 
While in most approaches %
the momentum forcing mechanism indeed depends upon
the state of the flow field (deterministic forcing), the methods
proposed by Eswaran and Pope\cite{eswaran:88} and by
Alvelius\cite{alvelius:99} construct the forcing term randomly in a
manner which is independent of the flow field. 

In simulations of particulate flow with fully-resolved
phase-interfaces there are several additional points to be considered
when choosing a turbulence forcing scheme. 
First, it should be recalled that the particle motion is
naturally 
two-way coupled to the fluid flow, which means that the 
question of numerical stability of the discretized Navier-Stokes
equations involving a turbulent forcing scheme  
arises. 
Second, in the presence of inertial particles and non-zero gravity,
the particle phase experiences a mean gravitational acceleration,
which in turn through the mean drag provides a source term to the
fluid's kinetic energy equation. As a consequence, the coupled
system's kinetic energy budget is significantly modified as compared
to the zero-gravity (or neutrally-buoyant) case. 
Finally, it should be kept in mind that the numerical requirements for
fully-resolved particulate flow can be substantially harder to meet than
for single-phase turbulence. This is particularly true for the case
with finite gravity, since mean particle settling places additional
requirements upon the small-scale resolution and upon the
computational domain size in the vertical direction. 
Therefore, in the present context we require the following properties
of a turbulence forcing method:  
\begin{enumerate}[(I)]
\renewcommand{\itemsep}{0ex}
\item In single-phase flow it should allow for simulations which yield
  small-scale statistics in agreement with previous studies. 
\item It should allow for stable long-time integration in the presence
  of particles which are fully two-way coupled to the flow field, even
  in the presence of finite gravity. 
\item It should be efficient to compute in the framework of a 
  numerical method which is not based upon Fourier transform (i.e.\
  not pseudo-spectral).  
\item It should allow for a reasonable a priori estimation of the
  parameter values which need to be set in order to attain the target
  values of the turbulence scales. 
\end{enumerate}
A number of forcing schemes which have been proposed in the past do
not fulfill the entire set of the above requirements.
In particular, the linear forcing method of
Lundgren\cite{lundgren:03,rosales:05} -- although very attractive
in view of requirement (III) -- does not fulfill requirement (II), as
already noted by Naso and Prosperetti\cite{naso:10}. 
Our own experience confirms this observation: applying Lundgren's
linear forcing to two-way coupled particulate flow problems leads to
unbounded growth of the kinetic energy (Doychev and Uhlmann,
unpublished results).  

Forcing schemes which potentially seem to adapt to all of the above
requirements are those which are based upon external random
processes\cite{eswaran:88,alvelius:99}. In this category the force term
which is added to the momentum equation does not depend upon the
velocity field itself, therefore eliminating a source of instability
when applied to particulate flow. 

The first numerical study of forced homogeneous-isotropic turbulence
seeded with finite-size resolved particles has been performed by Ten
Cate et al.\cite{tencate:04} who employed a lattice-Boltzmann method
and maintained  statistically stationary turbulence with the aid of 
Alvelius' random forcing scheme\cite{alvelius:99}. These authors
pre-computed a number of realizations of the forcing field and
imposed it in a random sequence upon the flow field. 
Homann and Bec\cite{homann:10} used a Fourier pseudo-spectral
Navier-Stokes solver in conjunction with a penalty method to represent
individual finite-size particles, while 
keeping the energy content of two small-wavenumber shells constant.  
Yeo et al.\cite{YeoClimentMaxey:10} represented the action of finite-size
particles upon the flow by means of the force-coupling method of Lomholt
and Maxey\cite{lomholt:03}; in their simulations turbulence was maintained
statistically stationary with the random forcing method of Eswaran and
Pope\cite{eswaran:88}. 
In all of these studies\cite{tencate:04,homann:10,YeoClimentMaxey:10} the
particles' submerged weight was zero (either because of a density
ratio of unity or because of zero gravity). 
Therefore, 
the problems and open questions associated with mean
particle settling in a turbulent environment still remain largely
unsolved.  

The aim of the present paper is to discuss the feasibility of
investigating with fully-resolved simulations 
the motion of particles which are
larger than the Kolmogorov scale, have a higher density than
the fluid, and which are subjected to finite gravity. 
The random forcing scheme of Eswaran and Pope\cite{eswaran:88} is
employed to generate a turbulent background field, and the
immersed-boundary technique of Uhlmann\cite{uhlmann:04} is used to
force the no-slip condition at the interface between the fluid and the
solid particles. 
The purpose of the present contribution is two-fold. 
On the one hand we present the methodology of turbulence forcing in
the framework of two-way particulate flow with gravity. A study of this
configuration, for which case several adaptations of the basic forcing
method are necessary, has to our knowledge not been published to the
present date.  
We feel that it is important to ensure that the chosen forcing
strategy satisfies the four requirements set forth above, in
particular regarding numerical stability.  
Here we apply the turbulence forcing scheme to obtain isotropic
turbulence in vertically-elongated boxes; with a slight modification 
this method allows us to simulate heavy particles which settle with
respect to the forcing field. 
On the other hand the evolution of the kinetic energy in this system
which exhibits two source terms (related to 
gravity and to large-scale forcing) 
has not been investigated in detail.   
Here we analyze the energy budget derived from the momentum equation
solved in the simulations, 
which includes two volume force terms (the large-scale turbulence forcing
and the forcing of the no-slip interface condition).  
Results are presented from medium-sized simulations of dilute
suspensions at low turbulent Reynolds number and intermediate Galileo
number. 
Where possible the present results are compared to available
experimental and numerical reference data from previous studies. 
\section{Numerical method}
\label{sec-num}
We present in this section the numerical method as well as the main
statistical quantities considered in this study. 
The equations which describe the motion of an incompressible 
fluid with constant density $\rho_f$ and constant kinematic viscosity
$\nu$ can be written as follows
\begin{subequations}\label{equ-navier-stokes-general}
  \begin{eqnarray}\label{equ-navier-stokes-general-mom}
    \frac{\partial\mathbf{u}}{\partial t}
    +(\mathbf{u}\cdot\nabla)\mathbf{u} 
    +\frac{1}{\rho_f}\nabla p 
    &=&
    \nu\nabla^2\mathbf{u}
    +{\boldsymbol{\mathcal F}}
    \,,
    \\\label{equ-navier-stokes-general-conti}
    \nabla\cdot\mathbf{u}&=&0
    \,,
  \end{eqnarray}
\end{subequations}
where $\mathbf{u}$ denotes the fluid velocity vector 
and $p$ the hydrodynamic pressure. 
The vector ${\boldsymbol{\mathcal F}}$ (which will be further specified
below) collects all specific volume force terms associated with: 
(a) generating and maintaining homogeneous turbulence 
(denoted as $\mathbf{f}^{(t)}$ in the following),  
and 
(b) the representation of the solid particles ($\mathbf{f}^{(p)}$) in
the framework of an immersed boundary technique. 
Therefore, the total volume force term in the most general case
considered here is given by the sum of these two contributions:
\begin{equation}\label{equ-total-volume-force}
  {\boldsymbol{\mathcal F}}
  =
  \mathbf{f}^{(t)}
  +
  \mathbf{f}^{(p)}
  \,.
\end{equation}
\subsection{Forcing homogeneous-isotropic turbulence}
\label{sec-num-turb-forcing}
We consider a forcing method which is expressed as a momentum source term
$\mathbf{f}^{(t)}(\mathbf{x},t)$ introduced into 
(\ref{equ-navier-stokes-general-mom}) 
through (\ref{equ-total-volume-force}). 
The formulation is based upon random processes driving the time
evolution of a number of large scales (i.e.\ small wavenumber Fourier
modes). 
Two variants of this category of forcing methods have been proposed
and employed by various authors in the past: 
the scheme of Eswaran \& Pope\cite{eswaran:88}
(henceforth denoted as ``EP'') 
and the scheme by Alvelius\cite{alvelius:99}. 
The principle difference between these two approaches lies in the
characteristic time-scale over which the random processes are
correlated:  
in EP an explicit forcing time-scale can be prescribed, 
whereas Alvelius' method does not involve such time-scale 
(i.e.\ the random processes are uncorrelated from step to step). 
In the present work we have chosen EP's method, since it allows us to
investigate the influence of the forcing time scale. If required, 
essentially uncorrelated random processes can be recovered by choosing
the forcing time-scale equal to the numerical time step, as will be
made more precise below. 

We summarize below the main characteristics of EP forcing; 
for more details we refer the reader to the original
work\cite{eswaran:88}. 
The EP forcing method is first formulated in Fourier space. 
Let $\boldsymbol{\kappa}=(\kappa_1,\kappa_2,\kappa_3)$ denote the 
wavenumber vector 
which can be defined in a cuboidal domain with side-lengths 
$({\cal L}_x, {\cal L}_y, {\cal L}_z)$ in the three physical space
directions. 
The forcing term in Fourier space is denoted by
$\hat{\mathbf{f}}^{(t)}(\boldsymbol{\kappa},t)$. 
It is non-zero only in a low-wavenumber band for which 
$|\boldsymbol{\kappa}|\leq\kappa_f$, where $\kappa_f$ is a parameter
to be determined. 
Note that the origin $\boldsymbol{\kappa}=0$ is not forced. 
The vector of complex Fourier coefficients
$\hat{\mathbf{f}}^{(t)}(\boldsymbol{\kappa},t)$ 
is computed from six independent Uhlenbeck-Ornstein processes (the
real and imaginary parts of each one of the three vector components)
for each of the forced wavenumbers. 
The components of the complex vector describing the 
random 
process, denoted as $\hat{\mathbf{b}}(\boldsymbol{\kappa},t)$, are 
given by the following finite-difference equation: 
\begin{equation}\label{UO_process}
  \hat{b}_i(\boldsymbol{\kappa},t+\Delta t) 
  =
  \hat{b}_i(\boldsymbol{\kappa},t) 
  \left(1-\frac{\Delta t}{T_L} \right)
  + 
  e_i(\boldsymbol{\kappa},t) 
  \left( 2 \sigma^2 \frac{\Delta t}{T_L} \right)^{1/2} 
  \,,
  \quad
  \forall\,\,i=1,2,3
  \,.
\end{equation}
In (\ref{UO_process}) the symbol $\Delta t$ denotes the numerical time
step, $e_i(\boldsymbol{\kappa},t)$ is a complex random number drawn 
(at each time step and for each wavenumber) 
from a standardized Gaussian distribution (with zero mean and variance
unity),  
$T_L$ 
is the characteristic time scale of the random
process, and $\sigma^2$ its variance. It can be seen that choosing 
$T_L=\Delta t$ 
corresponds to a random process which is uncorrelated
in time. As shown by EP, in the limit as $\Delta t$ tends to zero the
process (\ref{UO_process}) has zero mean
($\langle\hat{\mathbf{b}}(\boldsymbol{\kappa},t)\rangle=0$, the
angular brackets indicating statistical averaging), 
and its temporal correlation indeed exhibits an exponential decay,
viz. 
\begin{equation}\label{UO_correlation}
  \langle
  \hat{b_i}(\boldsymbol{\kappa},t)
  \hat{b_j}^{*}(\boldsymbol{\kappa},t+s)
  \rangle 
  =
  2\sigma^2\delta_{ij}\exp(-s/T_L) 
  \,,
\end{equation}
where an asterisk denotes complex conjugation. 

The volume force term $\hat{\mathbf{f}}^{(t)}(\boldsymbol{\kappa},t)$
is obtained by orthogonal projection of the vector describing the
random processes, 
\begin{equation}\label{EswaranPope_Projection}
  \hat{\mathbf{f}}^{(t)}(\boldsymbol{\kappa},t+\Delta t)
  =
  \hat{\mathbf{b}}(\boldsymbol{\kappa},t+\Delta t)
  - 
  \boldsymbol{\kappa}
  (\boldsymbol{\kappa} \cdot 
  \hat{\mathbf{b}}(\boldsymbol{\kappa},t+\Delta t)
  )
  /
  (\boldsymbol{\kappa} 
  \cdot 
  \boldsymbol{\kappa})
  \,,
\end{equation}
thereby guaranteeing zero divergence. 

It can be seen that a total of three parameters are introduced by the
EP scheme: $T_L$, $\kappa_f$ and $\sigma$. 
In practice the amplitude parameter $\sigma$ is not directly
prescribed; instead a value for the combined quantity
$\epsilon^\ast=\sigma^2\,T_L$ 
is imposed.   
Doing so (i.e.\ choosing the values for $T_L$ and $\epsilon^\ast$
independently) ensures that in the limit of vanishing $T_L$ the mean
energy input does not trivially tend to zero\cite{eswaran:88}. 

Since our Navier-Stokes solver is based upon finite differences, the
Fourier series involving the coefficients given by
(\ref{EswaranPope_Projection}) needs to be evaluated at the discrete
grid points in physical space in order to determine the force field
$\mathbf{f}^{(t)}(\mathbf{x},t)$.    
In order to avoid the cost associated with a complete fast Fourier
transform step (which for massively parallel computations mainly stems
from the necessary communication overhead), we perform a direct
evaluation of the Fourier series.  
Note that when splitting up the transform into three successive
one-dimensional transforms, this operation is computationally
not expensive, as described in appendix~\ref{sec-app-fou-trans}.   
Since each parallel process simulates the entire set
of random processes redundantly in local memory, this procedure does
not involve any inter-processor communication. Furthermore, since the
number of forced Fourier modes (controlled by the parameter
$\kappa_f$) is typically small 
(below $100$) 
the computational cost due to the turbulence forcing scheme remains
small compared to the overall cost of a Navier-Stokes time-step. 
In practice we have observed an increase in the necessary wall-clock
time per time step of roughly 10-15\% 
for the cases considered in this work. 
\subsection{Representing particles with an immersed boundary method}  
\label{sec-num-particles}
The numerical method employed in the present simulations 
has been described in detail in Ref.~\citenum{uhlmann:04}. 
It has been previously used for the direct
numerical simulation of various particulate flow configurations  
\citep{uhlmann:08a,chan-braun:11a,villalba:12,kidanemariam:13,
chan-braun:13,uhlmann:13a,uhlmann:14a,kidanemariam:14a,kidanemariam:14b}.

The incompressible Navier-Stokes equations are solved by a fractional
step approach with an implicit treatment of the viscous terms
(Crank-Nicolson) and a low-storage, three-step Runge-Kutta scheme for
the non-linear terms. The spatial discretization employs second-order
central finite-differences on a staggered mesh which is uniform and
isotropic (i.e.\ $\Delta x=\Delta y=\Delta z=cst.$). 

The no-slip condition at the surface of moving solid particles is
imposed by means of the specifically designed immersed boundary
technique of Uhlmann\citep{uhlmann:04}. 
This gives rise to an additional volume force term in the momentum
equation which we denote as $\mathbf{f}^{(ibm)}$ 
(defined in appendix~\ref{sec-app-time-scheme}). 
The appendix also gives the details of the presently chosen temporal
integration of the Navier-Stokes equations in the presence of solid
particles and artificial turbulence forcing. 
Furthermore, in order to be able to obtain a statistically stationary
state when simulating settling bodies in a triply-periodic domain, it
is necessary to compensate the average force exerted by the particles
upon the fluid\cite{fogelson:88,hoefler:00} 
(cf.\ appendix~\ref{sec-app-time-scheme}). 
As a consequence, the particle-related contribution to the volume
force field (\ref{equ-total-volume-force}) which is added to the 
right-hand-side of the momentum equation
(\ref{equ-navier-stokes-general-mom}) reads: 
\begin{equation}\label{equ-def-particle-related-vol-force}
  \mathbf{f}^{(p)}(\mathbf{x},t)
  =
  \mathbf{f}^{(ibm)}(\mathbf{x},t)
  -
  \langle\mathbf{f}^{(ibm)}\rangle_{\Omega}(t)
  \,,
\end{equation}
where $\langle\,\rangle_\Omega$ indicates an average over the entire
computational domain comprising the fluid and the solid phase (cf.\
definition in appendix~\ref{sec-app-avg-ops}). 

On the other hand, the motion of the particles is computed from the
Newton equations for linear and angular motion of rigid bodies, driven
by buoyancy, hydrodynamic force/torque and contact forces (in case of
collisions). 
Since the particle suspensions under consideration here are dilute,
collisions are occurring infrequently. Therefore, in the present work
they are treated by a simple repulsive force mechanism
\citep{glowinski:99} formulated such as to keep colliding particles
from overlapping non-physically. In the case of dense particle
arrangements, the discrete element method of
Ref.~\citenum{kidanemariam:14a} can be employed instead. 
\subsection{The basic kinetic energy balance}
\label{sec-num-tke-balance}
Defining the instantaneous kinetic energy
$E_k(\mathbf{x},t)=\mathbf{u}\cdot\mathbf{u}/2$, 
we can derive its transport equation from the momentum equation 
(\ref{equ-navier-stokes-general-mom}).  
Substituting (\ref{equ-total-volume-force}) and
(\ref{equ-def-particle-related-vol-force}), 
and integrating the result over a triply-periodic spatial domain
$\Omega$ yields the following evolution equation:
\begin{equation}\label{equ-ek-box-avg}
  \frac{\mbox{d}}{\mbox{d}t}\langle E_k\rangle_\Omega
  =
  -\varepsilon_\Omega
  +\Psi^{(t)}
  +\Psi^{(p)}
  \,,
\end{equation}
where we have defined the instantaneous box-averaged dissipation rate
$\varepsilon_\Omega$, viz. 
\begin{equation}\label{equ-def-eps-box-avg}
  \varepsilon_\Omega(t)=2\nu\langle S_{ij}S_{ij}\rangle_\Omega
  \,,
\end{equation}
with $S_{ij}=(u_{i,j}+u_{j,i})/2$, the work done by the turbulence 
forcing, $\Psi^{(t)}$, 
\begin{equation}\label{equ-def-energy-input-box-avg}
  \Psi^{(t)}(t)
  =
  \langle\mathbf{u}\cdot\mathbf{f}^{(t)}\rangle_\Omega
  \,,
\end{equation}
as well as the fluid-particle coupling term, $\Psi^{(p)}$, 
\begin{equation}\label{equ-def-two-way-coupling-box-avg}
  \Psi^{(p)}(t)
  =
  \langle\mathbf{u}\cdot\mathbf{f}^{(ibm)}\rangle_\Omega
  -
  \langle\mathbf{u}\rangle_\Omega
  \cdot
  \langle\mathbf{f}^{(ibm)}\rangle_\Omega
  \,. 
\end{equation}
The fluid-particle coupling term can be brought into a
physically more insightful form by making use of the Newton-Euler
equations for rigid body motion. This reformulation will be deferred 
until section~\ref{sec-two-phase-time-evol} below. 

Let us now define some further quantities which will be used in the
following. 
The fluid phase velocity is decomposed into an average (over the
region filled with fluid) and a fluctuation, viz.
\begin{equation}\label{equ-def-decomp-fluid-only}
  \mathbf{u}(\mathbf{x},t)
  =
  \langle\mathbf{u}\rangle_{\Omega_f}(t)
  +
  \mathbf{u}^\prime(\mathbf{x},t)
  \,,
\end{equation}
where the averaging operator $\langle\cdot\rangle_{\Omega_f}$ is defined in 
appendix~\ref{sec-app-avg-ops}. 
Then the kinetic energy of the fluctuations, averaged over the fluid
phase, is defined as 
\begin{equation}\label{equ-ek-fluct-fluid-only}
  k(t)
  =
  \frac{1}{2}
  \langle \mathbf{u}^\prime\cdot \mathbf{u}^\prime \rangle_{\Omega_f} 
  =
  \frac{3}{2}
  \,u_{rms}^2(t)
  \,,
\end{equation}
where the characteristic velocity scale $u_{rms}(t)$ has been defined
simultaneously. 
We also define a dissipation rate averaged over the fluid phase, viz. 
\begin{equation}\label{equ-def-eps-box-avg-fluid-only}
  \varepsilon(t)=2\nu\langle S_{ij}^\prime S_{ij}^\prime\rangle_{\Omega_f}
  \,.
\end{equation}
From these quantities we can compute the Kolmogorov length scale 
$\eta=(\nu^3/\varepsilon)^{1/4}$, 
the Taylor micro-scale 
$\lambda=(15\nu u_{rms}^2/\varepsilon)^{1/2}$, 
the large-eddy length-scale 
$L=k^{3/2}/\varepsilon$, 
the large-eddy turn-over time 
$T_e=u_{rms}^2/\varepsilon$, 
the Kolmogorov time 
$\tau_\eta=(\nu/\varepsilon)^{1/2}$ 
and the vorticity fluctuation amplitude 
$\omega_{rms}=(\varepsilon/\nu)^{1/2}$. 
\section{Simulation of single-phase turbulence}
\label{sec-single-phase}
The aim of the present section is to provide a validation of the EP
forcing technique in a finite-difference context, and to discuss its
application to elongated computational domains. 
\subsection{Parameters, setup and flow characterization}
\label{sec-single-phase-setup}
The single-phase simulations were run with the numerical parameters
specified in table~\ref{tab-single-phase-num-param}. Let us first
focus on cases A and B which feature cubic domains with $256^3$ and
$512^3$ grid nodes, respectively (the cases AL, BL featuring elongated
boxes will be discussed in \S~\ref{sec-single-phase-elongated} below). 
The forcing parameters $\kappa_f$, $T_L$, and $\varepsilon^{*}$ are
chosen such that a very good small-scale resolution $\eta/\Delta x$
can be obtained, 
while a reasonable ratio between the box-size ${\cal L}_x$
and the integral length scale can be maintained (in the worst case
this ratio is approximately equal to $7$).  
With the given moderate number of grid points, the target Reynolds
number $Re_\lambda=\lambda u_{rms}/\nu$ 
(based upon the Taylor micro-scale 
$\lambda$) 
is approximately 60 and 140, respectively. 
The procedure for choosing the forcing parameters which yield the
target Reynolds number and small-scale resolution is specified in
appendix~\ref{sec-single-phase-parametrization}. 

Some simulations were first started on a coarser
grid in order to speed up the computation during the initial 
transient. They were subsequently interpolated onto the final grid,
and the simulations were continued. Averages have been computed (on
the finest grid) in the statistically stationary regime over intervals
larger than $25$ large-eddy time scales ($T_e$), cf.\
Table~\ref{tab-single-phase-phys-param}.  
The table
also shows the obtained Reynolds
number, length scales, the non-dimensional dissipation rate as well as
the skewness of the velocity gradients derived from these statistical
results. 
\begin{table} %
  \centering
  \renewcommand{\arraystretch}{1.25}
  \setlength{\tabcolsep}{1ex}
  \begin{tabular}{lllcccc}
    \hline\\[-1.ex]
    \multicolumn{1}{c}{case} & 
    \multicolumn{1}{c}{$\Omega=\mathcal{L}_x\times\mathcal{L}_y\times\mathcal{L}_z$} &
    $N_x \times N_y \times N_z$ 
    & $\kappa_f/\kappa_{x,1}$ 
    & $N_F$ 
    & $T_L\nu/{\cal L}_x^2$ 
    & $\varepsilon^{*}{\cal L}_x^4/\nu^3$
    \\[.5ex] 
    \hline\\[-1.ex]
    A & $\mathcal{L}_x=\mathcal{L}_y=\mathcal{L}_z$ & $256^3$ 
    & $2.3$ & $56$ &
    $2.12\cdot10^{-4}$ & $3.72\cdot10^{7}$
    \\
    B & $\mathcal{L}_x=\mathcal{L}_y=\mathcal{L}_z$ & $512^3$ 
    & $2.5$ & $80$ &
    $5.94\cdot10^{-5}$ & $3.52\cdot10^{9}$
    \\
    AL & $\mathcal{L}_x=\mathcal{L}_y=\mathcal{L}_z/2$ & $256^2 \times 512$ 
    & $2.3$ & $56$ &
    $2.12\cdot10^{-4}$ & $3.72\cdot10^{7}$
    \\
    BL & $\mathcal{L}_x=\mathcal{L}_y=\mathcal{L}_z/2$ & $512^2\times 1024$ 
    & $2.5$ & $80$ &
    $5.94\cdot10^{-5}$ & $3.52\cdot10^{9}$
    \\
    \hline
  \end{tabular}
  \caption{Imposed parameters in the present single-phase simulations. The
    size of the computational domain $\Omega$ is given in terms of its
    linear dimensions ${\cal L}_x$, ${\cal L}_y$, ${\cal L}_z$. The
    number of grid points along the coordinate directions are denoted by
    $N_x$, $N_y$, $N_z$. The forcing cut-off wavenumber $\kappa_f$ is
    normalized by the smallest wavenumber in the $x$-coordinate
    direction ($\kappa_{x,1}$); 
    $N_F$ denotes the total number of Fourier modes which
    are forced.
  }
  \label{tab-single-phase-num-param}
\end{table}  
\begin{table} %
  \centering
  \renewcommand{\arraystretch}{1.25}
  \setlength{\tabcolsep}{1ex}
  \begin{tabular}{lrlccccccc}
    \hline\\[-1.ex]
    case
    & \multicolumn{1}{l}{$Re_\lambda$} 
    & $L/{\cal L}_x$ 
    & $\lambda/{\cal L}_x$ 
    & $\eta/{\cal L}_x$ 
    & $\eta/\Delta x$ 
    & $T_e \omega_{rms}$
    & $\varepsilon L_f/u_{rms}^3$
    & $S(u_{,x})$ 
    & $T_{obs}/T_e$ 
    \\[.5ex] 
    \hline\\[-1.ex]
    A
    & $65.5$      & $0.5970$     & $0.0744$ 
    & $4.673 \cdot10^{-3}$      & $1.20$      & $16.92$
    & $1.3378$
    & $-0.5109$ 
    & $79.7$
    \\[.5ex] 
    B
    & $143.0$      & $0.5665$      & $0.0323$ 
    & $1.374 \cdot10^{-3}$      & $0.704$      & $37.26$
    & $1.2971$
    & $-0.5341$ 
    & $80.5$
    \\[.5ex] 
    AL
    & $65.0$     & $0.5934$      & $0.0743$ 
    & $4.678 \cdot10^{-3}$      & $1.20$      & $16.93$
    & $1.3461$
    & $-0.5064$ 
    & $33.9$
    \\[.5ex] 
    BL
    & $141.6$     & $0.5604$      & $0.0323$ 
    & $1.379 \cdot 10^{-3}$      & $0.706$      & $36.57$
    & $1.3114$
    & $-0.5341$ 
    & $60.8$
    \\[.5ex] 
    \hline
  \end{tabular}
  \caption{%
    Physical parameters of the present single-phase
    simulations. $L=k^{3/2}/\varepsilon$ denotes the large-eddy length scale, 
    $\lambda=(15\nu u_{rms}^2/\varepsilon)^{1/2}$ is
    the Taylor micro-scale with $Re_\lambda=\lambda u_{rms}/\nu$ the
    corresponding Reynolds number, 
    $\eta=(\nu^3/\varepsilon)^{1/4}$
    the Kolmogorov length scale, 
    $\Delta x$ the grid width, 
    $T_e=u_{rms}^2/\varepsilon$ the large-eddy turn-over
    time, 
    $L_f=2\pi/\kappa_f$ the forcing length scale, 
    $S(u_{,x})$ the skewness of the velocity derivatives 
    and 
    $T_{obs}$ is the total simulated time.
    The vorticity fluctuation amplitude
    $\omega_{rms}=(\varepsilon/\nu)^{1/2}$ has been used as a
    time-scale for the normalization of $T_e$. 
    All quantities were evaluated from time-averages after the
    statistically stationary regime was reached in the respective
    simulation. 
  }
  \label{tab-single-phase-phys-param}
\end{table}

\begin{figure} %
  \begin{minipage}{3ex}
    \rotatebox{90}{%
      $k/k_{ref}$
    }
  \end{minipage}
  \begin{minipage}{0.45\linewidth}
    \centerline{$(a)$}
    \includegraphics[width=\linewidth]
    {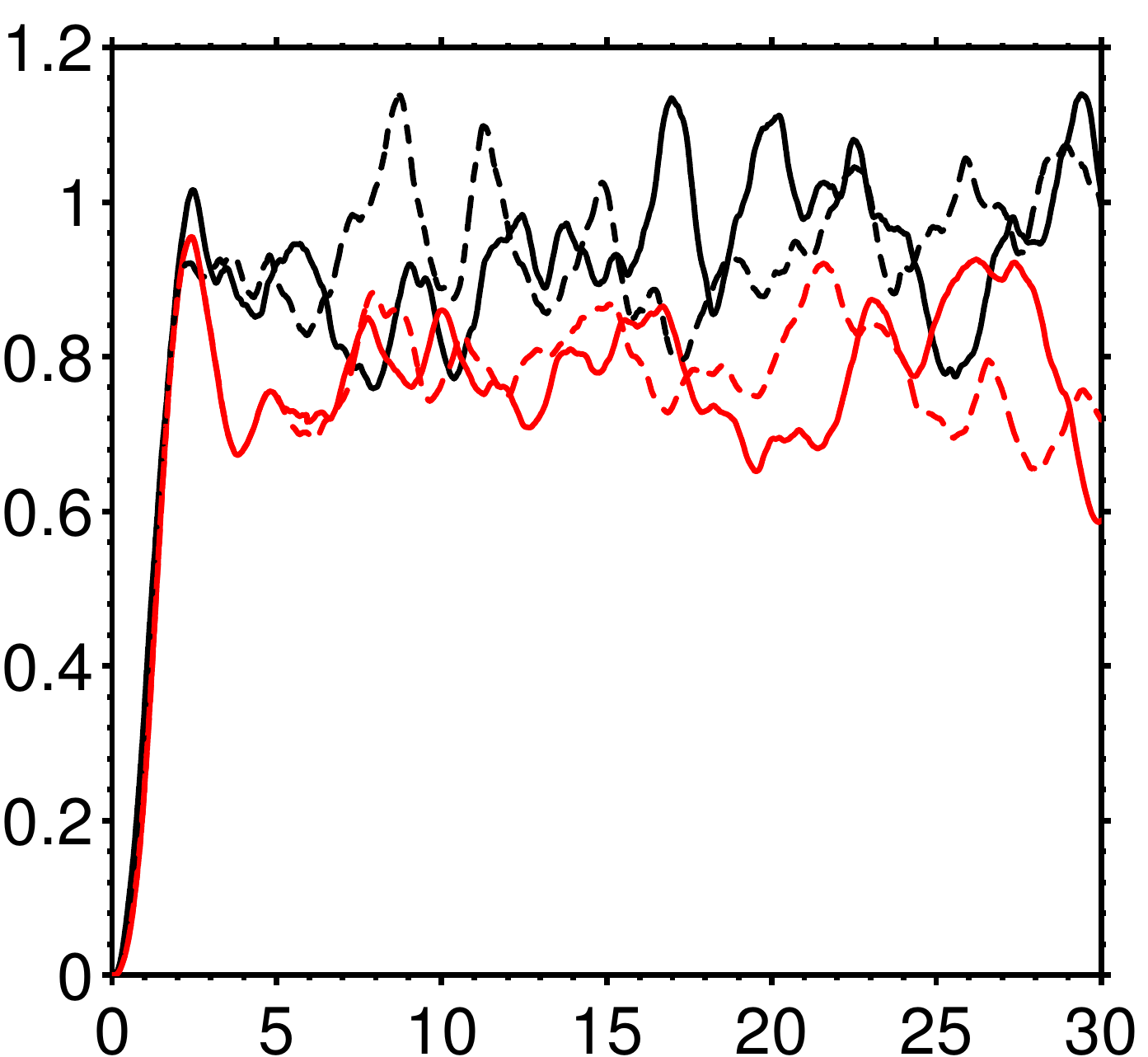}
    \\
    \centerline{$t/T_e$}
  \end{minipage}
  \hfill
  \begin{minipage}{3ex}
    \rotatebox{90}{%
      $\varepsilon/\varepsilon_{ref}$
    }
  \end{minipage}
  \begin{minipage}{0.45\linewidth}
    \centerline{$(b)$}
    \includegraphics[width=\linewidth]
    {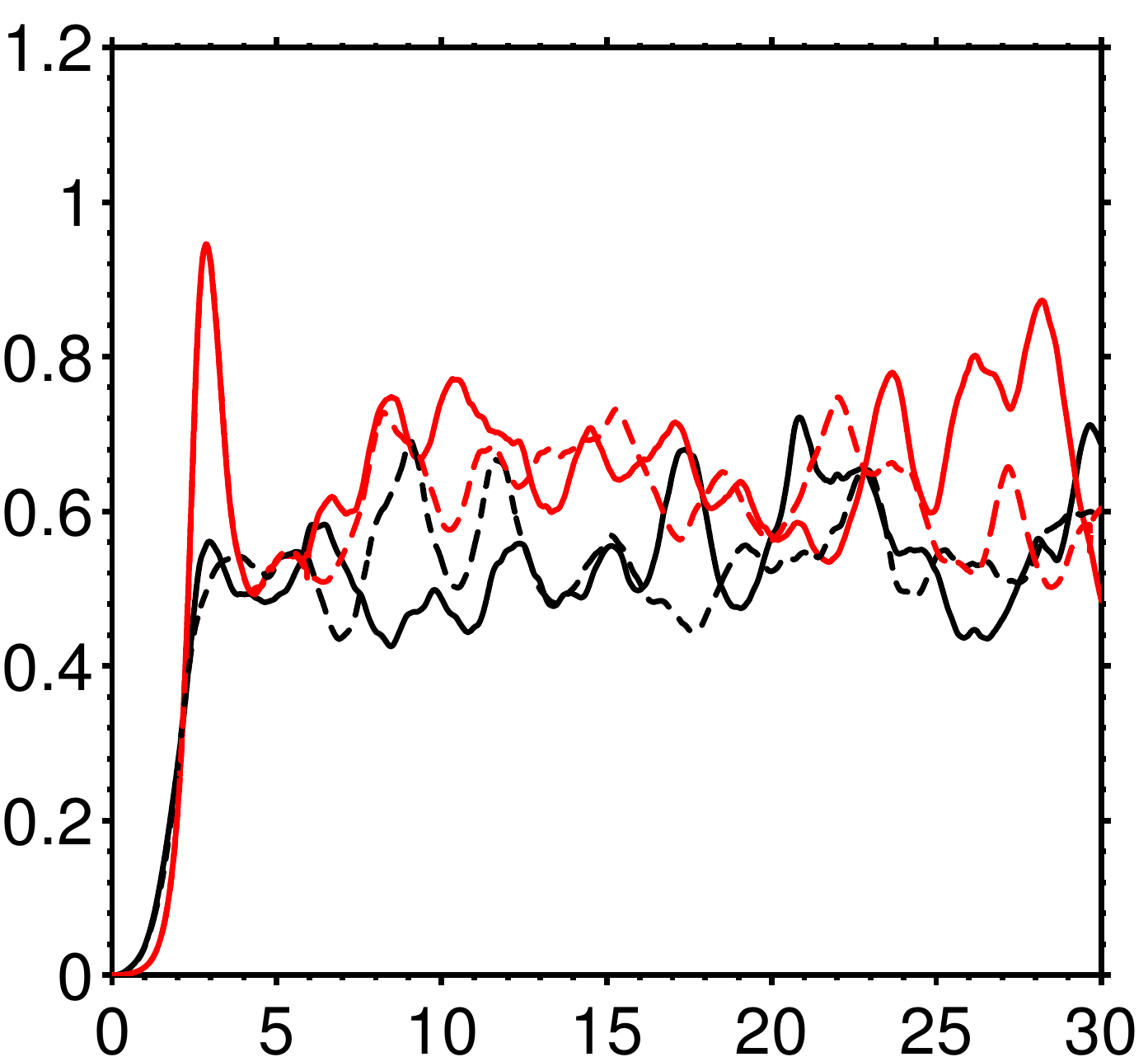}
    \\
    \centerline{$t/T_e$}
  \end{minipage}
  \\
  \centering
  \begin{minipage}{3ex}
    \rotatebox{90}{%
      $Re_\lambda$
    }
  \end{minipage}
  \begin{minipage}{0.45\linewidth}
    \centerline{$(c)$}
    \includegraphics[width=\linewidth]
    {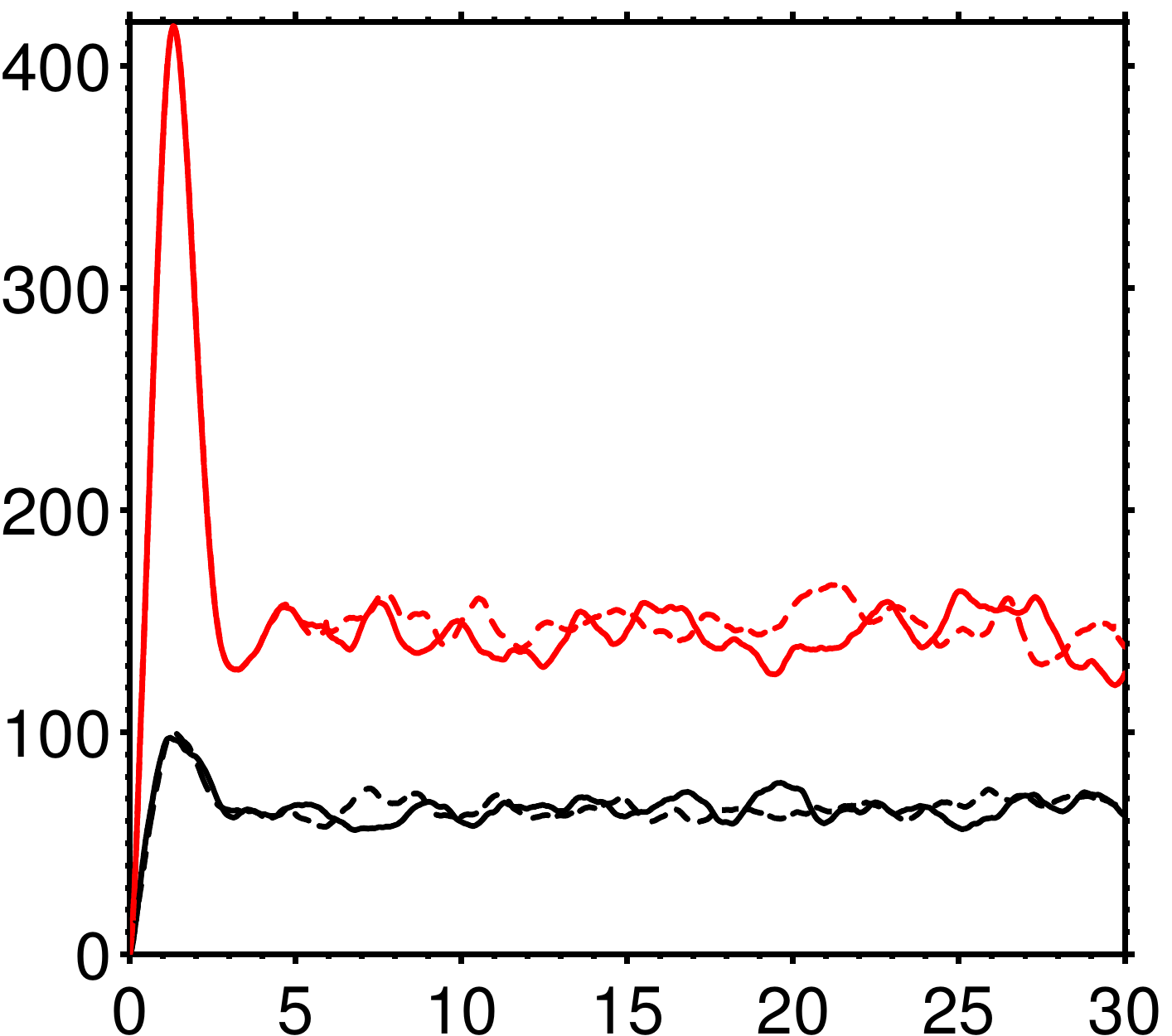}
    \\
    \centerline{$t/T_e$}
  \end{minipage}
  \caption{%
    Temporal evolution of box-averaged quantities: 
    $(a)$ kinetic energy of the fluctuations, 
    $(b)$ dissipation rate, 
    $(c)$ Reynolds number. 
    The lines styles correspond to: 
    {\color{black}\solidthick} case A, 
    {\color{red}\solidthick} case B, 
    {\color{black}\dashed} case AL,
    {\color{red}\dashed} case BL.
    The definition of the reference scales $k_{ref}$ and
    $\varepsilon_{ref}$ is given in the text. 
    Note that only a sub-set of the simulated time interval is shown
    for clarity. 
  }
  \label{fig-single-phase-time-evol-1}
\end{figure}

\begin{figure} %
   \centering
   \begin{minipage}{4ex}
     \rotatebox{90}{%
       $\varepsilon/\langle\varepsilon\rangle_t$, 
       $\Psi^{(t)}/\langle\varepsilon\rangle_t$
     }
   \end{minipage}
   \begin{minipage}{0.49\linewidth}
      \includegraphics[width=\linewidth]
      {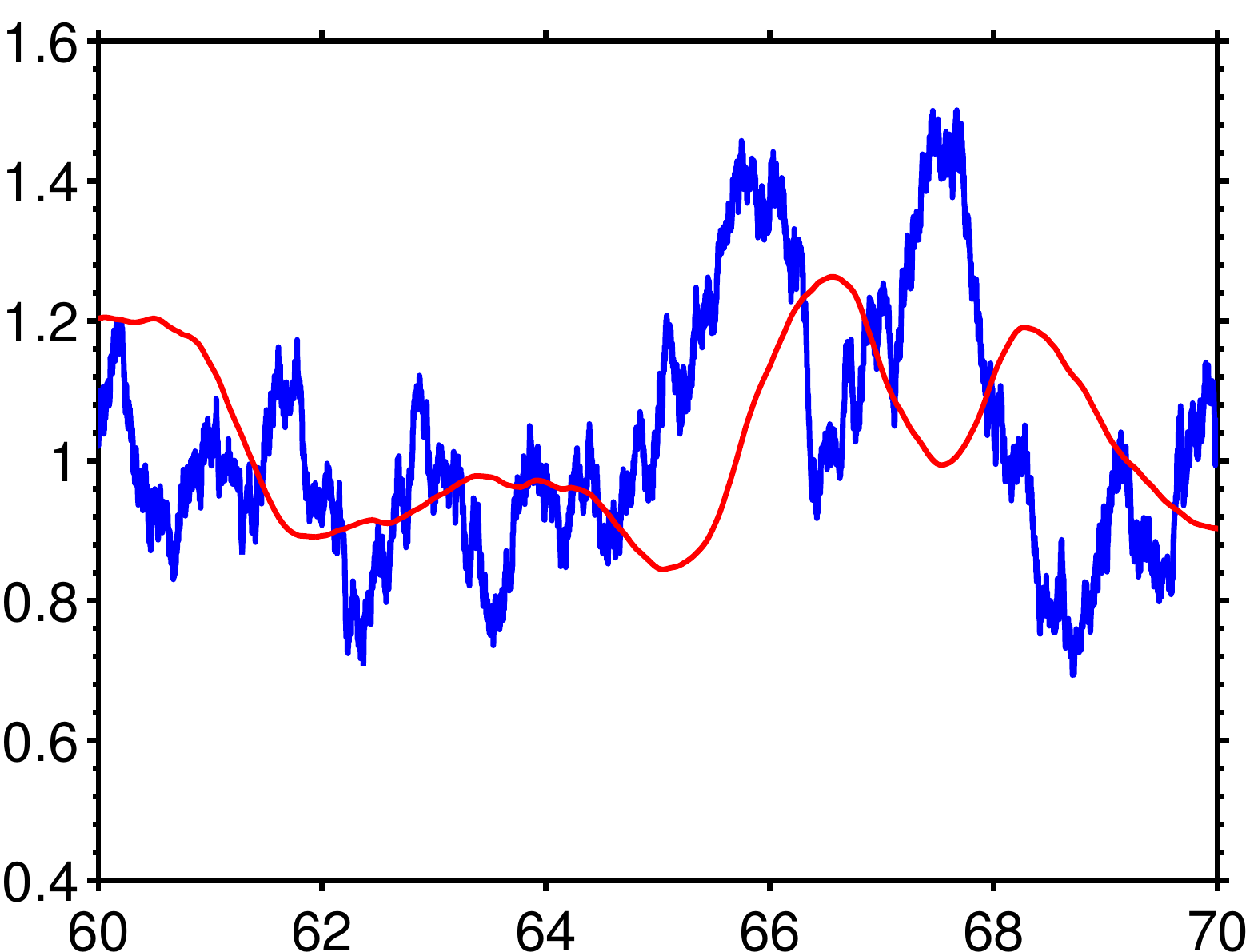}
      \\
      \centerline{$t/T_e$}
   \end{minipage}
   \caption{Time evolution of volume-averaged viscous dissipation
     $\varepsilon$ ({\color{red}\solidthick}) 
      and of turbulence forcing power input $\Psi^{(t)}$
      ({\color{blue}\solidthick}) in case~B, both scaled by the  
      time-average value of the dissipation rate
      $\langle\varepsilon\rangle_t$. 
      Note that an arbitrarily chosen time interval in the
      statistically stationary regime is shown. 
      }
   \label{fig-single-phase-time-evol-2}
\end{figure}

\begin{figure} %
   \centering
   \begin{minipage}{2ex}
     \rotatebox{90}{$z/{\cal L}_x$}
   \end{minipage}
   \begin{minipage}{0.4\linewidth} %
    \centerline{$(a)$}
      \includegraphics[width=\linewidth]
      {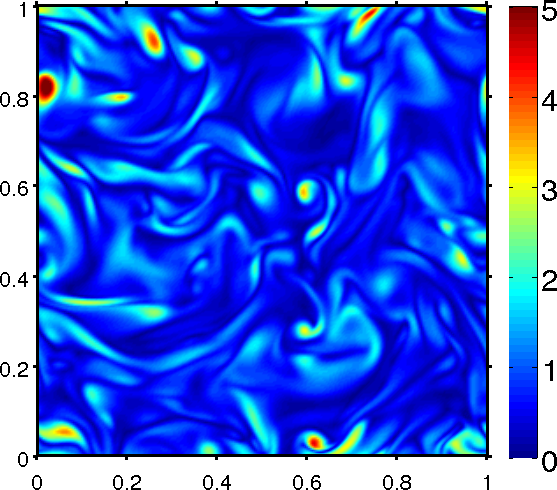}
      \\
      \centerline{$x/{\cal L}_x$}
   \end{minipage}
   \hfill
   \begin{minipage}{2ex}
     \rotatebox{90}{$z/{\cal L}_x$}
   \end{minipage}
   \begin{minipage}{0.4\linewidth} %
    \centerline{$(b)$}
      \includegraphics[width=\linewidth]
      {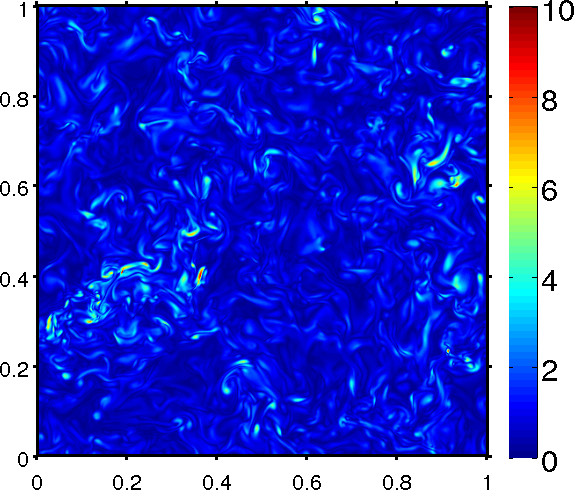}
      \\
      \centerline{$x/{\cal L}_x$}
   \end{minipage}
   \\
   \begin{minipage}{2ex}
     \rotatebox{90}{$z/{\cal L}_x$}
   \end{minipage}
   \begin{minipage}{0.4\linewidth} %
    \centerline{$(c)$}
      \includegraphics[width=\linewidth]
      {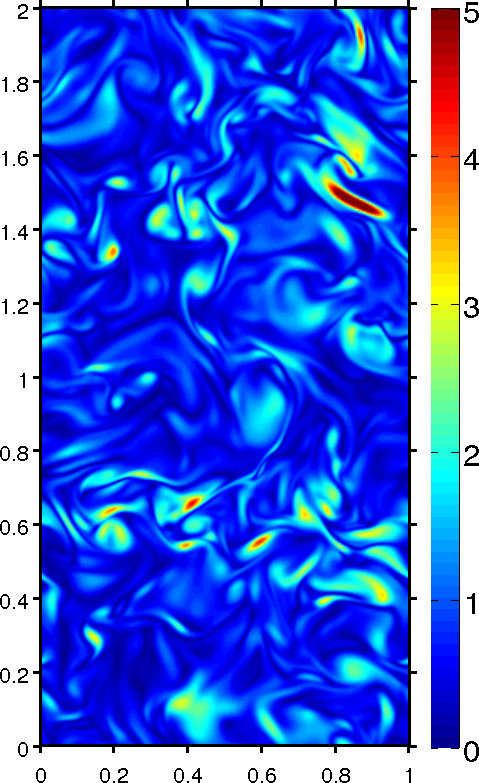}
      \\
      \centerline{$x/{\cal L}_x$}
   \end{minipage}
   \hfill
   \begin{minipage}{2ex}
     \rotatebox{90}{$z/{\cal L}_x$}
   \end{minipage}
   \begin{minipage}{0.4\linewidth} %
    \centerline{$(d)$}
      \includegraphics[width=\linewidth]
      {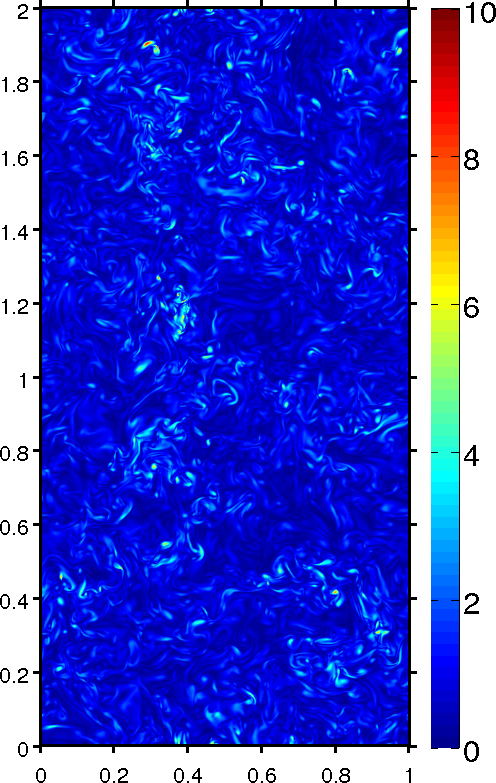}
      \\
      \centerline{$x/{\cal L}_x$}
   \end{minipage}
   \caption{Snapshots of the modulus of vorticity normalized by
     $\omega_{rms}$, shown in one plane. 
     $(a)$~case~A, 
     $(b)$~case~B, 
     $(c)$~case~AL, 
     $(d)$~case~BL. 
   }
   \label{fig-single-phase-vort-snap}
\end{figure}

\begin{figure} %
   \begin{minipage}{1ex}
     \rotatebox{90}{pdf}
   \end{minipage}
   \begin{minipage}{0.46\linewidth}
    \centerline{$(a)$}
      \includegraphics[width=\linewidth]
      {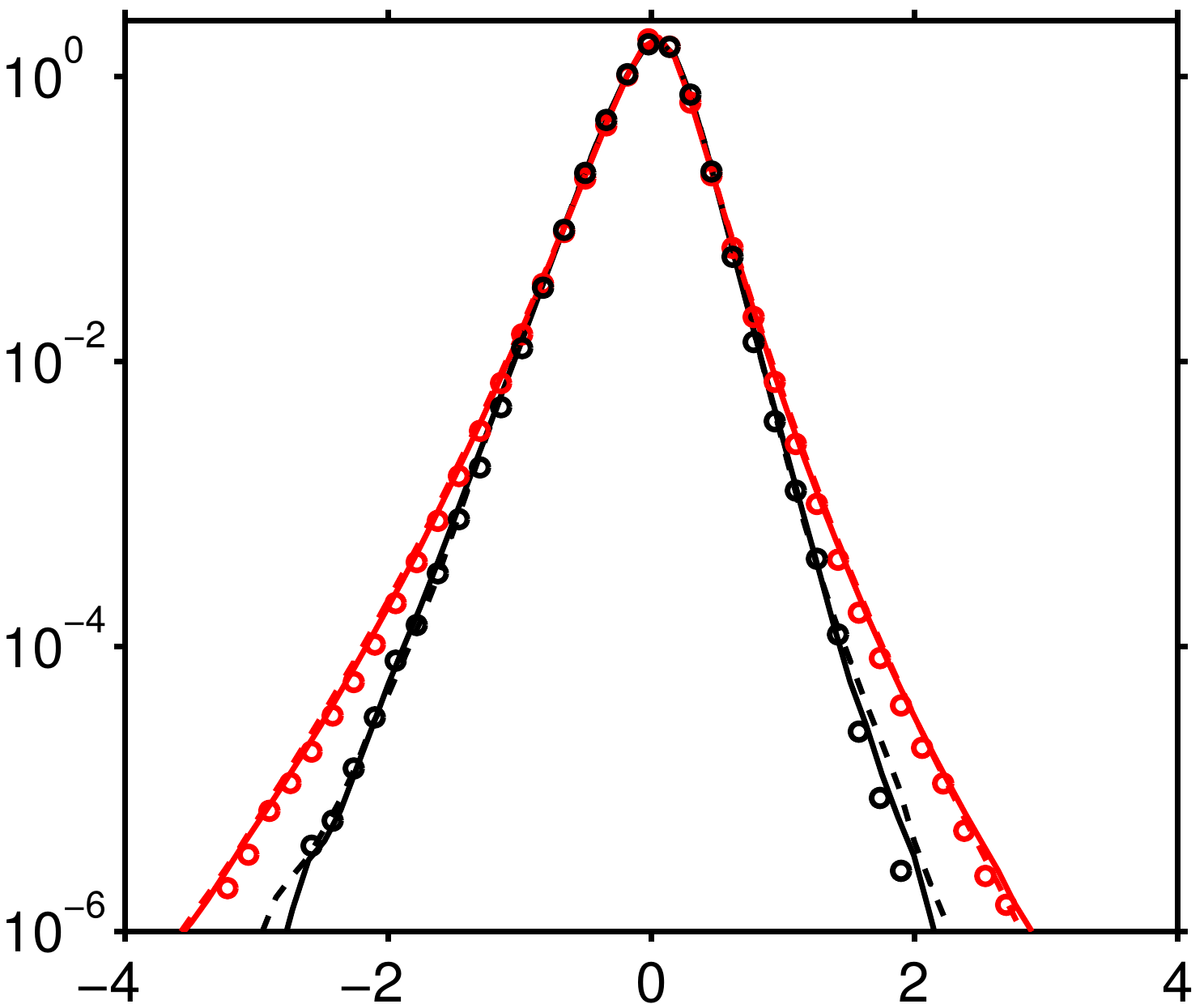}
    \centerline{$(\partial u_\alpha/\partial x_\alpha)/\omega_{rms}$}
   \end{minipage}
   \hfill
   \begin{minipage}{1ex}
     \rotatebox{90}{pdf}
   \end{minipage}
   \begin{minipage}{0.46\linewidth}
    \centerline{$(b)$}
      \includegraphics[width=\linewidth]
      {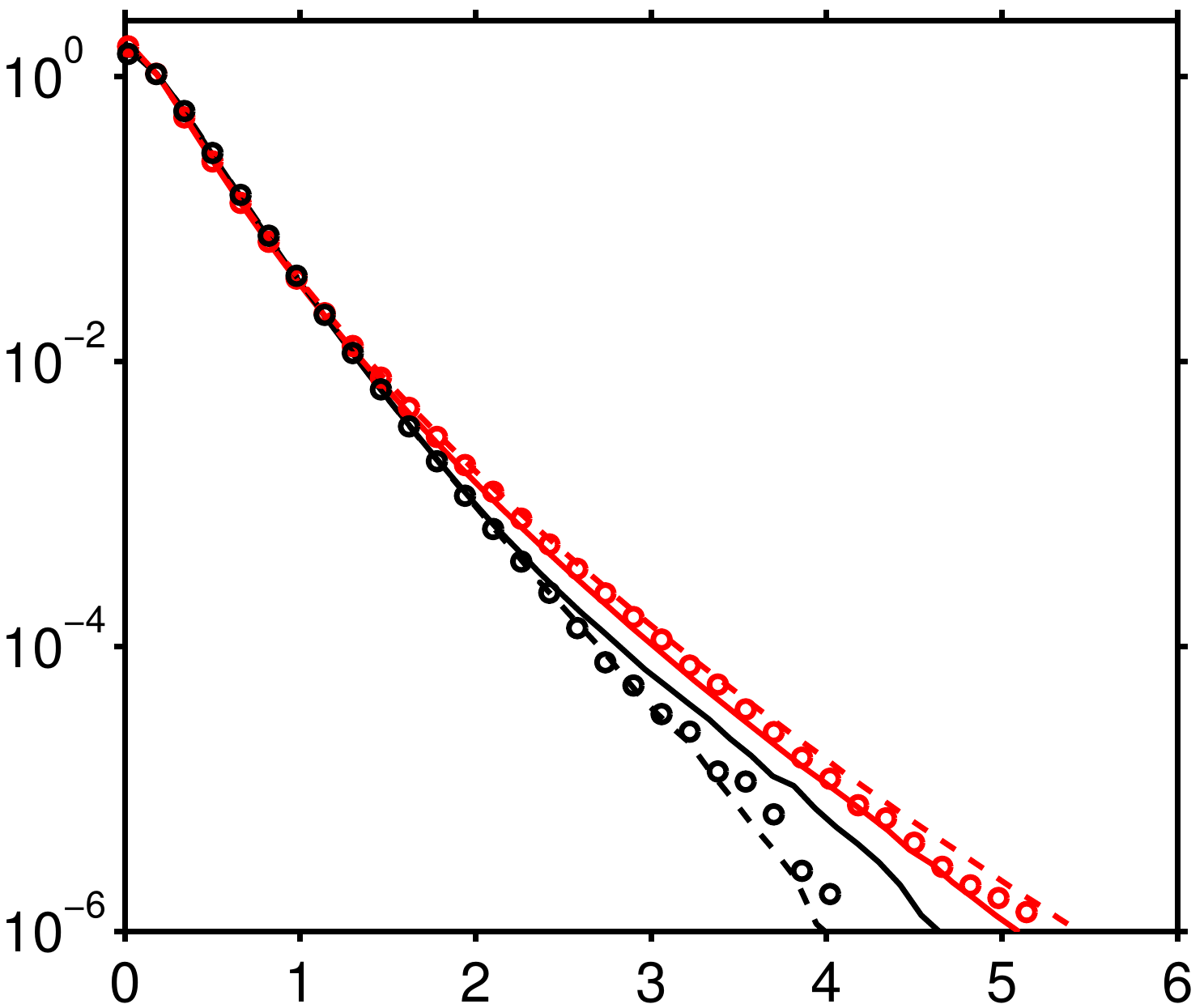}
    \centerline{$(\partial u_\alpha/\partial x_\beta)/\omega_{rms}$}
   \end{minipage}
   \caption{%
     Normalized probability density functions of velocity gradients: 
     $(a)$ longitudinal gradients, 
     $(b)$ transverse gradients. 
     The lines and symbols correspond to: 
     {\color{black}\solidthick}~case~A; 
     {\color{black}\dashed}~case~AL; 
     {\color{red}\solidthick}~case~B; 
     {\color{red}\dashed}~case~BL; 
     `{\color{black}$\boldsymbol{\circ}$}'~Jim\'enez et
     al.\cite{jimenez:93}, $Re_\lambda=62$; 
     `{\color{red}$\boldsymbol{\circ}$}'~Jim\'enez et
     al.\cite{jimenez:93}, $Re_\lambda=142$. 
   }
   \label{fig-single-phase-pdf-uf-derivative}
\end{figure}

The time evolution of box-averaged fluctuation energy $k$ and 
dissipation rate $\varepsilon$ as well as the Reynolds number
$Re_\lambda$ is shown in figure~\ref{fig-single-phase-time-evol-1}. 
Therein we use the forcing parameters $T_L$ and $\varepsilon_T^\ast$
(cf.\ appendix~\ref{sec-single-phase-parametrization}) to construct  
reference values for the fluctuation energy,
$k_{ref}=3T_L\varepsilon_T^\ast/2$, and for the dissipation rate,
$\varepsilon_{ref}=\varepsilon_T^\ast$.  
It can be observed that a statistically stationary state is reached
after roughly $6T_e$  
during which the energy cascade is forming. For
later times the flow state oscillates around well-defined
time-averages for each quantity. 
Figure~\ref{fig-single-phase-time-evol-2} additionally shows for
case~B the only two source terms of the kinetic energy budget
(\ref{equ-ek-box-avg}) which are non-zero in the single-phase
context. It can be seen that the 
energy input $\Psi^{(t)}$ is oscillating on a scale which corresponds
to the imposed parameter value $T_L$ 
($T_L/T_e=0.84$ in case~B).
The dissipation rate data \revision{exhibits}{exhibit} a smoother
temporal evolution, with 
an overall similar shape as the power input, albeit with a certain
time-lag. 
Snapshots of the modulus of vorticity in a slice through the
computational domain are shown in
figure~\ref{fig-single-phase-vort-snap}, 
where the typical intermittent multi-scale patterns featuring various
size eddies as well as very thin filaments can be distinguished.
Let us proceed to a quantitative comparison with respect to reference
data of Jim\'enez et al.\cite{jimenez:93}
(also available in electronic form in Ref.\citenum{agard:345})
which \revision{was}{were} obtained through DNS with the aid of a pseudo-spectral
method.  
Those authors assured energy input to the large scales through a
negative viscosity in a small-wavenumber band; the amplitude was
dynamically adjusted such as to maintain the small-scale
resolution constant. Their series of simulations covers the present
range of parameters, with two simulations featuring nearly
identical Reynolds number values as presently simulated
($Re_\lambda=62$ and $142$). 
Figure~\ref{fig-single-phase-pdf-uf-derivative} shows the probability 
density (p.d.f.) of longitudinal and transverse velocity gradients. It
can be seen that the statistics of the present simulations closely
match the data of Ref.\citenum{jimenez:93} all the way down to extreme
events with six orders of magnitude smaller probability than the
maximum.  
\revision{Note that the p.d.f.\ of the velocity flucutations themselves (figure
omitted) exhibits the well known Gaussian behavior.}{%
Note that the p.d.f.\ of the velocity flucutations themselves (cf.\
appendix~\ref{sec-single-phase-fluid-vel-pdf}) exhibits the well known
Gaussian behavior.} 

\begin{figure} %
   \centering
   \begin{minipage}{1.5ex}
     \rotatebox{90}{$E(\kappa)/(\varepsilon^{2/3}\eta^{5/3})$}
   \end{minipage}
   \begin{minipage}{0.46\linewidth}
    \centerline{$(a)$}
      \includegraphics[width=\linewidth]
      {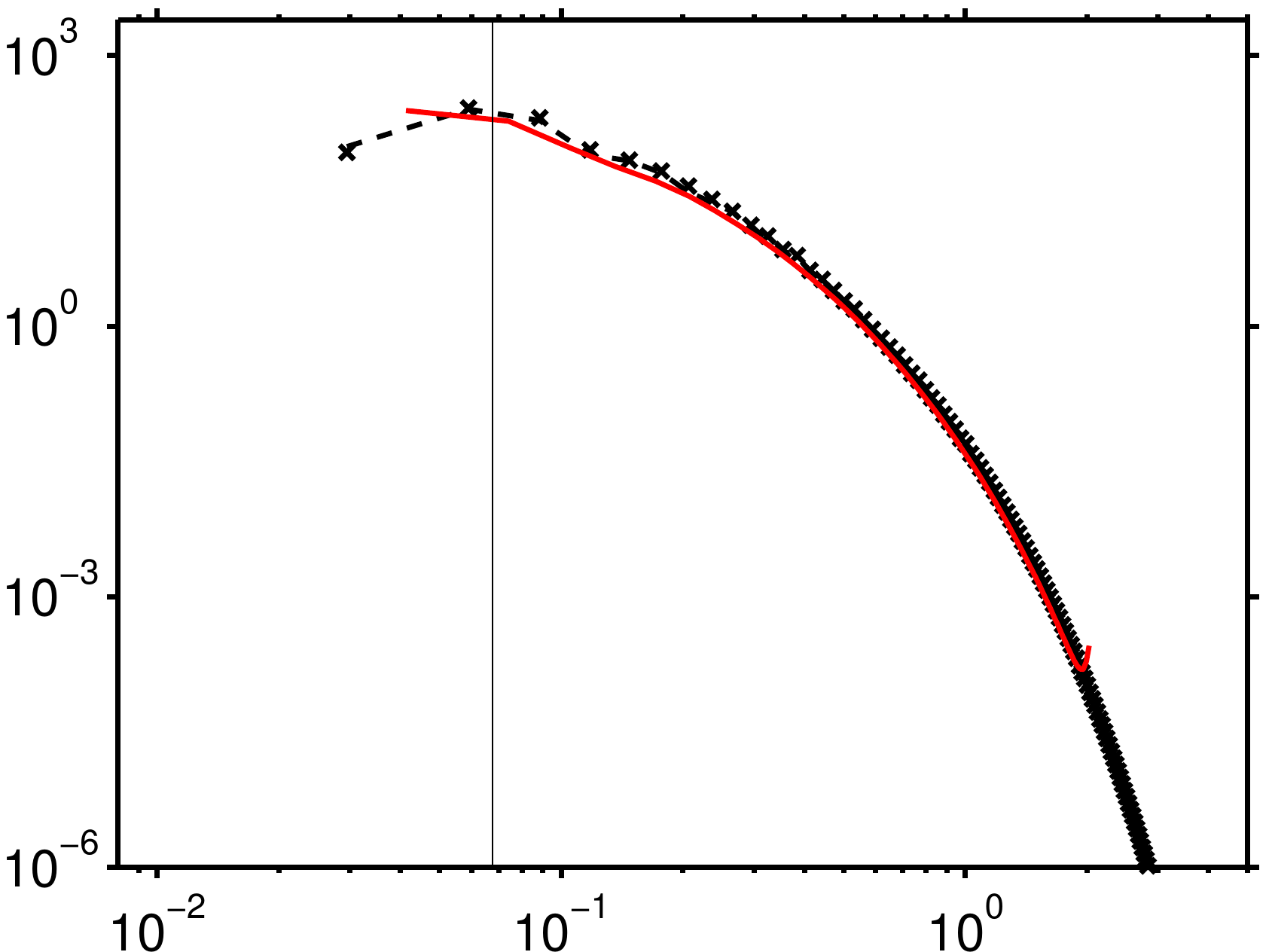}
    \centerline{$\kappa\,\eta$}
   \end{minipage}
   \hfill
   \begin{minipage}{1.5ex}
     \rotatebox{90}{$E(\kappa)/(\varepsilon^{2/3}\eta^{5/3})$}
   \end{minipage}
   \begin{minipage}{0.46\linewidth}
    \centerline{$(b)$}
      \includegraphics[width=\linewidth]
      {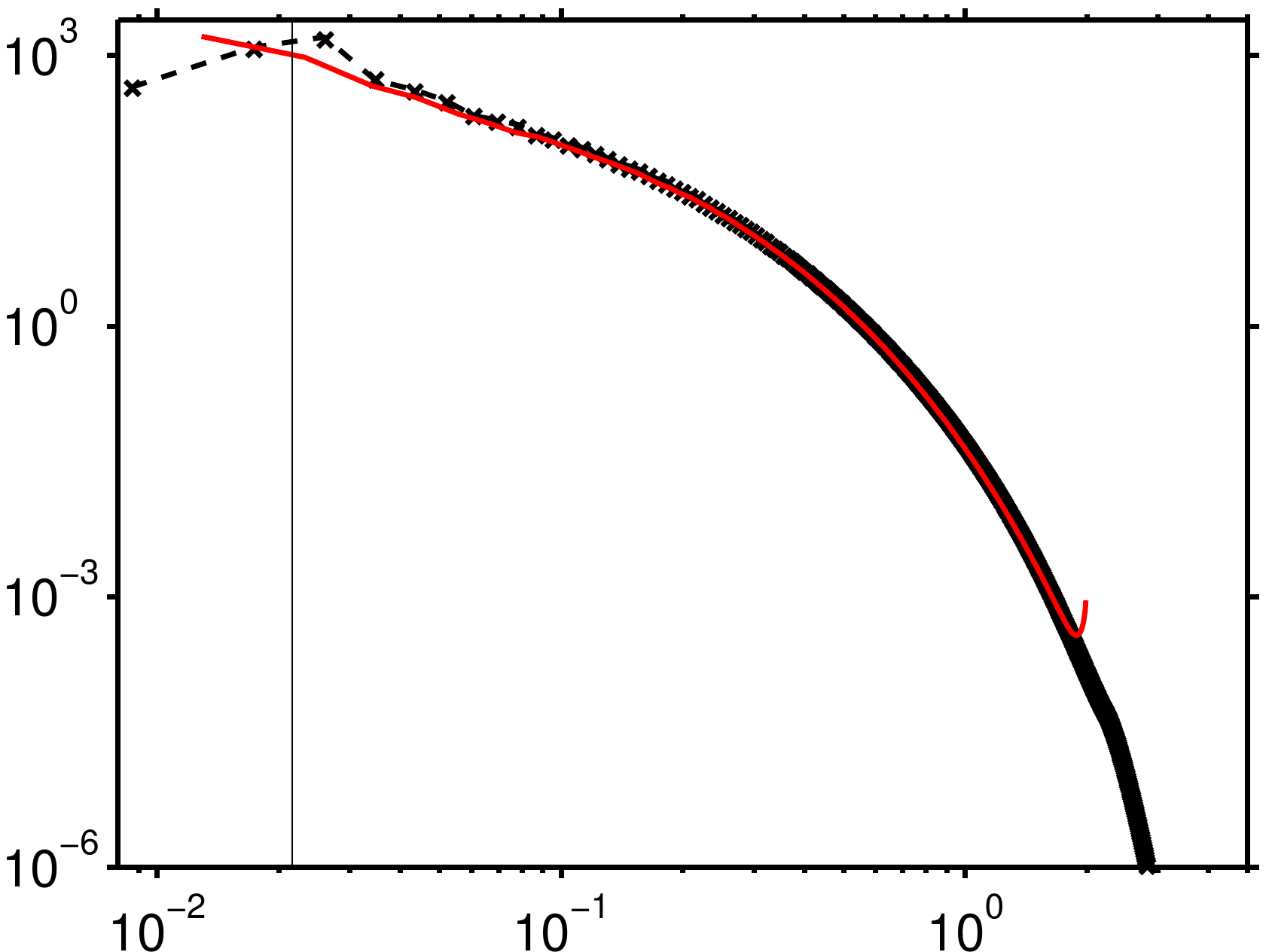}
    \centerline{$\kappa\,\eta$}
   \end{minipage}
   \caption{%
     Three-dimensional energy spectra computed from at least 10
     flow fields in the observation interval.
     Data at $Re_\lambda \approx 60$ \revision{is}{are} shown in $(a)$: 
     {\color{ablack}\dashed}, case A, 
     {\color{ablack}\crossed}, case AL. 
     The graph in $(b)$ collects the data at $Re_\lambda \approx 140$: 
     {\color{ablack}\dashed}, case B, 
     {\color{ablack}\crossed}, case BL.
     The solid lines ({\color{red}\solidthick}) correspond to the
     reference data of Jim\'enez et al.\cite{jimenez:93} at
     $Re_\lambda=62$  and at 
     $Re_\lambda=142$, respectively. 
     The thin vertical lines indicate the respective cut-off
     wavenumber of the turbulence forcing scheme, $\kappa_f$.  
   }
   \label{fig-single-phase-spec-3d}
\end{figure}

\begin{figure} %
   \begin{minipage}{2ex}
     \rotatebox{90}{$E_{\alpha\alpha}(\kappa_\beta)/(\varepsilon^{2/3}\eta^{5/3})$}
   \end{minipage}
   \begin{minipage}{0.46\linewidth}
    \centerline{$(a)$}
      \includegraphics[width=\linewidth]
      {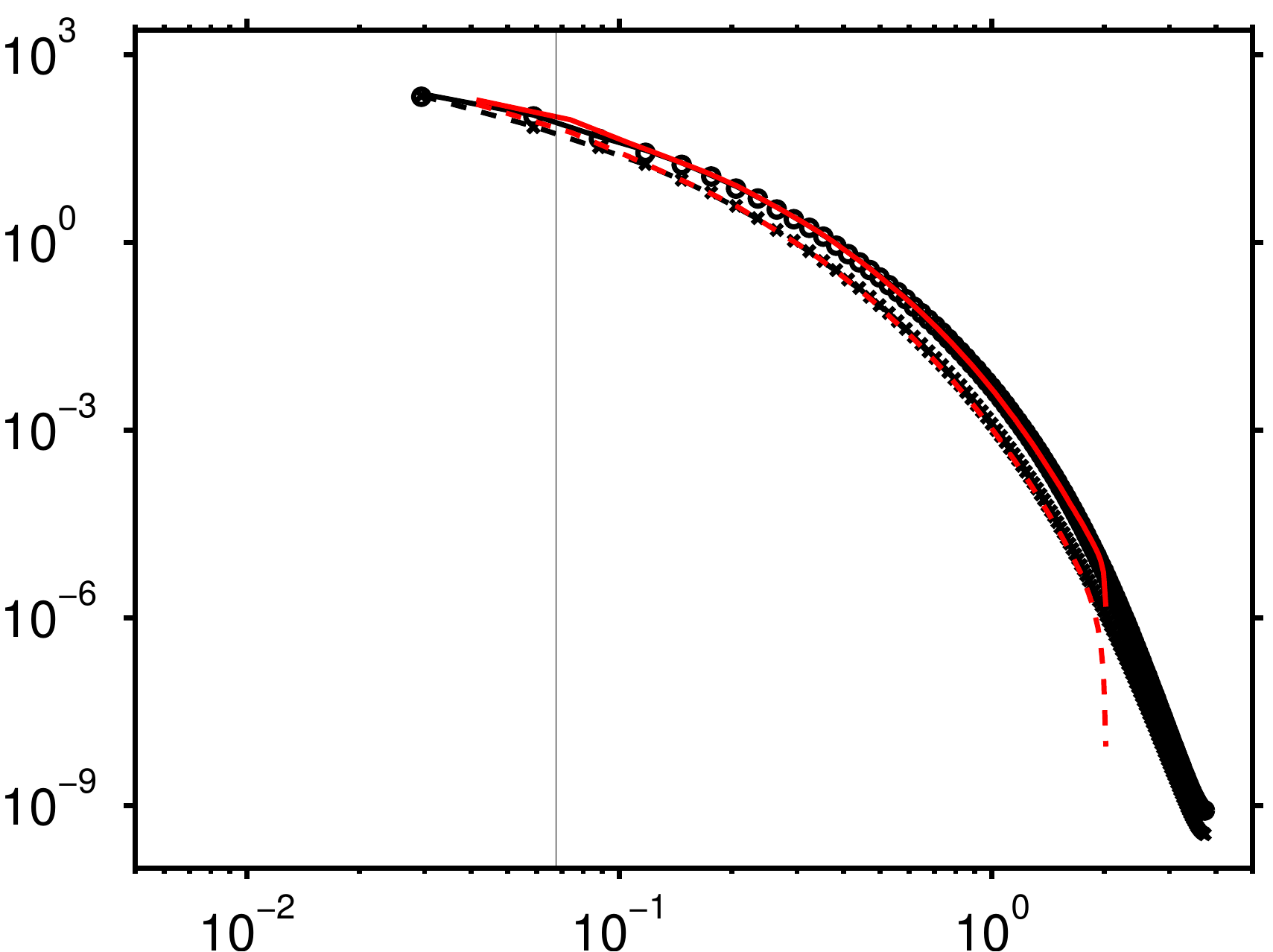}
    \centerline{$\kappa_\beta\,\eta$}
   \end{minipage}
   \hfill
   \begin{minipage}{2ex}
     \rotatebox{90}{$E_{\alpha\alpha}(\kappa_\beta)/(\varepsilon^{2/3}\eta^{5/3})$}
   \end{minipage}
   \begin{minipage}{0.46\linewidth}
    \centerline{$(b)$}
      \includegraphics[width=\linewidth]
      {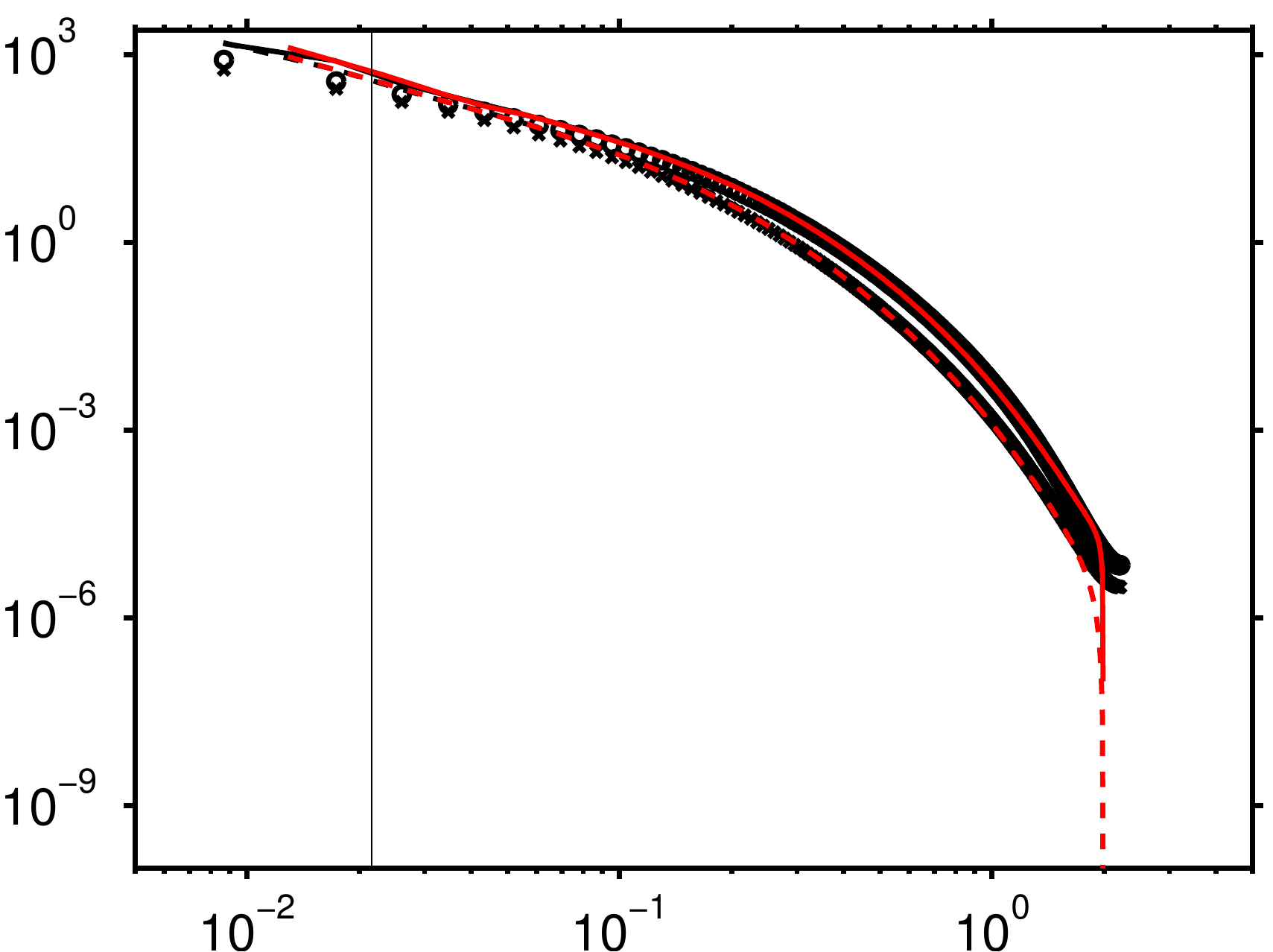}
    \centerline{$\kappa_\beta\,\eta$}
   \end{minipage}
   \caption{%
     One-dimensional energy spectra. 
     Data at $Re_\lambda \approx 60$ \revision{is}{are} shown in $(a)$: 
     {\color{black}\dashed}, case A longitudinal, 
     {\color{black}\solidthick}, case A transverse,
     {\color{black}\crossed}, case AL longitudinal,
     {\color{black}\opencircle}, case AL transverse, 
     {\color{red}\dashed}, Jim\'enez et
     al.\cite{jimenez:93} longitudinal,
     {\color{red}\solidthick}, Jim\'enez et
     al.\cite{jimenez:93} transverse.
     The graph in $(b)$ collects the data at $Re_\lambda \approx 140$: 
     {\color{black}\dashed}, case B longitudinal, 
     {\color{black}\solidthick}, case B transverse,
     {\color{black}\crossed}, case BL longitudinal,
     {\color{black}\opencircle}, case BL transverse,
     {\color{red}\dashed}, Jim\'enez et
     al.\cite{jimenez:93} longitudinal,
     {\color{red}\solidthick}, Jim\'enez et
     al.\cite{jimenez:93} transverse.
     The thin vertical lines indicate the respective cut-off
     wavenumber of the turbulence forcing scheme, $\kappa_f$.  
      }
   \label{fig-single-phase-spec-1d}
\end{figure}

The time-averaged energy spectra are shown in
figure~\ref{fig-single-phase-spec-3d} and
\ref{fig-single-phase-spec-1d}. Both three-dimensional and
one-dimensional spectra practically collapse with the data of
Jim\'enez et al.\cite{jimenez:93} at the corresponding Reynolds
number. Only slight differences can be discerned at the smallest
wavenumbers, as can be expected in view of the different forcing
methods employed. 
It should be remarked that the small-scale resolution employed here
(in terms of $\eta/\Delta x$, cf.\
table~\ref{tab-single-phase-phys-param})  
is very good compared to studies which employed similar second-order,
finite-difference based discretization of the Navier-Stokes
equations\cite{lucci:10}. 
\begin{figure} %
  \centering 
   \begin{minipage}{1.5ex}
     \rotatebox{90}{$I(\kappa_1)$}
   \end{minipage}
   \begin{minipage}{0.45\linewidth}
      \includegraphics[width=\linewidth]
      {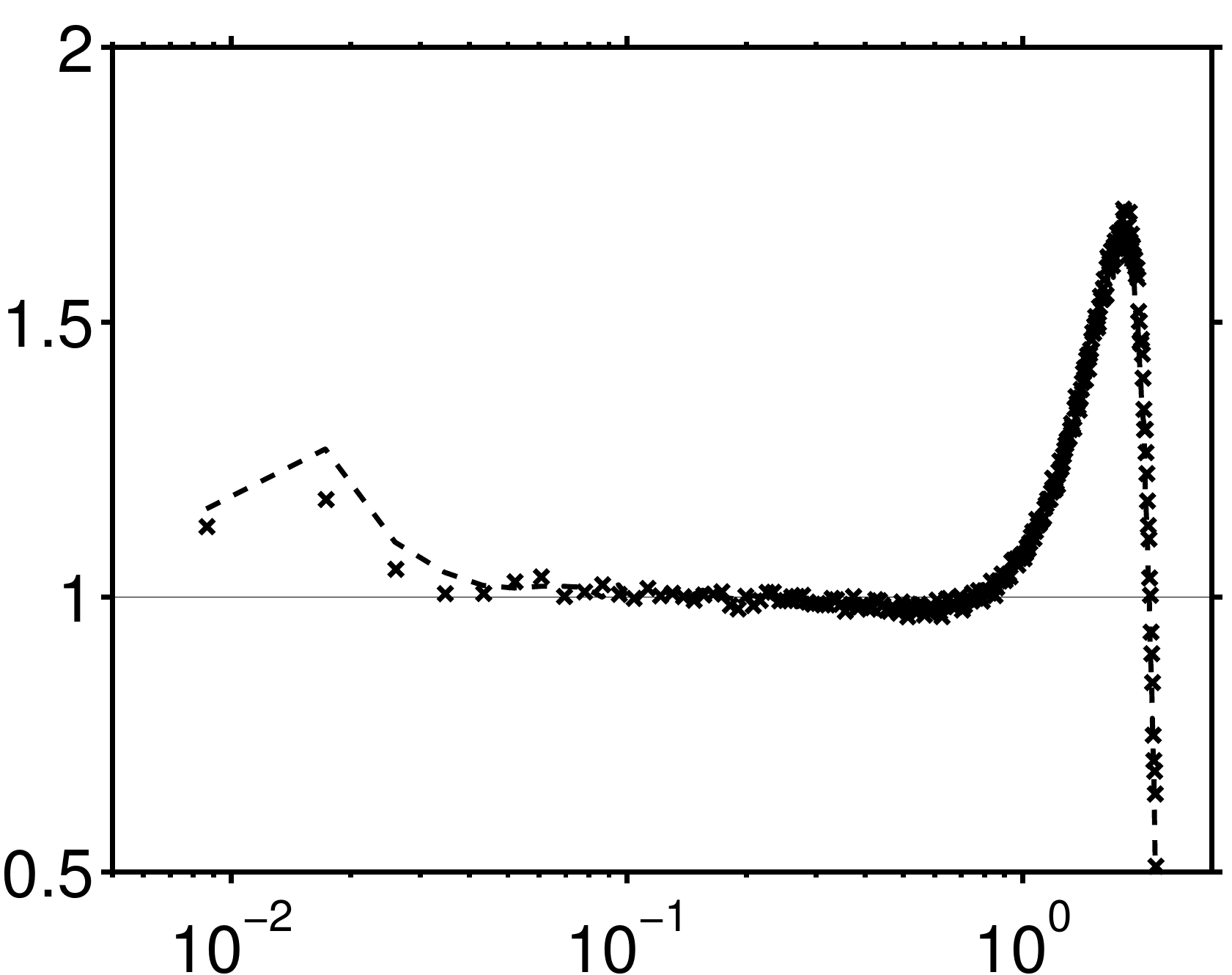}
    \centerline{$\kappa_1\,\eta$}
   \end{minipage}
   \caption{%
     The isotropy parameter $I$ defined in
     (\ref{equ-def-isotropy-param}) as a function of the wavenumber
     $\kappa_1$, shown for case B ({\color{ablack}\dashed}) 
     and BL ({\color{ablack}$\boldsymbol{\times}$}).
      }
   \label{fig-single-phase-isotropy-param-2p-corr}
\end{figure}
\begin{figure} 
   \begin{minipage}{2ex}
     \rotatebox{90}{$\kappa_\alpha E_{\alpha\alpha}(\kappa_\alpha)/(\varepsilon^{2/3}\eta)$}
   \end{minipage}
   \begin{minipage}{0.45\linewidth}
      \includegraphics[width=\linewidth]
      {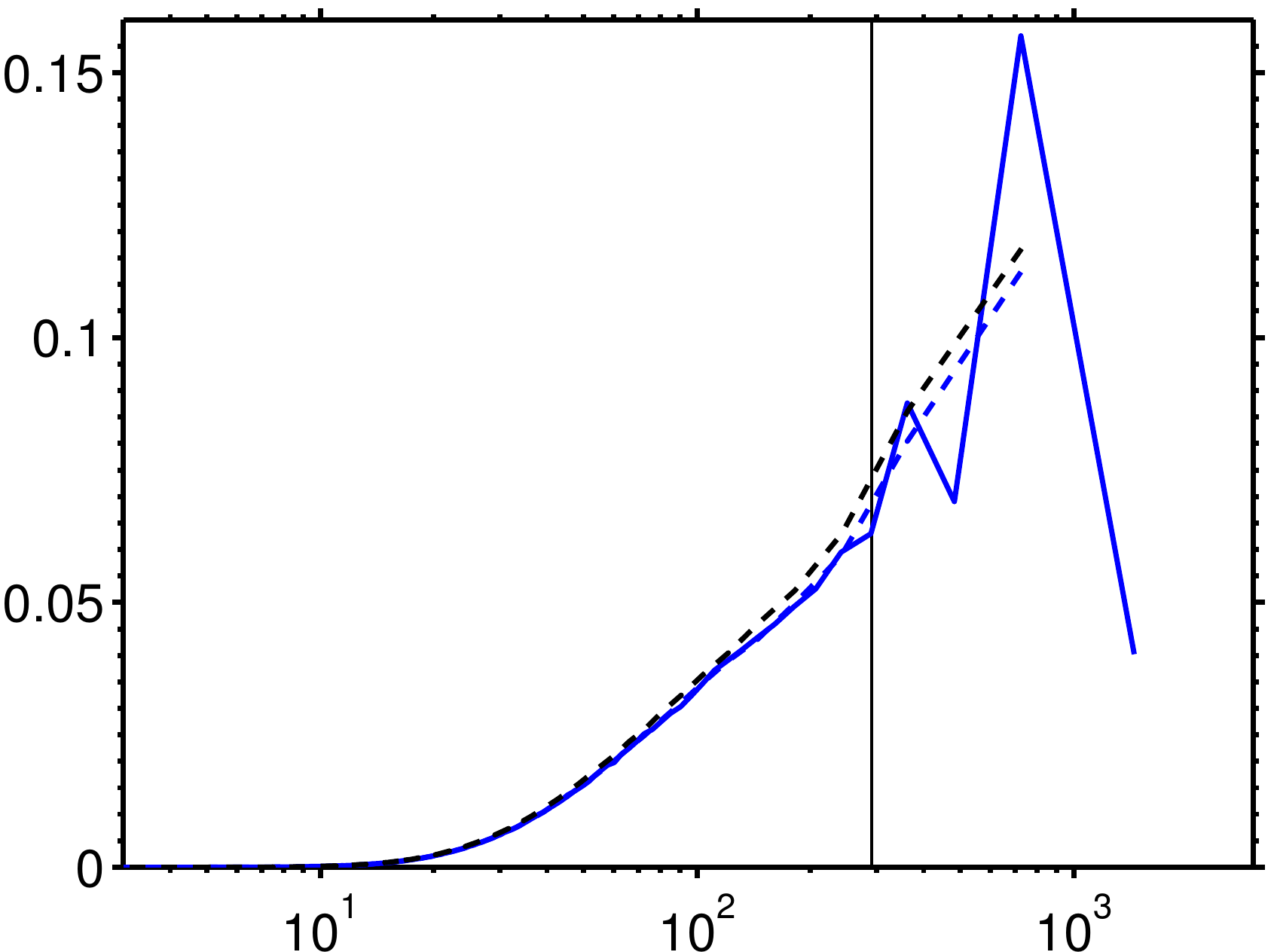}
    \centerline{$\lambda_\alpha/\eta$}
   \end{minipage}
   \hfill
   \begin{minipage}{1ex}
     \rotatebox{90}{$R_{ww}$}
   \end{minipage}
   \begin{minipage}{0.45\linewidth}
      \includegraphics[width=\linewidth]
      {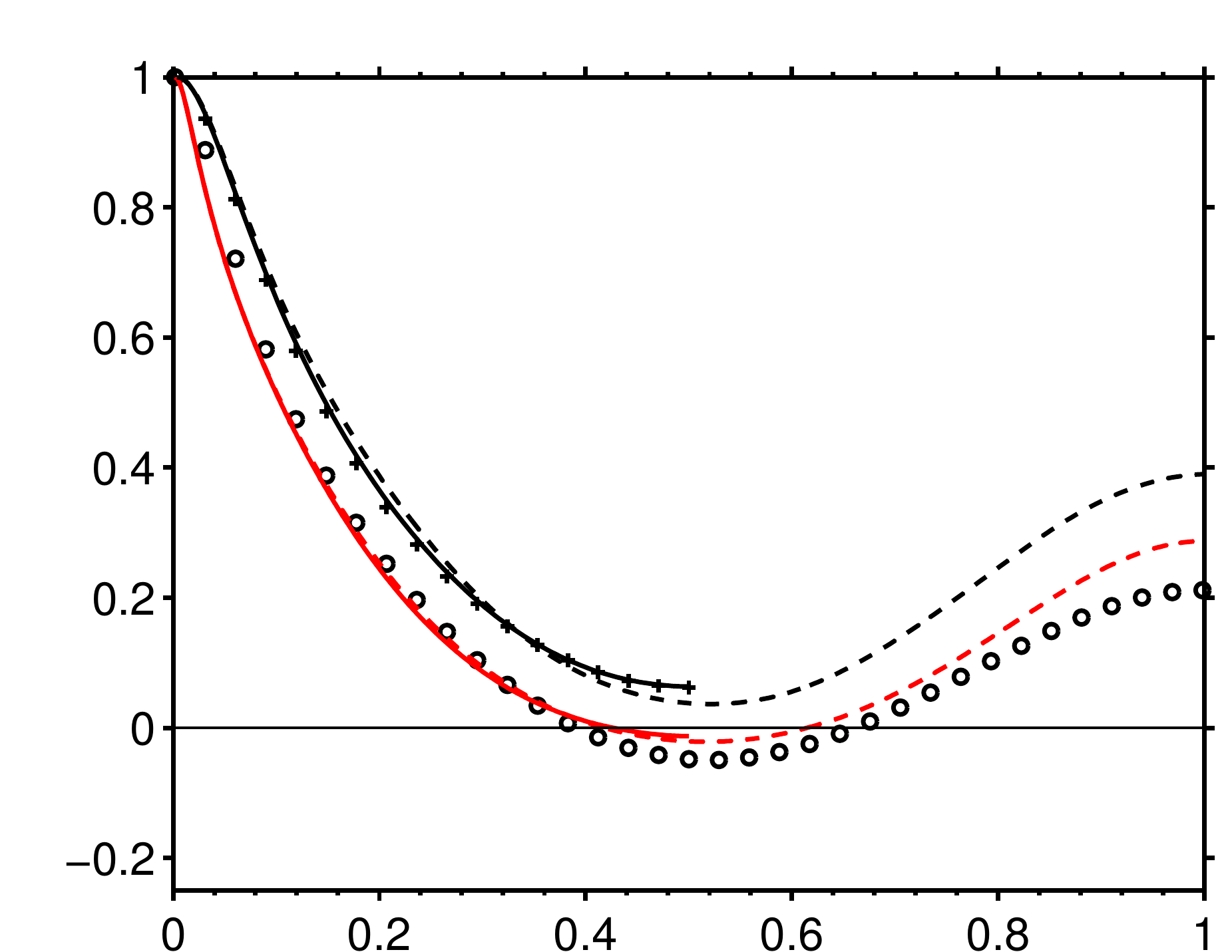}
    \centerline{$r_z/{\cal L}_x$}
   \end{minipage}
   \caption{%
     $(a)$ Premultiplied longitudinal energy spectrum at $Re_\lambda\approx140$,
     similar to some of the information in
     figure~\ref{fig-single-phase-spec-1d}$(b)$. 
     {\color{black}\dashed}~case B; 
     {\color{blue}\dashed}~case BL, short sides ($\alpha=1,2$); 
     {\color{blue}\solid}~case BL, elongated side ($\alpha=3$).
     Note that the horizontal axis shows the wavenumber
     $\lambda_\alpha=2\pi/\kappa_\alpha$. 
     The thin vertical line delimits the wavelength corresponding to
     the forcing cut-off $\kappa_f$. 
     $(b)$ Longitudinal two-point correlation function in the
     $z$-coordinate direction, $R_{ww}(r_z)$. 
     The lines correspond to: 
     {\color{black}\solidthick}~case~A; 
     {\color{black}\dashed}~case~AL; 
     {\color{red}\solidthick}~case~B; 
     {\color{red}\dashed}~case~BL. 
     The two particulate flow cases of section~\ref{sec-two-phase} are
     indicated by the following symbols: 
     `{\color{black}$\boldsymbol{+}$}'~case~A-G0; 
     `{\color{black}$\boldsymbol{\circ}$}'~case~AL-G120. 
      }
   \label{fig-single-phase-spec-ELONG-2p-corr}
\end{figure}
\subsection{Isotropic turbulence in elongated boxes}
\label{sec-single-phase-elongated}
In the case of gravity-induced particle settling in a
vertically-periodic computational domain unphysical results may be 
caused by the two following situations: 
(i) particles encounter their own wakes after performing one trip
around the vertical box-size (${\cal L}_z$), 
and/or 
(ii) particles interact with the same flow structures after one
vertical round-trip. 
Situation (i) occurs if the vertical domain size is smaller than the
axial distance after which the velocity deficit in the wake has
decayed to a negligible amplitude. In some sense the case of
sedimenting particles without background turbulence\cite{uhlmann:14a}
is the ``worst case'', since the decay of the velocity deficit is
accelerated by the presence of background
turbulence\cite{amoura:10,bagchi:04,wu:94b,legendre:06,villalba:12}. 
Situation (ii) occurs when the particle return time (defined as the
vertical domain size divided by the average settling velocity) is
smaller than the largest life-time of all turbulent eddies. 
Fede et al.\cite{fede:07} recommend that the particle return time
should be four times larger than the integral time scale in order to
avoid statistical bias of Lagrangian particle data caused by vertical 
periodicity. 

Therefore, in the case of two-phase flow with settling particles it is
generally desirable to employ computational domains that are
sufficiently long in the vertical direction. 
In practice this requirement leads to the use of non-cubic domains
which are elongated in the direction of gravity. 
At the same time it is desirable to maintain the spatial isotropy of
the forced turbulence. Note that we refer here to the isotropy of the
turbulent flow field in the absence of settling particles, since
the effect of particle settling will immediately break the statistical
symmetry.  

In order to achieve an approximately statistically isotropic flow, we
have formulated the forcing term
$\hat{\mathbf{f}}^{(t)}(\boldsymbol{\kappa})$ in a perfectly isotropic
manner. 
This is automatically assured in the case that the side-lengths of the
cuboidal domain in physical space, 
$({\cal L}_x, {\cal L}_y, {\cal L}_z)$, 
are integer multiples of one of them (here arbitrarily taken as ${\cal
  L}_x$), viz.\ 
${\cal L}_y=n{\cal L}_x$, 
and ${\cal L}_z=m{\cal L}_x$. 
If the definition of the discrete wavenumbers appearing in the forcing
scheme (as defined in equation~\ref{equ-def-wavevector-elt}) is simply
replaced by the following one
\begin{equation}\label{equ-def-wavevector-elt-iso-elong}
  \kappa_{\alpha,i}^{(iso)}
  =
  \frac{2\pi i}{{\cal L}_x}
  \,,
  \quad
  \forall\,\,
  i=-N_f\ldots N_f
  \,,
\end{equation}
the forcing term $\hat{\mathbf{f}}^{(t)}$ 
defined in (\ref{UO_process}-\ref{EswaranPope_Projection}) is 
statistically isotropic. 

In order to verify whether the resulting flow field remains isotropic
in case of a non-cubic computational box, we have performed additional
simulations with twice elongated boxes in the $z$-coordinate direction
(i.e.\ $n=1$, $m=2$) while maintaining the grid width $\Delta x=\Delta
y=\Delta z$. These cases are denoted as ``AL'' and ``BL'',
where all other parameters remain unchanged with respect to the
corresponding simulations in cubic boxes (cf.\
table~\ref{tab-single-phase-num-param}).  
Note that elongation of the domain has introduced new Fourier modes
into the system, but due to (\ref{equ-def-wavevector-elt-iso-elong})
we do not force them. 
As can be seen from table~\ref{tab-single-phase-phys-param} we do not
observe a statistically significant influence of this elongation on
the resulting flow scales. 
The time evolution of energy and dissipation, as well as flow
visualization and pdfs (cf.\ 
figures~\ref{fig-single-phase-time-evol-1}-\ref{fig-single-phase-pdf-uf-derivative}) 
all confirm that we essentially obtain the same flow irrespective of
the elongation of the domain. 

A more stringent test of isotropy can be performed 
by considering the isotropy parameter $I$ introduced in
Ref.\citenum{jimenez:93}, viz. 
\begin{equation}\label{equ-def-isotropy-param}
  I(\kappa_1)
  =
  \frac{
    E_{11}(\kappa_1)
    -\kappa_1\,\partial E_{11}(\kappa_1)/\partial \kappa_1
  }{
    2E_{22}(\kappa_1)
  }
  \,,
\end{equation}
which is equal to unity for all wavenumbers $\kappa_1$ in the case of
a strictly homogeneous-isotropic velocity field
(Ref.\citenum{batchelor:53}, p.\ 50).  
From figure~\ref{fig-single-phase-isotropy-param-2p-corr} it can be seen that
there exists indeed a considerable range of wavenumbers over which the
flow can be judged as isotropic. As in previous studies (e.g.\
Ref.\citenum{jimenez:93}) this excludes the largest scales (those
which are directly forced); 
likewise, the smallest scales start to deviate from the
isotropic value for $\kappa_1\eta\geq1$. 

We now turn to the effect of elongating the box in the $z$-direction
upon the energy spectra. 
Figure~\ref{fig-single-phase-spec-ELONG-2p-corr}$(a)$ shows the
longitudinal one-dimensional energy spectrum in case~BL, where the
average of the data along the short directions ($E_{11}(\kappa_1)$,
$E_{22}(\kappa_2)$) is compared to the 
corresponding spectrum along the elongated direction, i.e.\
$E_{33}(\kappa_3)$. Note that the data \revision{is}{are} presented in form of 
premultiplied spectra as function of the wavelength
$\lambda_\alpha=2\pi/\kappa_\alpha$ in order to facilitate the
discussion.  
It can be seen that the spectral energy density along short directions
is practically identical to the corresponding spectrum evaluated in the
cubical box in case~B.   
The counterpart represented along the elongated direction
($\alpha=3$), on the other hand, exhibits an alternating behavior in 
the range of wavelengths larger than the forcing cut-off
$\lambda>2\pi/\kappa_f$: those modes which directly receive power
input from the forcing scheme have a larger energy content than in the
cubical case, and those which are interspersed (i.e.\ the unforced
modes) have a significantly smaller amplitude. 
The implications of the large-scale energy distribution for the flow
field in elongated boxes are best discussed in terms of two-point 
correlations.
Figure~\ref{fig-single-phase-spec-ELONG-2p-corr}$(b)$  
shows the longitudinal correlation function in the $z$-coordinate
direction, $R_{ww}(r_z)$. 
In the cases with cubic domains we obtain an
approximate decorrelation at the maximum separation (equal to half the
box size, $r_z/{\cal L}_x=0.5$), with the correlation values 
$R_{ww}(r_z/{\cal L}_x=0.5)=0.06$ and $-0.01$ in cases~A and B,
respectively. 
In the cases with a twice elongated domain the longitudinal
correlation functions in the direction of elongation are practically
identical to the corresponding cubical cases up to $r_z/{\cal
  L}_x=0.5$, and then they increase monotonically up to $0.39$ ($0.29$) at
separations equal to half the elongated side length in case~AL
(BL). 
This means that the flow field in adjacent cubical sub-domains retains 
a non-negligible correlation -- a consequence of the fact that the
forcing is applied only to every $n$th Fourier mode in a direction
which is elongated by an integer factor $n$
Note that an exact $n$-fold copy of
a given cubical domain would result in the two-point correlation
$R_{ww}(r_z)$ to be exactly symmetric with respect to the point
$r_z={\cal L}_x/n$, i.e.\ it would increase again up to the value of
unity at a separation of $r_z/{\cal L}_x=1$. 
Therefore, elongating the computational domain in a simulation of
single-phase 
\revision{}{isotropic} 
turbulence does not promise any substantial benefit. 
In the two-phase case with settling particles, however, it is still a
useful technique in view of the two situations which can cause a
non-physical bias of the Lagrangian particle statistics mentioned at
the beginning of the present section. This is because the particle
positions themselves will not present any periodicity other than the
fundamental one. Therefore, their wakes tend to be less correlated
over the elongated domain size and the particles' effect upon the flow
field (two-way coupling) will tend to reduce the observed correlation
of the single-phase case. 
Figure~\ref{fig-single-phase-spec-ELONG-2p-corr}$(b)$ indeed
confirms this for the particulate flow case~AL-G120 with finite
settling velocity which will be discussed in detail in
section~\ref{sec-two-phase}.  
It can be seen that the correlation value at the largest separation is
decreased by 46\% with respect to case~AL, yielding $R_{ww}(r_z/{\cal
  L}_x=1)=0.21$. 
\revision{}{%
  Note that in the particulate flow cases
  the velocity field was first continued inside the space occupied by
  the particles by means of linear interpolation involving only grid
  nodes located inside the fluid; then, as in the single-phase cases,
  the two-point correlation functions were computed with the aid of
  Fourier transform. }
\section{Simulation of forced turbulence in the presence of
  finite-size particles} 
\label{sec-two-phase}
In the previous section it was shown that basic tests with
single-phase flow provide good agreement with the literature, 
and elongating the computational domain in one direction appears to be
an option for the case with settling particles. 
We also verified that the forcing method considered here satisfies
points~I, III and IV of the requirements set forth in the
introduction, 
and we now apply this forcing scheme to the case of particle-laden
flows in order to ensure that requirement~II is also satisfied, and
that statistics are in agreement with data from the literature in the
two-phase case.  

For this purpose we are considering the flow in a 
triply-periodic domain seeded with $N_p$ solid, spherical particles of
equal diameter $D$ and density $\rho_p$. 
The system experiences a constant gravitational acceleration
$\mathbf{g}$ which is directed into the negative $z$-coordinate
direction.  
Simultaneously, the random forcing procedure described above is
used in order to generate 
additional turbulent fluctuations. 
\subsection{Parameters of the problem and setup of the simulations}
\label{sec-two-phase-params}
Whereas in the single-phase counterpart a single dimensionless
group (i.e.\ a Reynolds number such as $Re_\lambda$) adequately describes the system,
dimensional analysis shows that the multi-particle case requires five
such parameters. 
The additional four non-dimensional parameters can be taken as the
following ones:  
the solid volume fraction $\phi_s=N_pV_p/({\cal L}_x{\cal L}_y{\cal
  L}_z)$ (where $V_p=\pi D^3/6$ is the volume occupied by a single
particle); 
the density ratio $\rho_p/\rho_f$; 
the length scale ratio $D/\eta$; 
the Galileo number $Ga=u_gD/\nu$, where a gravitational velocity is
defined as $u_g=(|\rho_p/\rho_f-1||\mathbf{g}|D)^{1/2}$. 

\begin{table} %
   \centering
   \renewcommand{\arraystretch}{1.25}
   \setlength{\tabcolsep}{1ex}
   \begin{tabular}{lccccccccc}
     \hline\\[-1.ex]
      {case}& 
      $\phi_s$ & 
      $\rho_p/\rho_f$ & 
      $Ga$ &
      $D/\eta$ &
      $D/\Delta x$ &
      $\mathcal{L}_x\times\mathcal{L}_y\times\mathcal{L}_z$ & 
      $N_p$ &
      $Nx\times N_y\times N_z$&
      $T_{obs}/T_e$ 
      \\[.5ex] 
      \hline\\[-1.ex]
      A-G0
      & $0.005$ & $1.5$ & $0$ & $6.7$ & 16 &
      $32D \times 32D \times 32D $ & $313$ &
      $512^3$ &
      $49.7$ 
      \\ 
      AL-G120
      & $0.005$ & $1.5$ & $120$ & $6.7$ & 16 &
      $32D \times 32D \times 64D $ & $616$ &
      $512^2\times 1024$ & 
      $27.6$ 
      \\ 
      \hline
   \end{tabular}
   \caption{Imposed parameters in the particle-laden simulations. Note
     that in both cases the turbulence forcing parameters are
     identical to the single-phase case~A (cf.\
     Table~\ref{tab-single-phase-phys-param}).
     For the purpose of normalization in this table the time-averaged
     values of $\eta$ and $T_e$ in the single-phase simulation are
     used. 
   }
   \label{tab-two-phase-num-param}
\end{table}
\begin{table} %
  \centering
  \renewcommand{\arraystretch}{1.25}
  \setlength{\tabcolsep}{1ex}
  \begin{tabular}{lccc}
    \hline\\[-1.ex]
    case
    & $Re_{\tilde{\lambda}}=\tilde{\lambda}u_{rms}/\nu$ 
    & $I=u_{rms}/w_{rel}$
    & $\langle Re_p\rangle_t=|\langle w_{rel}\rangle_t|D/\nu$
    \\[.5ex] 
    \hline\\[-1.ex]
    A-G0
    & $65.4$      
    & --
    & --
    \\[.5ex] 
    AL-G120
    & $66.2$     
    & $0.197$  %
    & $137.6$ 
    \\[.5ex] 
    \hline
  \end{tabular}
  \caption{%
    Some additional physical parameters of the particle-laden
    simulations.  
    All quantities were evaluated from time-averages after the
    statistically stationary regime was reached in the respective
    simulation. 
  }
  \label{tab-two-phase-phys-param}
\end{table}
Two cases are considered, with and without gravity, under otherwise
identical conditions. 
For this purpose we have chosen the turbulence
forcing parameters of cases~A and AL of section~\ref{sec-single-phase}
which (in the absence of particles) yield a moderate Reynolds number
of $Re_\lambda=65$ in both cases (cf.\
tables~\ref{tab-single-phase-num-param} and  
\ref{tab-single-phase-phys-param}), the only difference being the
vertical domain size. 
Here we use the cubical domain of the single-phase case~A for the
simulation of a zero-gravity particle-laden case which we denote as
case~A-G0; 
likewise we use the vertically elongated domain of
single-phase case~AL in a particle-laden case with the Galileo number
set to $120$, henceforth denoted as case~AL-G120. 
An isolated heavy particle at the latter Galileo number ($Ga=120$, for
any density ratio $\rho_p/\rho_f>1$)  
settles steadily on a vertical path, featuring an axisymmetric
wake\cite{jenny:04,uhlmann:13a}. 
The remaining imposed non-dimensional parameters 
are identical in both particulate cases (cf.\
Table~\ref{tab-two-phase-num-param}):  
the suspension can be considered as dilute with a solid volume
fraction $\phi_s=0.005$, while the density ratio is set to
$\rho_p/\rho_f=1.5$ (which corresponds e.g.\ to some plastic materials 
such as PVC in water), and the particle diameter is larger than (but
comparable to) the Kolmogorov length scale, $D/\eta=6.7$; note that
this value as well as the following length scale ratios are based upon
the average dissipation rate $\langle\varepsilon\rangle_t$ obtained in
the corresponding single-phase simulations.   
In both cases A-G0 and AL-G120 the particle diameter corresponds to
$0.42$ Taylor length scales and to $0.05$ large eddy length scales.
It is known that in the absence of turbulence the dilute suspension
with the  parameter set ($\phi_s=0.005$, $\rho_p/\rho_f=1.5$,
$Ga=120$) as chosen in the present case~AL-G120 does not form
significant wake-induced particle clusters\cite{uhlmann:14a}.  

\begin{figure} %
  \begin{minipage}{.45\linewidth}
    \centerline{$(a)$}
    \includegraphics[width=\linewidth]
    {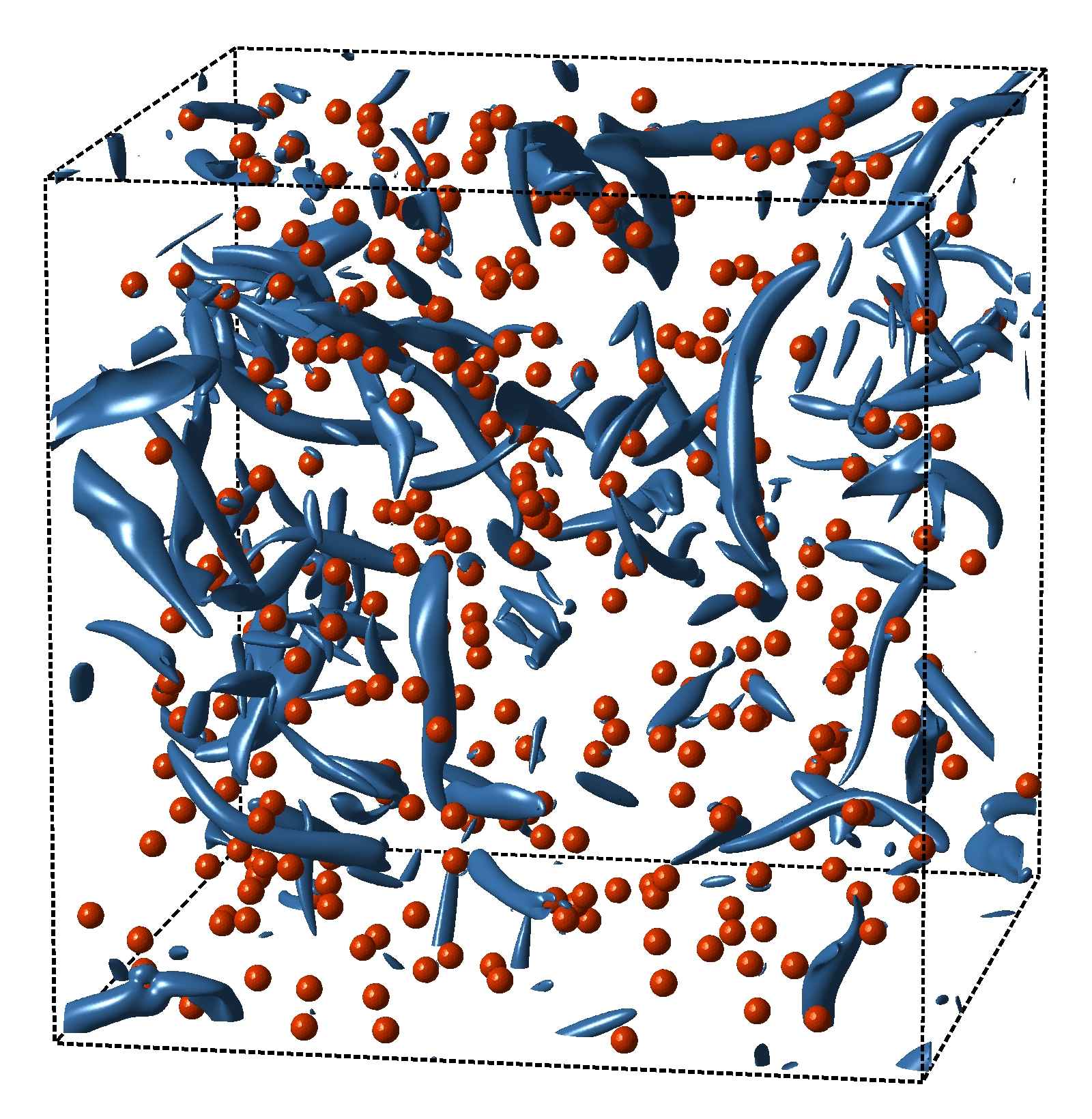}
  \end{minipage}
  \hfill
  \begin{minipage}{.45\linewidth}
    \centerline{$(b)$}
    \includegraphics[width=\linewidth]
    {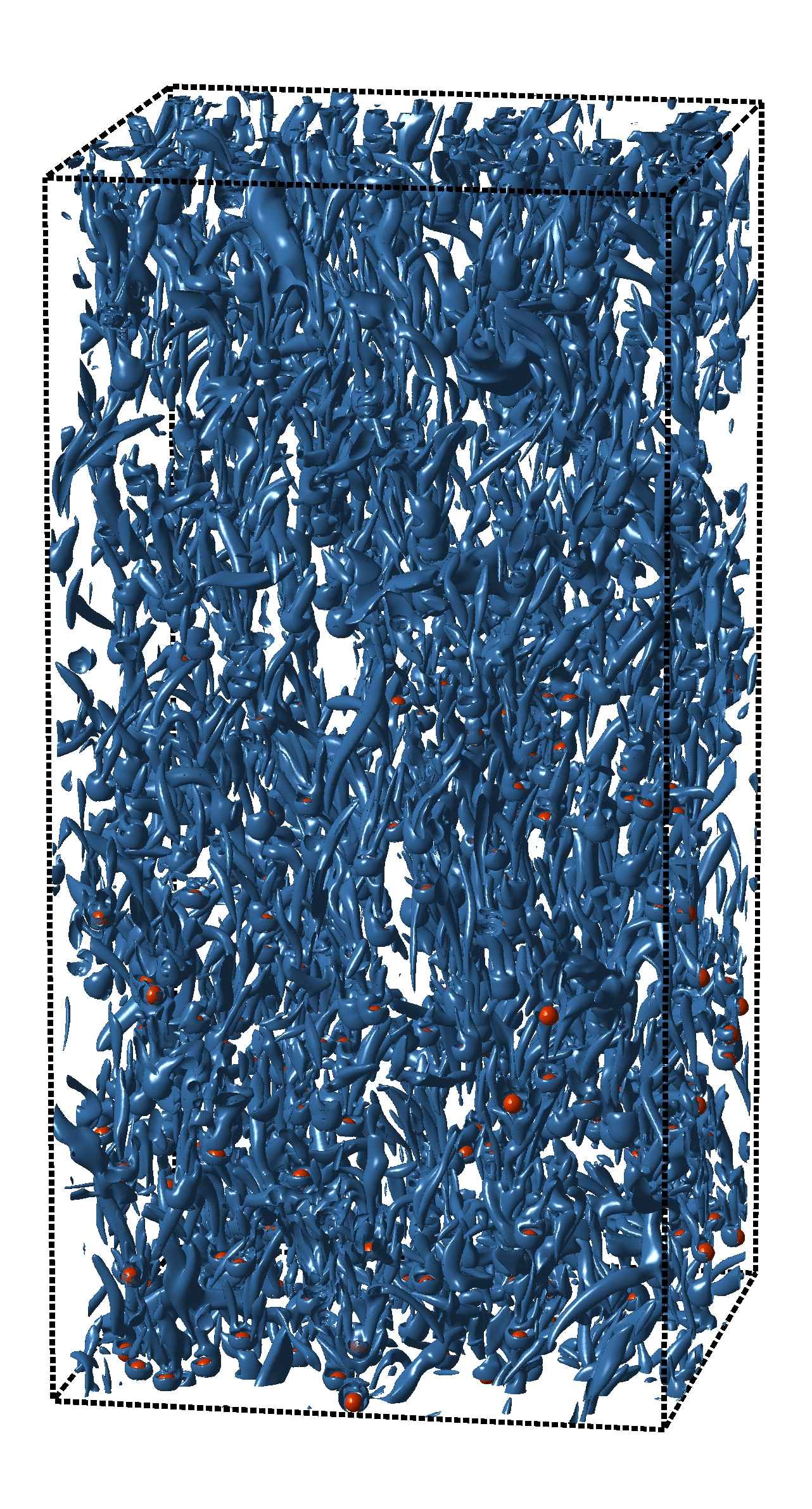}
  \end{minipage}
   \caption{%
     Visualization of instantaneous vortical structures identified with
     the q-criterion of Hunt et al.\cite{hunt:88} (blue iso-surfaces, 
     $q/\omega_{rms}^2=1$, where $\omega_{rms}$ is taken from
     single-phase flow) and particle positions (red spheres): 
     $(a)$ case A-G0; 
     $(b)$ case AL-G120.
   }
   \label{fig-two-phase-visu-3d}
\end{figure}

In the particulate flow problem a number of additional time scales can
be defined. First, the particle response time (based upon Stokes drag)
is given by $\tau_p=D^2\rho_p/(18\nu\rho_f)$. 
In the case of finite gravity, a gravitational time scale can be
inferred from the gravitational velocity $u_g$ and the particle
diameter, viz.\ $\tau_g=D/u_g$. 
In the simulations A-G0 and AL-G120 the ratio between the Stokes drag
time scale and the Kolmogorov time (i.e.\ the typical small-scale
Stokes number) is larger than unity ($\tau_p/\tau_\eta=3.73$), where
the time-scale of the flow is based upon single-phase data from case~A
(cf.\ table~\ref{tab-single-phase-phys-param}); 
the Stokes number is smaller than unity when based upon the integral
time scale ($\tau_p/T_e=0.22$). Therefore, we can expect the particles
to respond at least partially to the hydrodynamic forces caused by the
turbulent background flow.  
In the finite gravity case~AL-G120 the gravitational (settling) time
scale is 
\revision{more than twice as large as the Kolmogorov time}{%
less than half as large as the Kolmogorov time} 
($\tau_g/\tau_\eta=0.37$, again based upon single-phase data). At the
same time the standard deviation of the turbulent background velocity
field (measured in the absence of particles) is nearly five times
smaller than the gravitational velocity scale
($u_{rms}/u_g=0.22$). This indicates that the vertical velocity will
be the dominant contribution to the particle motion, but that the
influence of turbulence will not be negligibly small.  
Flow visualisation (such as shown in
figure~\ref{fig-two-phase-visu-3d}) indeed demonstrates the effect of 
gravity, i.e.\ the formation of significant vorticity in the wakes of
the settling particles. 

The start-up of the present simulations is again performed in a
step-wise manner. First, the grid of the single-phase cases~A and AL
is used in cases A-G0 and AL-G120, respectively, and a developed
single-phase flow field is used as initial condition, while particles
are added at random positions. The particles are first held fixed in
space. After several integral time scales have elapsed, the grid is
refined (by linear interpolation) to its final resolution of $D/\Delta
x=16$ (cf.\ validation in Refs.\citenum{uhlmann:13a,uhlmann:14a}), and
the simulation is further advanced for several integral time scales
(the particles still being fixed in space). 
Finally, particles are released with an initial velocity equal to zero. 
We arbitrarily set the time to zero at the moment of release.  

In the finite-gravity case~AL-G120 a constant (positive) vertical 
velocity $w_{sh}$ was added to the initial flow field taken
from the single-phase simulation case~AL (at time $t=t_{sh}$), in
order to impose a mean relative velocity while particles are initially
held fixed with respect to the computational grid. 
The value of $w_{sh}$ was determined from the anticipated settling
velocity of an isolated sphere\cite{uhlmann:14a}. 
Since the turbulent background flow in this case is effectively
transported with velocity $\mathbf{u}_{sh}=(0,0,w_{sh})$, we apply a 
phase-shift to the forcing term by a corresponding amount, i.e.\ we
modify $\hat{\mathbf{f}}^{(t)}$ given in 
(\ref{EswaranPope_Projection}) by the following expression:
\begin{equation}\label{equ-forcing-term-phase-shifted}
  \hat{\mathbf{f}}_{sh}^{(t)}(\boldsymbol{\kappa},t)
  =
  \hat{\mathbf{f}}^{(t)}(\boldsymbol{\kappa},t)
  \,
  \exp(-I(t-t_{sh})\boldsymbol{\kappa}\cdot\mathbf{u}_{sh})
  \,,
\end{equation}
(where $I=(-1)^{1/2}$), then replace $\hat{\mathbf{f}}^{(t)}$ in
(\ref{equ-total-volume-force}) by $\hat{\mathbf{f}}_{sh}^{(t)}$.  
\subsection{Temporal evolution and the kinetic energy budget}
\label{sec-two-phase-time-evol}
\begin{figure} %
   \begin{minipage}{2ex}
     \rotatebox{90}{%
       budget
     }
   \end{minipage}
   \begin{minipage}{0.45\linewidth}
     \centerline{$(a)$}
     \includegraphics[width=\linewidth]
     {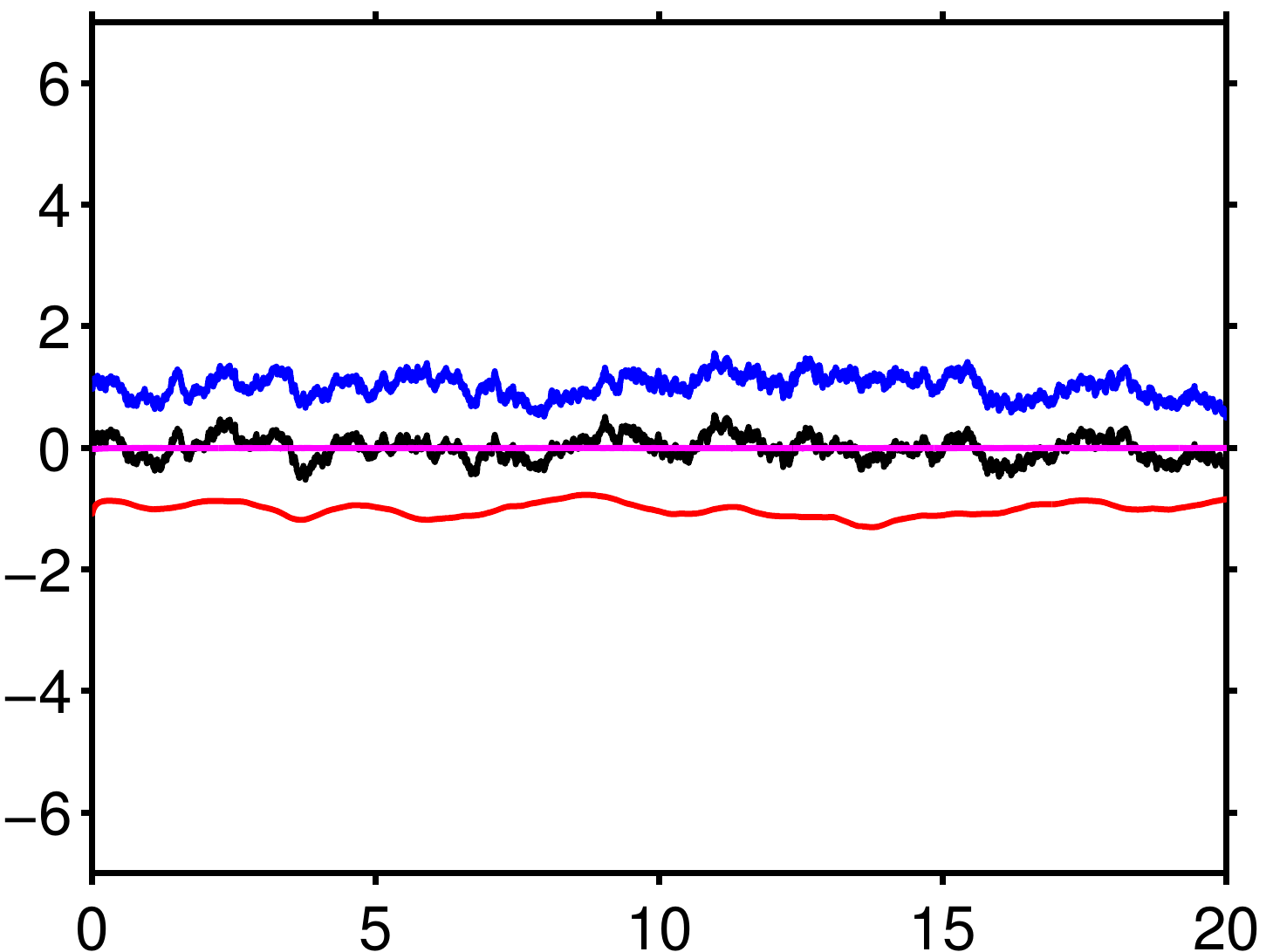}
     \\
     \centerline{$t/T_e$}
   \end{minipage}
   \hfill
   \begin{minipage}{2ex}
     \rotatebox{90}{%
       budget
     }
   \end{minipage}
   \begin{minipage}{0.45\linewidth}
     \centerline{$(b)$}
     \includegraphics[width=\linewidth]
     {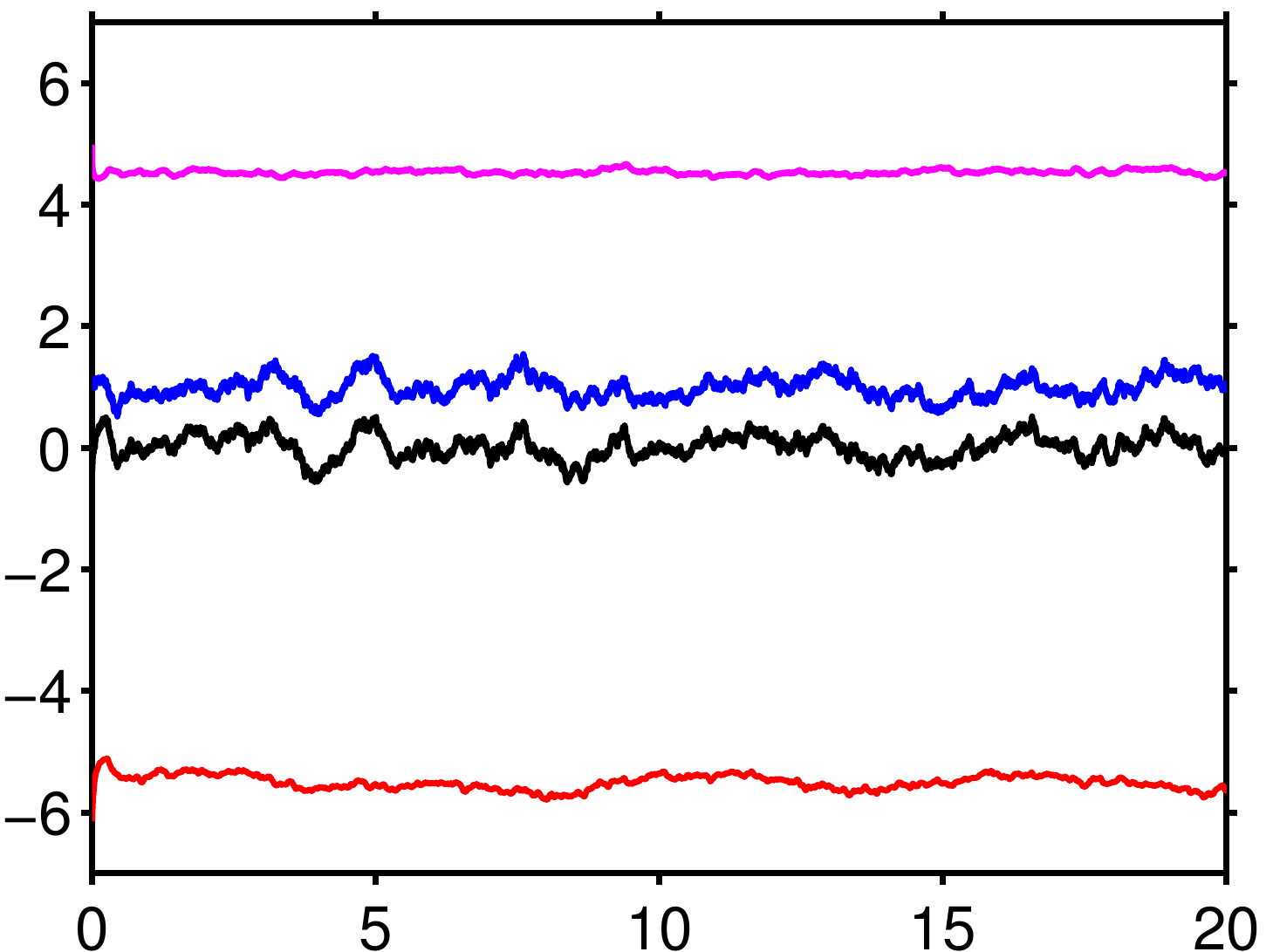}
     \\
     \centerline{$t/T_e$}
   \end{minipage}
   \caption{Time evolution of the terms in the budget of
     volume-averaged kinetic energy (\ref{equ-ek-box-avg}): 
     $(a)$~case~A-G0, 
     $(b)$~case~AL-G120. 
     Line styles: 
     {\color{blue}\solidthick}, turbulence forcing 
     ($\Psi^{(t)}$); 
     {\color{magenta}\solidthick}, fluid-particle coupling
     ($\Psi^{(p)}$);  
     {\color{red}\solidthick}, dissipation ($-\varepsilon_\Omega$); 
     {\color{black}\solidthick}, time rate-of-change ($-\mbox{d}\langle
     E_k\rangle_\Omega/\mbox{d}t$). 
     All terms are normalized by the  
     time-average value of the dissipation rate
     in the corresponding single-phase case. 
   }
   \label{fig-two-phase-budget-time-evol-1}
\end{figure}
The temporal evolution of the four individual terms of the
box-averaged kinetic energy equation (\ref{equ-ek-box-avg}) are shown in
figure~\ref{fig-two-phase-budget-time-evol-1}. 
Recall that in the particulate case the time-rate-of-change of $\langle
E_k\rangle_\Omega$ is the result of a balance between turbulence-forcing
power input ($\Psi^{(t)}$), dissipation rate ($\varepsilon_\Omega$)
and fluid-particle coupling term ($\Psi^{(p)}$). 
\revision{%
The figure shows that in the zero-gravity case the particles have
essentially no influence upon the global kinetic energy balance, since
the fluid-particle coupling term is found to be negligible and the
magnitude of the dissipation rate is practically identical to the
single-phase counterpart with the same external turbulence forcing
parameters (case~A). 
Globally speaking one can therefore qualify case
A-G0 as ``one-way coupled'', as might actually be expected on the
grounds of the small solid volume fraction and the absence of average 
relative velocity.}{%
The figure shows that in the zero-gravity case 
the fluid-particle coupling term is negligibly small 
(its time-average is equal to $0.002$ times the mean single-phase
dissipation rate value), 
and that both the magnitude of the dissipation rate and that of the
turbulence forcing power input are of similar values as the 
single-phase counterparts in case~A which features the same external
turbulence forcing parameters (the relative differences of the
time-averaged values amount to 2\% and 3.5\%, respectively). 
This result might actually be expected on the
grounds of the small solid volume fraction and the absence of average 
relative velocity.
In fact, in the simulations with decaying homogeneous-isotropic
turbulence performed by Lucci et al.\cite{lucci:10,lucci:11} the two
particle sets with the smallest volume fraction (cases ``B'' in
Ref.\citenum{lucci:10} and Ref.\citenum{lucci:11}, both at
$\phi_s\approx0.01$ and $\rho_p/\rho_f\approx2.5$) exhibit a similarly
small effect upon the temporal evolution of the
turbulent kinetic energy and upon the dissipation rate as observed in
the present case~A-G0.}  

Concerning the finite-gravity case AL-G120, we observe in
figure~\ref{fig-two-phase-budget-time-evol-1}$(b)$ that while the
magnitude of the power input due to turbulence forcing is similar to
the single-phase counterpart, the fluid-particle coupling term now
dominates the balance, while the magnitude of the dissipation is
likewise increased by nearly a factor of six. 
It can be seen that the kinetic energy budget is fundamentally
different in the case with non-zero gravity even at a relatively small
solid volume fraction, mainly due to the increased dissipation in the
wakes. 

Let us mention that the box-averaged kinetic energy balance
(\ref{equ-ek-box-avg}) is fulfilled by the simulation data up to the
contribution due to numerical error. In the present case A-G0
(AL-G120) the maximum residual amounts to $0.005$ ($0.059$) times the
time average of the dissipation rate
($\langle\varepsilon_\Omega\rangle_t$).  
Grid refinement by a factor $3/2$ in case AL-G120 leads to a reduction
of the maximum residual to a value of
$0.032\langle\varepsilon_\Omega\rangle_t$.  
The fulfillment of the numerically evaluated kinetic energy balance
can therefore be considered as an additional grid convergence test
which to our knowledge has previously not been reported in
interface-resolved particulate flow simulations. 

In order to further elucidate the kinetic energy budget, we rewrite
the fluid-particle coupling term in
(\ref{equ-def-two-way-coupling-box-avg}) through use 
of the Newton-Euler equation for rigid body motion of spherical
particles as follows:
\begin{equation}\label{equ-def-two-way-coupling-box-avg-after-subs-newton}
  \Psi^{(p)}(t)
  =
  \phi_s\left(\frac{\rho_p}{\rho_f}-1\right)
  \mathbf{u}_{rel,\Omega}
  \cdot\mathbf{g}
  +\Psi^{(p)}_{accel}(t)
  +\Psi^{(p)}_{coll}(t)
  \,, 
\end{equation}
where the contributions due to particle acceleration
($\Psi^{(p)}_{accel}$) and due to inter-particle collisions
($\Psi^{(p)}_{coll}$) are defined in
appendix~\ref{sec-app-ekin}. Furthermore, an 
apparent slip velocity with respect to the box-averaged velocity,
defined as $\mathbf{u}_{rel,\Omega}=
\langle\mathbf{u}_p^{(i)}\rangle_p
-
\langle\mathbf{u}\rangle_\Omega$, 
has been used in setting up
(\ref{equ-def-two-way-coupling-box-avg-after-subs-newton}).  
Since the two-way coupling term is negligibly small in the
zero-gravity case A-G0, we now focus on case AL-G120. 
The data in the latter case \revision{shows}{show} that the sum of the  particle
acceleration and particle collision terms 
($\Psi^{(p)}_{accel}+\Psi^{(p)}_{coll}$) 
is negligibly small compared to the gravity term in
(\ref{equ-def-two-way-coupling-box-avg-after-subs-newton}). 
Substituting
(\ref{equ-def-two-way-coupling-box-avg-after-subs-newton}) into
(\ref{equ-ek-box-avg}), 
performing statistical averaging (or time averaging over a 
statistically stationary interval) and neglecting the small terms then
yields the following approximate relation:
\begin{equation}\label{equ-ek-box-avg-stat-avg-approx-1}
  \tilde{\varepsilon}
  \equiv
  \langle\varepsilon_\Omega\rangle_t
  -\phi_s\left(\frac{\rho_p}{\rho_f}-1\right)
  \langle\mathbf{u}_{rel,\Omega}\rangle_t
  \cdot\mathbf{g}
  \approx
  \langle\Psi^{(t)}\rangle_t
  \,.
\end{equation}
From this balance it can be argued that the difference between the
mean dissipation rate and the gravity term, which we henceforth denote
as $\tilde{\varepsilon}$, represents the dissipation rate of the 
``background'' turbulence. In other words this amounts to removing the
added dissipation of the work done by the gravitational potential. 
We will use the quantity $\tilde{\varepsilon}$ in the definition of
a modified Taylor micro-scale 
\begin{equation}\label{equ-def-taylor-scale-modif}
  \tilde{\lambda}=(15\nu u_{rms}^2/\tilde{\varepsilon})^{1/2}
  \,,
\end{equation}
from which a modified Reynolds number
$Re_{\tilde{\lambda}}=\tilde{\lambda}u_{rms}/\nu$ can be formed. 
Due to the anisotropy of the flow in the presence of
gravity, the surrogate Taylor micro-scale in
(\ref{equ-def-taylor-scale-modif}) does obviously not correspond to
its original definition in homogeneous-isotropic turbulence (e.g.\
Ref.\citenum{pope:00}).   
It is here simply used in order to provide a means of computing a
value for the Reynolds number based on this length scale, as this is
the most widely used one in homogeneous turbulence. 
Indeed, since the work done by the turbulence 
forcing is only little affected by the presence of the settling
particles, the Reynolds number $Re_{\tilde{\lambda}}$ based upon the
definition (\ref{equ-def-taylor-scale-modif}) is practically the same
in case AL-G120 and in case A-G0 (cf.\
table~\ref{tab-two-phase-phys-param}), and the value closely matches
the one of the corresponding single-phase cases
(cf.\ table~\ref{tab-single-phase-phys-param}).  

\begin{figure} %
   \begin{minipage}{2ex}
     \rotatebox{90}{%
       $k/k_{ref}$
     }
   \end{minipage}
   \begin{minipage}{0.45\linewidth}
     \centerline{$(a)$}
      \includegraphics[width=\linewidth]
      {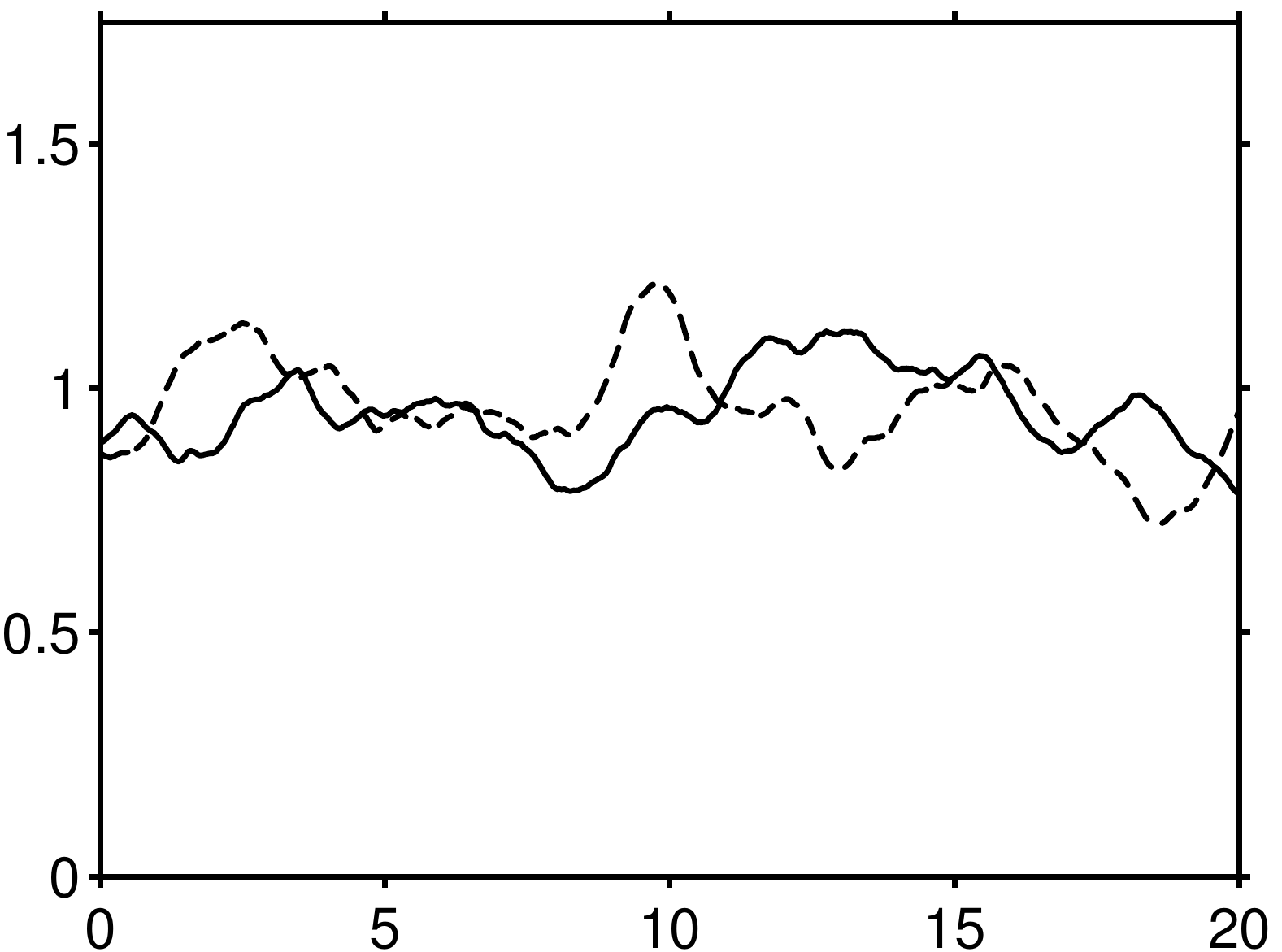}
      \\
      \centerline{$t/T_e$}
   \end{minipage}
   \hfill
   \begin{minipage}{2ex}
     \rotatebox{90}{%
       $k_p/k_{ref}$
     }
   \end{minipage}
   \begin{minipage}{0.45\linewidth}
     \centerline{$(b)$}
      \includegraphics[width=\linewidth]
      {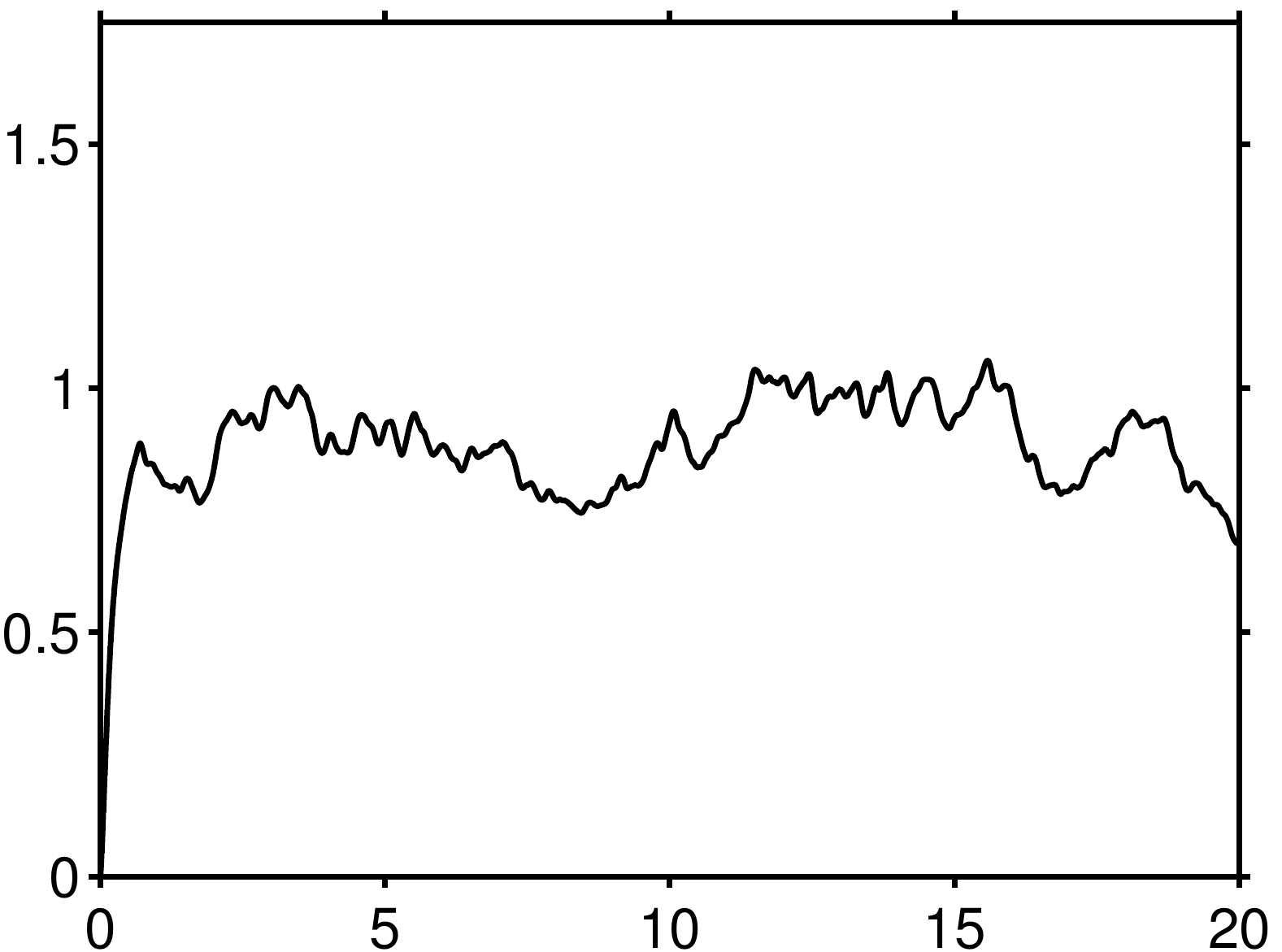}
      \\
      \centerline{$t/T_e$}
   \end{minipage}
   \caption{%
     The energy associated with the velocity fluctuations of both
     phases in case~A-G0: 
     $(a)$ fluid phase turbulent kinetic energy $k$ defined in 
     (\ref{equ-ek-fluct-fluid-only});  
     $(b)$ solid phase kinetic energy of the fluctuations 
     $k_p\equiv\langle\mathbf{u}_p^\prime\cdot\mathbf{u}_p^\prime\rangle_p/2$.  
     The dashed line in $(a)$ corresponds to the evolution of $k$ in
     the single-phase case~A. 
     Note that only part of the simulated interval is shown for clarity. 
   }
   \label{fig-two-phase-vel-fluct-g0}
\end{figure}
\begin{figure} %
   \begin{minipage}{2ex}
     \rotatebox{90}{%
       $k/k_{ref}$
     }
   \end{minipage}
   \begin{minipage}{0.45\linewidth}
     \centerline{$(a)$}
      \includegraphics[width=\linewidth]
      {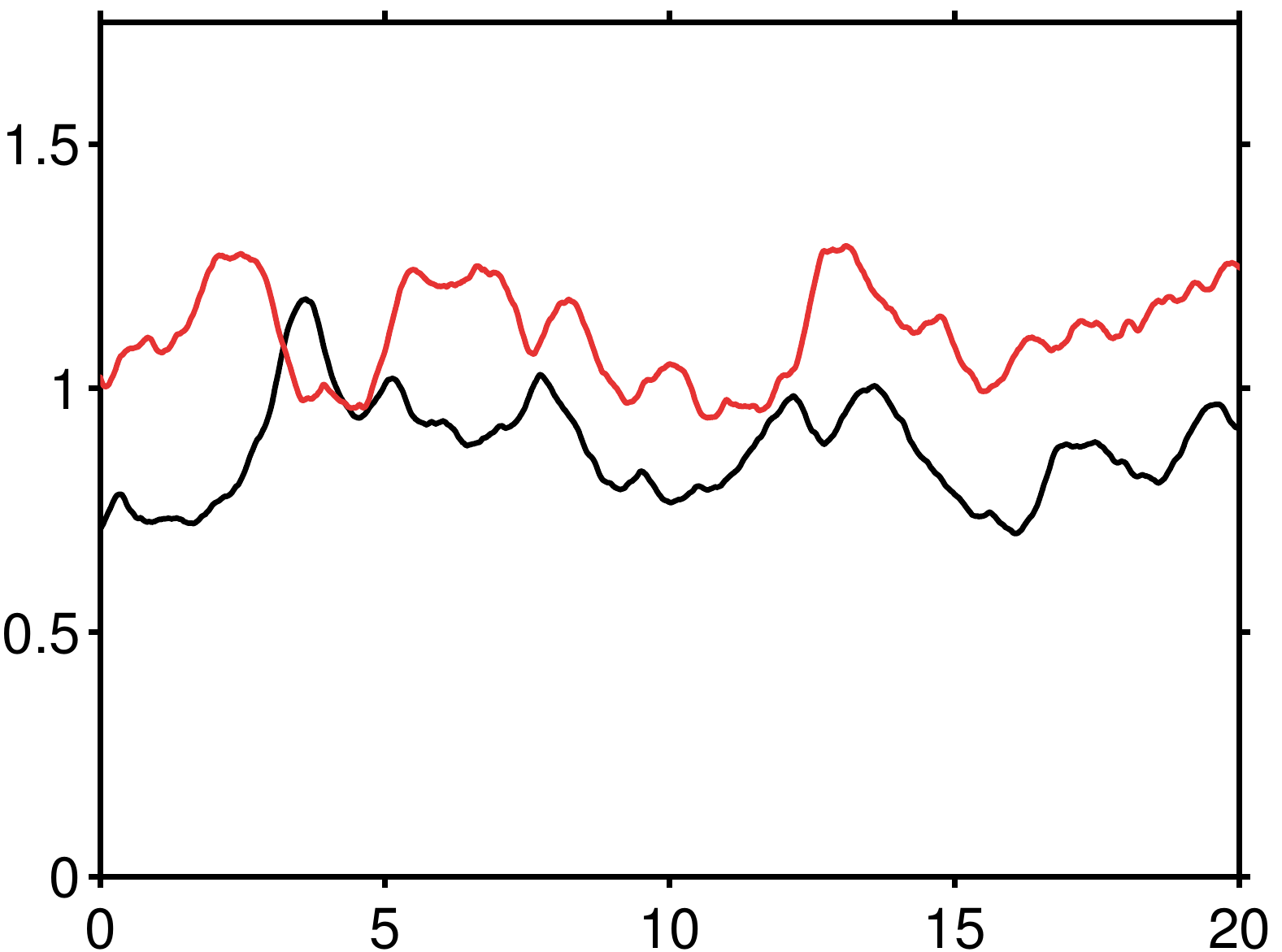}
      \\
      \centerline{$t/T_e$}
   \end{minipage}
   \hfill
   \begin{minipage}{2ex}
     \rotatebox{90}{%
       $k_p/k_{ref}$
     }
   \end{minipage}
   \begin{minipage}{0.45\linewidth}
     \centerline{$(b)$}
      \includegraphics[width=\linewidth]
      {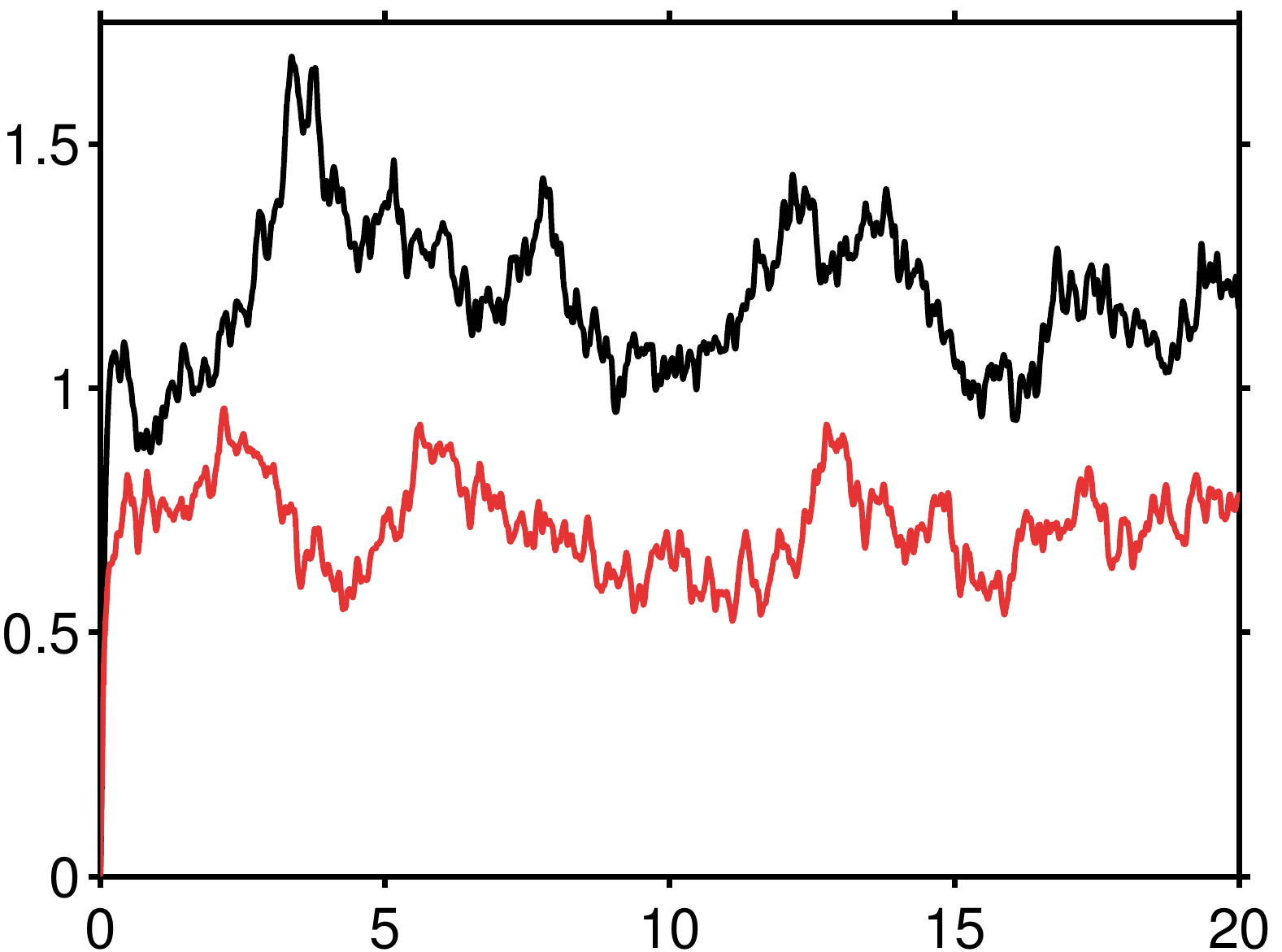}
      \\
      \centerline{$t/T_e$}
   \end{minipage}
   \caption{%
     The energy associated with the velocity fluctuations of both
     phases in case~AL-G120, distinguishing between the horizontal
     and vertical directions. 
     $(a)$ fluid phase; 
     $(b)$ solid phase. 
     In $(a)$ the black solid line corresponds to the horizontal component 
     $k^{(h)}\equiv3(\langle u^\prime u^\prime\rangle_{\Omega_f}+\langle
     v^\prime v^\prime\rangle_{\Omega_f})/4$; 
     the red solid line shows the vertical component 
     $k^{(v)}\equiv3\langle w^\prime
     w^\prime\rangle_{\Omega_f}/2$. 
     Note that with this definition $k=(2k^{(h)}+k^{(v)})/3$, and that
     in particular for an ideally isotropic velocity field
     $k=k^{(h)}=k^{(v)}$ would hold.  
     In $(b)$ the corresponding quantities $k_p^{(h)}$ and $k_p^{(v)}$
     are plotted with the same line-styles as in $(a)$,
     where the fluid velocity fluctuations $\mathbf{u}^\prime$ are
     replaced by the particle velocity fluctuations 
     $\mathbf{u}_p^\prime$, and the averaging is over the number of
     particles ($\langle\,\rangle_p$ instead of
     $\langle\,\rangle_{\Omega_f}$).   
     The reference scale $k_{ref}$ is the same as in
     figure~\ref{fig-single-phase-time-evol-1}. 
     Note that only part of the simulated interval is shown for clarity. 
   }
   \label{fig-two-phase-vel-fluct-g120}
\end{figure}
Next we examine the temporal evolution of the energy associated to
fluctuations of the fluid velocity field and the particles. 
Figure~\ref{fig-two-phase-vel-fluct-g0} shows the turbulent kinetic
energy of the fluid (defined in \ref{equ-ek-fluct-fluid-only}) in
case~A-G0 together with the same quantity in the single-phase case~A. 
Both signals exhibit similar features concerning time scales and
amplitude of the fluctuations, the time average of the particulate
case A-G0 being 2.2\% lower than the single-phase counterpart.
Note that the amplitude of the temporal fluctuations of $k$ is less
than 10\% of its time-average value when using the present forcing
scheme.  
The kinetic energy of the particulate phase, analogously defined as
$k_p\equiv\langle\mathbf{u}_p^\prime\cdot\mathbf{u}_p^\prime\rangle_p/2$,
quickly reaches a statistically stationary state after particle
release in case~A-G0; afterwards it fluctuates around a time-average
which is approximately 6\% lower than the fluid-phase kinetic
energy. Furthermore, the two quantities $k_p$ and $k$ are clearly
positively correlated in this case. 

Turning now to the finite-gravity case AL-G120
(figure~\ref{fig-two-phase-vel-fluct-g120}), we need to distinguish
the vertical from the horizontal directions. For this purpose we
separately define two kinetic energy contributions 
$k^{(v)}\equiv3\langle w^\prime
w^\prime\rangle_{\Omega_f}/2$ 
and 
$k^{(h)}\equiv3(\langle u^\prime u^\prime\rangle_{\Omega_f}+\langle
v^\prime v^\prime\rangle_{\Omega_f})/4$ 
which are constructed such that $k=(2k^{(h)}+k^{(v)})/3$, i.e.\
$k=k^{(h)}=k^{(v)}$ holds in case of isotropy. 
It can be observed from the figure that gravity indeed breaks the
isotropy, with the vertical component fluctuating around a larger
time-average value than the horizontal one (by 24\%). 
This result can be explained by the fact that particle wakes
predominantly constitute fluctuations of the vertical fluid velocity
component. 
The particle kinetic energy contributions $k_p^{(v)}$ and $k_p^{(h)}$,
which are analogously defined as $k^{(v)}$ and $k^{(h)}$ (cf.\
figure~\ref{fig-two-phase-vel-fluct-g120}$b$), on the other hand,
exhibit an inverse ordering, i.e.\ the horizontal component is larger
than the vertical one (on average by 71\%). 
This observation can be understood by recalling that the force
fluctuations of fixed spheres either swept by homogeneous-isotropic
turbulence\cite{bagchi:03} or towed through a homogeneous-isotropic
turbulent flow field\cite{homann:13} exhibit the same anisotropy,
i.e.\ the standard-deviation of the lateral force component is larger
than the axial one. It can therefore be expected that in the case of
mobile spheres the intensity of the lateral (horizontal) particle
velocity generated by these force fluctuations is larger than the
corresponding axial (vertical) one.  

\begin{figure} %
   \centering
   \begin{minipage}{2ex}
     \rotatebox{90}{%
       $Re_p$
     }
   \end{minipage}
   \begin{minipage}{0.45\linewidth}
      \includegraphics[width=\linewidth]
      {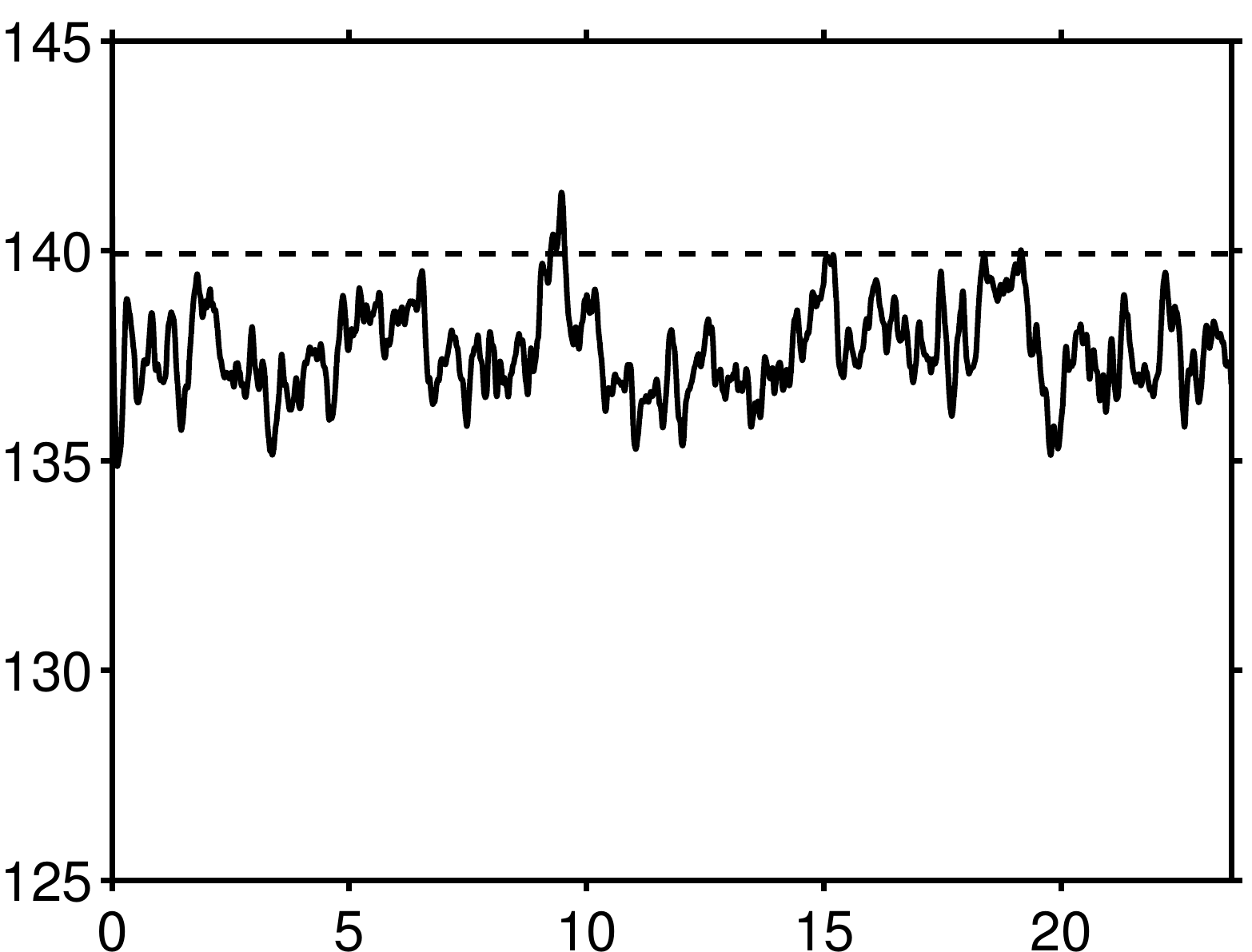}
      \\
      \centerline{$t/T_e$}
   \end{minipage}
   \caption{%
     Temporal evolution of the mean particle settling velocity
     normalized by viscous particle scales, 
     $Re_p=|w_{rel}|D/\nu$ 
     in case AL-G120. %
     The dashed line corresponds to the value for an isolated sphere
     in ambient fluid, which was extrapolated from the simulation with
     $Ga=121.24$ in Ref.\citenum{uhlmann:14a}, supposing that the
     ratio $|w_{rel}|/u_g$ remains constant.  
   }
   \label{fig-two-phase-average-particle-settling}
\end{figure}
A quantity of considerable interest for many applications is the
average settling velocity of the particles with respect to the global
average velocity of the fluid phase, viz.\
\begin{equation}\label{equ-def-global-urel}
  \mathbf{u}_{rel}
  =
  \langle\mathbf{u}_p^{(i)}\rangle_p
  -
  \langle\mathbf{u}\rangle_{\Omega_f}
  \,.
\end{equation}
Figure~\ref{fig-two-phase-average-particle-settling} shows the
absolute value of the vertical component of
(\ref{equ-def-global-urel}) in viscous particle scaling, i.e.\ a
settling Reynolds number $Re_p=|w_{rel}|D/\nu$, in the finite-gravity
case~AL-G120. 
The time-average of $Re_p$ measures $137.6$, which is
approximately 1.5\% smaller than the average settling Reynolds number
of an isolated sphere in ambient fluid at the same Galileo number and
density ratio 
(the reference value was extrapolated from the simulation ``S121''
with $Ga=121.24$ in Ref.\citenum{uhlmann:14a}, supposing that the
ratio $|w_{rel}|/u_g$ remains constant).  
Although collective effects cannot be
ruled out, they appear unlikely at the present solid volume fraction
and in the absence of any indication of particle clustering. 
\revision{Indeed Vorono\"i tesselation analysis shows that no significant 
departure from a random particle distribution is taking place in
either flow case (figure omitted). }{%
Indeed Vorono\"i tesselation analysis shows that no significant 
departure from a random particle distribution is taking place in
either flow case (cf.\
appendix~\ref{sec-two-phase-voronoi-cell-vol-stddev}). }
Therefore, the likely cause for the slightly reduced average settling
velocity lies in the turbulent background flow. 
One model for understanding reduced particle settling in a turbulent
environment is non-linear drag\cite{tunstall:68,bagchi:03}: 
if one assumes that the instantaneous drag on the particle can be
described by a non-linear function of some properly defined ``velocity
seen by the particle'', then any superposed fluctuations of this
fluid velocity can lead to an increase of the mean particle drag
force, even if the fluctuations are symmetrically distributed. 
Homann et al.\cite{homann:13} have shown that if one assumes an
isolated sphere is swept by a velocity composed of a constant
$w_{rel}$ plus a Gaussian random vector (whose components are
independently and identically distributed with standard-deviation
$u_{rms}$) and if a standard drag formula
(Schiller-Naumann\cite{schiller:33}) is invoked instantaneously, the  
mean particle drag increases by a relative amount which scales as
$\Delta F_D^{SN}\approx C(Re_p)\,I^2$ for small values of the relative
turbulence intensity $I=u_{rms}/w_{rel}$.  
Applying this result to our present case (i.e.\ invoking an
equilibrium between drag and submerged weight) yields a decrease of
the average settling velocity by 1.7\%. 
It is interesting to note that this value is relatively close to the
DNS results, despite the fact that the fluctuating flow field is
neither isotropic nor Gaussian distributed (cf.\ discussion in
section~\ref{sec-two-phase-steady-state}).  
\subsection{Validation of statistics in the stationary regime}
\label{sec-two-phase-steady-state}
In the following we attempt to compare our present results to the
available data for freely-mobile finite-size particles in homogeneous
turbulence without mean velocity gradients. 
The zero-gravity case~A-G0 is comparable to some of the configurations
simulated in Ref.\citenum{homann:10} and Ref.\citenum{YeoClimentMaxey:10}; 
for the finite-gravity case~AL-G120, on the other hand, no directly
comparable data \revision{exists}{exist} to our knowledge. 

Homann \& Bec\cite{homann:10} have simulated the motion of a single,
neutrally-buoyant particle in statistically stationary, 
homogeneous-isotropic turbulence at a relatively low Reynolds number
$Re_\lambda=32$. One of the particle diameters considered in that study
was equal to six times the Kolmogorov scale, which 
provides a relatively close match with the present value of
$D/\eta=6.7$.  
With this choice, those authors have individually addressed the effect
of finite particle size, while collective effects as well as particle
inertia and gravity were excluded. 

Yeo et.\ al.\cite{YeoClimentMaxey:10} used the so-called ``force-coupling
method''\cite{lomholt:03} to simulate the same configuration. 
The Reynolds number, however, was larger ($Re_\lambda=57$), and in
some cases the particles were denser than the fluid
($\rho_p/\rho_f=1.4$), albeit with gravity set to zero. In one
particular case (their label ``S2''), which we will exclusively
discuss in the following, the size ratio was set to $D/\eta=7$. It can
be seen that the parameter values so far are very close to the ones
chosen in our case A-G0.  
However, in Ref.\citenum{YeoClimentMaxey:10} the solid volume fraction was more than
ten times larger ($\phi_s=0.06$) than here. 

\begin{figure} %
   \begin{minipage}{2ex}
     \rotatebox{90}{pdf}
   \end{minipage}
   \begin{minipage}{0.45\linewidth}
     \centerline{$(a)$}
      \includegraphics[width=\linewidth]
      {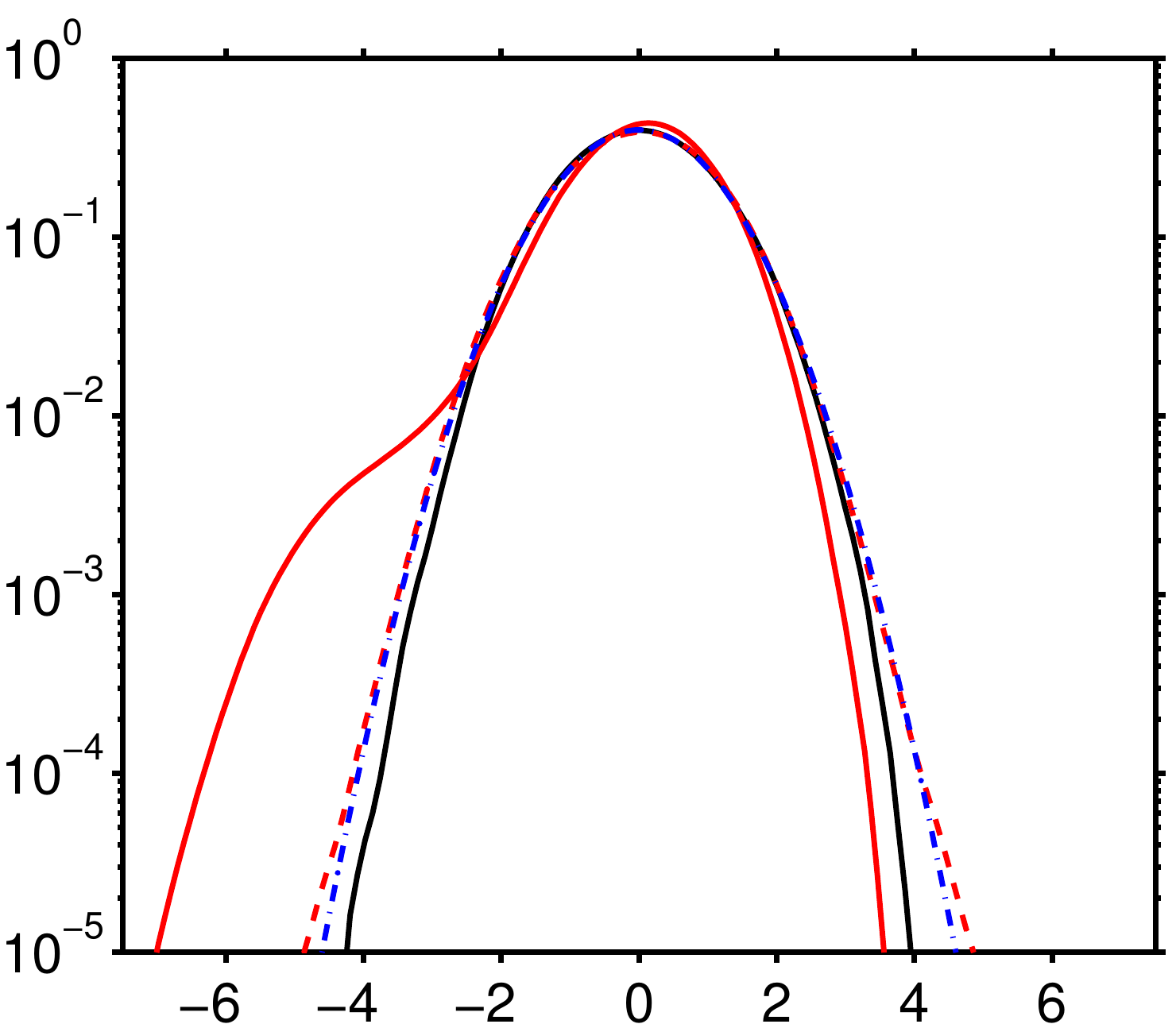}
      \\
      \centerline{$u^\prime/\sigma(u^\prime)$}
   \end{minipage}
   \begin{minipage}{2ex}
     \rotatebox{90}{pdf}
   \end{minipage}
   \begin{minipage}{0.45\linewidth}
     \centerline{$(b)$}
      \includegraphics[width=\linewidth]
      {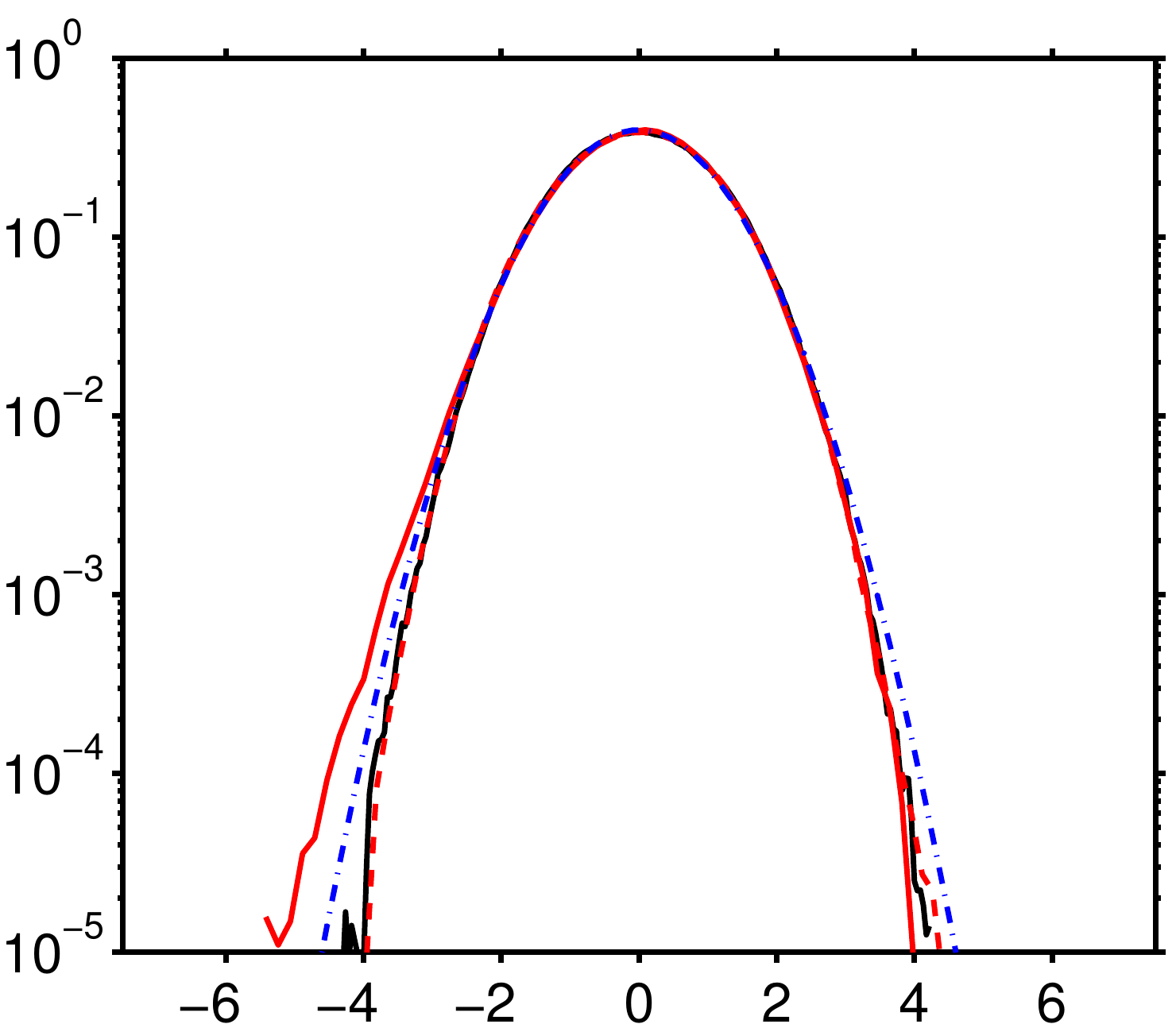}
      \\
      \centerline{$u_p^\prime/\sigma(u_p^\prime)$}
   \end{minipage}
   \caption{%
      Normalized probability density functions accumulated in the
      statistically stationary interval:  
      $(a)$ fluid velocity; 
      $(b)$ particle velocity.
      Line styles correspond to: 
      {\color{black}\solidthick},~case~A-G0; 
      {\color{red}\solidthick},~vertical velocity in case~AL-G120; 
      {\color{red}\dashed},~horizontal velocity in case~AL-G120; 
      {\color{blue}\chndot},~Gaussian.  
   }
   \label{fig-two-phase-fluid-vel-pdf}
\end{figure}
Figure~\ref{fig-two-phase-fluid-vel-pdf} shows the normalized
probability density functions (p.d.f.) of fluid and particle velocities,
accumulated over the statistically stationary interval. 
It can be seen that the distributions of the velocities of both phases
in the zero-gravity case~A-G0 are approximately Gaussian. 
For both phases in case~A-G0 we obtain a slightly smaller kurtosis 
value than for a Gaussian distribution, i.e.\ 
\revision{$K=2.86$}{$K=2.83$}
 for the fluid
and $K=2.82$ for the particle velocity. 
This result is fully consistent with case S2 of Yeo et.\
al.\cite{YeoClimentMaxey:10} who report a value of 2.81 for the particle velocity
kurtosis. 

Concerning the finite-gravity case~AL-G120, we observe that the
horizontal velocity components of both phases remain essentially
Gaussian distributed. The vertical fluid velocity component, on the
other hand, exhibits a significant negative skewness which amounts to
\revision{$S=-0.87$.}{$S=-0.78$.} 
This shape of the p.d.f.\ can be explained by the occurrence
of wakes downstream of the settling particles which constitute
predominantly negative fluctuations in the vertical fluid velocity
field. A similar effect has been previously reported in
Ref.\citenum{uhlmann:08a} for heavy particles settling in
vertically-oriented turbulent channel flow. 
The vertical particle velocity component in case~AL-G120 is also
negatively skewed, albeit to a lesser degree ($S=-0.10$). This latter
result implies that the particle motion is partially affected by the
skewed fluid velocity field, again consistent with observations in
channel flow\cite{uhlmann:08a}. 

\begin{figure} %
  \centering
  \begin{minipage}{2ex}
    \rotatebox{90}{pdf}
  \end{minipage}
  \begin{minipage}{0.45\linewidth}
    \includegraphics[width=\linewidth]
    {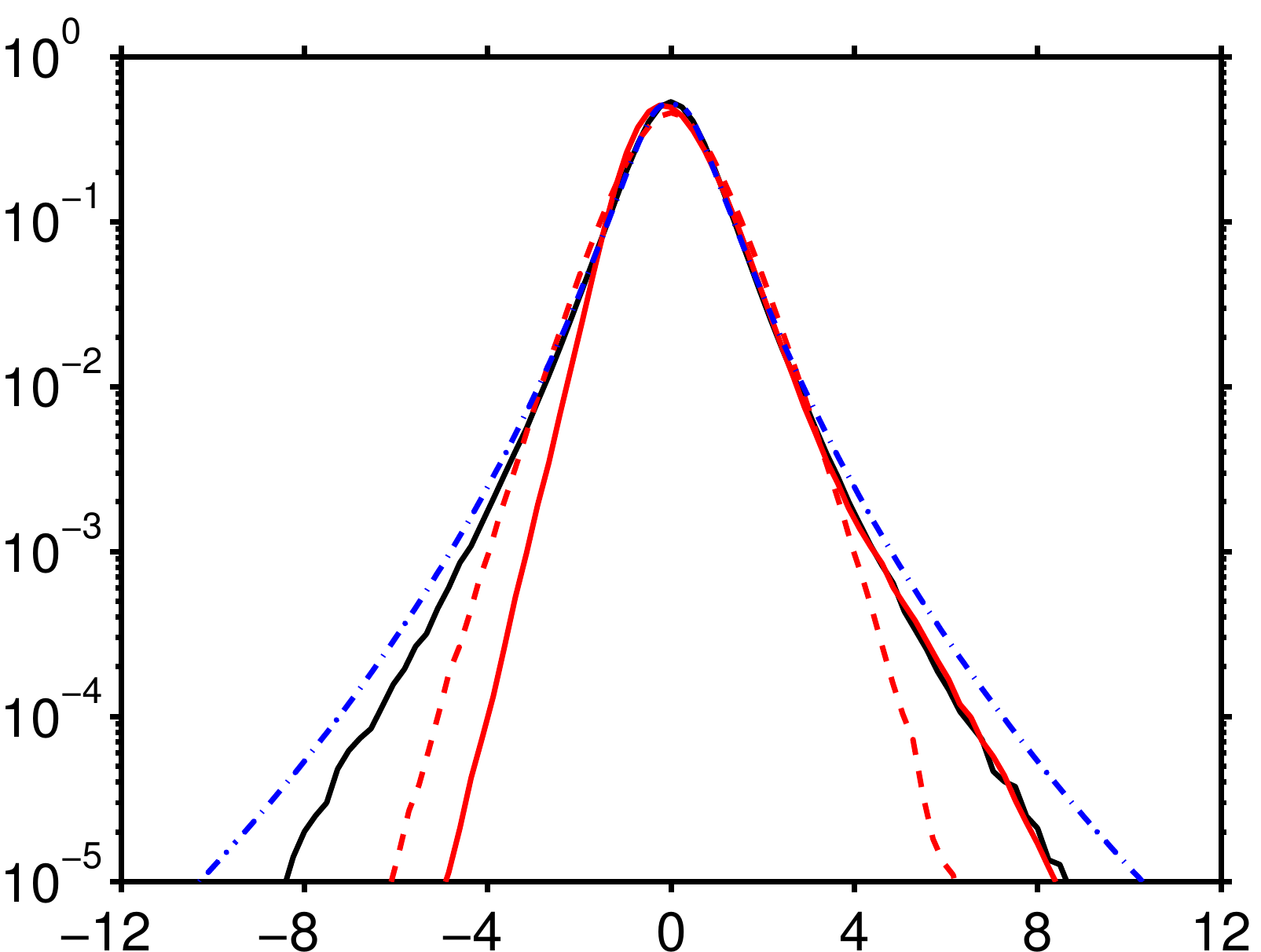}
    \\
    \centerline{$a_p^\prime/\sigma(a_p^\prime)$}
  \end{minipage}
  \caption{%
    Normalized probability density functions of particle
    acceleration. 
    Line styles correspond to: 
    {\color{black}\solidthick},~case~A-G0; 
    {\color{red}\solidthick},~vertical component in case~AL-G120; 
    {\color{red}\dashed},~horizontal component in case~AL-G120; 
    {\color{blue}\chndot},~log-normal fit of Qureshi et
    al.\cite{qureshi:08}.   
  }
  \label{fig-two-phase-fluid-acc-pdf}
\end{figure}
Let us now turn to the linear particle acceleration
$\mathbf{a}_p=(a_{p,1},a_{p,2},a_{p,3})$, which is obtained from the
following equation:
\begin{equation}\label{equ-def-particle-acc}
  \rho_pV_p\mathbf{a}_p
  =
  \mathbf{F}^{(H)}
  +
  \mathbf{F}^{(B)}
  +
  \mathbf{F}^{(C)}
  \,,
\end{equation}
where $\mathbf{F}^{(H)}$ is the hydrodynamic force, $\mathbf{F}^{(B)}$
the submerged weight and $\mathbf{F}^{(C)}$ the force contribution
from particle-particle contact (cf.\ appendix~\ref{sec-app-ekin} for
details of the present numerical implementation). 
The statistics of particle acceleration presented in the present
section have been computed after removing all samples corresponding to
colliding spheres\cite{villalba:12}. 
Let us note at this point that the mean collision-free interval for a
particle amounts to $4.2T_e$ ($2.7T_e$) in case~A-G0 (AL-G120). 

The particle acceleration variance normalized in Kolmogorov scales 
$A_\alpha\equiv
\langle a_{p,\alpha}^\prime
a_{p,\alpha}^\prime\rangle_t\eta^{2/3}/\varepsilon^{4/3}$  
(no summation over the index for the spatial direction $\alpha$)
measures $A_\alpha=1.07$ in case~A-G0.  
This value is close to the one in case S2 of Yeo et al.\cite{YeoClimentMaxey:10}
who obtain $A_\alpha=0.88$. The difference might be related to the
effect of inter-particle contact in those author's simulations
performed at a substantially higher solid volume fraction.
Note that the acceleration variance data of Homann \&
Bec\cite{homann:10} \revision{is}{are} not directly comparable to our
present data, since their simulation was run at much lower Reynolds
number, and it is well known at least for fluid
particles\cite{vedula:99} that the value of $A_\alpha$ exhibits a
significant Reynolds-number dependence.   

The p.d.f.\ of particle acceleration is shown in
figure~\ref{fig-two-phase-fluid-acc-pdf} along with a log-normal
distribution as first proposed by Mordant et al.\cite{mordant:04} and
fitted to experimental data by Qureshi et al.\cite{qureshi:07} with a
value of the kurtosis of $K=8.37$; 
this fit was later shown\cite{qureshi:08} to provide a
reasonable match of experimentally determined particle acceleration
p.d.f.s over a range of relative particle sizes %
and density ratios. %
In the figure it can be seen that the acceleration data of our
case~A-G0 \revision{is}{are} consistent with the experimental fit. Similar to Yeo et
al.\cite{YeoClimentMaxey:10}, however, we obtain a smaller value of the kurtosis
of $K=6.46$ (their reported value is $K=6.24$). 
The data of Homann \& Bec\cite{homann:10} for neutrally-buoyant single
particles \revision{suggests}{suggest} a decrease of the kurtosis from values of
$K\approx8$ to $K\approx6$ taking place at $D/\eta\approx7$; 
however, due to the smaller number of samples in that study
their data \revision{does}{do} not allow for a more precise comparison. 

When gravity is non-zero, the particle acceleration statistics become
non-isotropic. The vertical (horizontal) component in case~AL-G120
exhibits a kurtosis of $K=5.44$ ($K=3.94$), i.e.\ values which are
both smaller than in the zero-gravity case. 
In addition the vertical component is positively skewed with
$S=0.63$. 
While the reduction of the kurtosis in the presence of gravity
deserves further analysis, the skewness of the vertical component of
particle acceleration can be linked to the non-linear drag effect
already discussed in section~\ref{sec-two-phase-time-evol}. A similar
observation has been made in the context of particulate channel
flow\cite{villalba:12}.  

\begin{figure} %
   \begin{minipage}{2ex}
     \rotatebox{90}{$\rho_a(\tau)$}
   \end{minipage}
   \begin{minipage}{0.45\linewidth}
     \centerline{$(a)$}
     \centerline{$\tau/\tau_{\eta}$}
      \includegraphics[width=\linewidth]
      {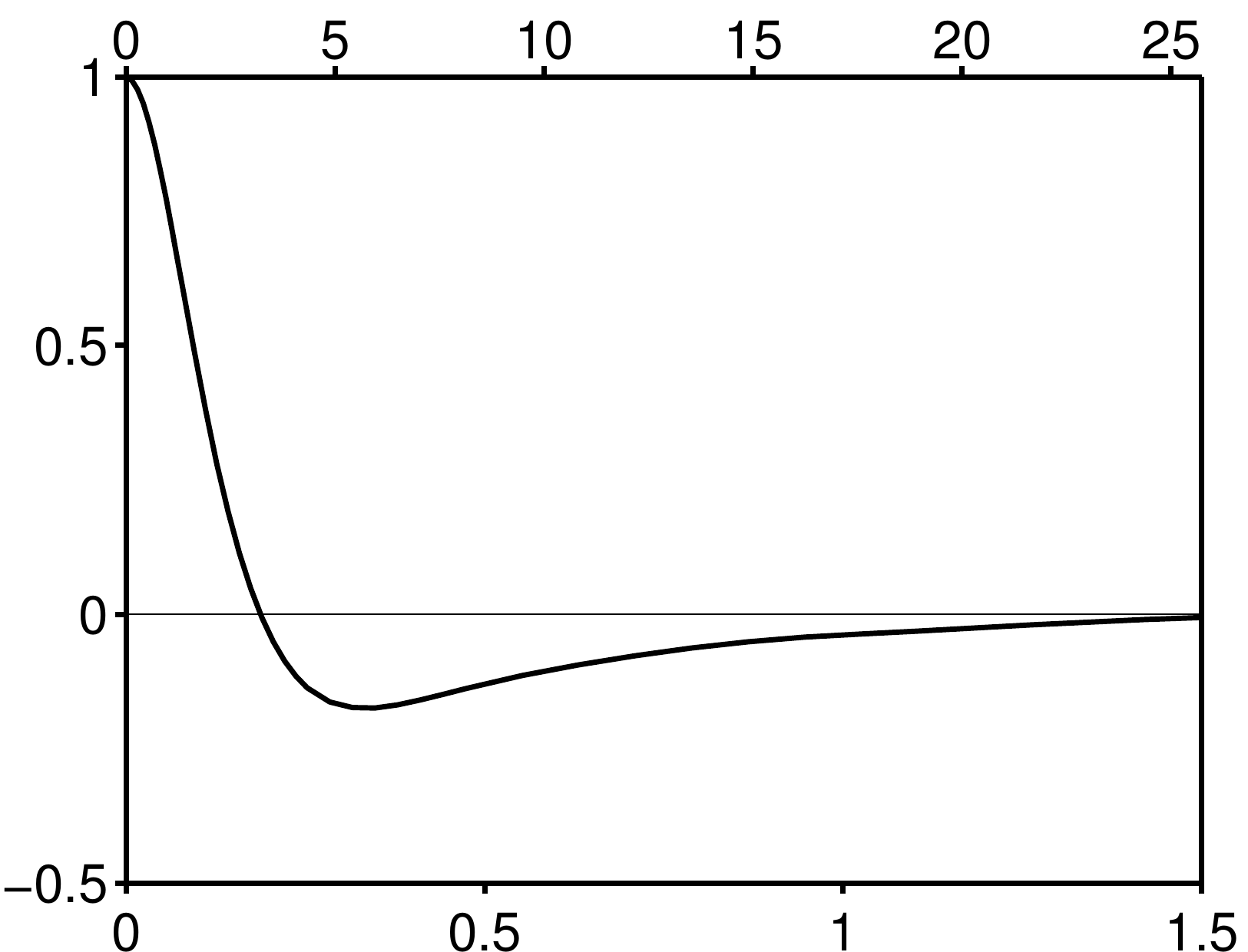}
      \\
      \centerline{$\tau/T_e$}
   \end{minipage}
   \begin{minipage}{2ex}
     \rotatebox{90}{$\rho_a(\tau)$}
   \end{minipage}
   \begin{minipage}{0.45\linewidth}
     \centerline{$(b)$}
     \centerline{$\tau/\tau_{\tilde{\eta}}$}
      \includegraphics[width=\linewidth]
      {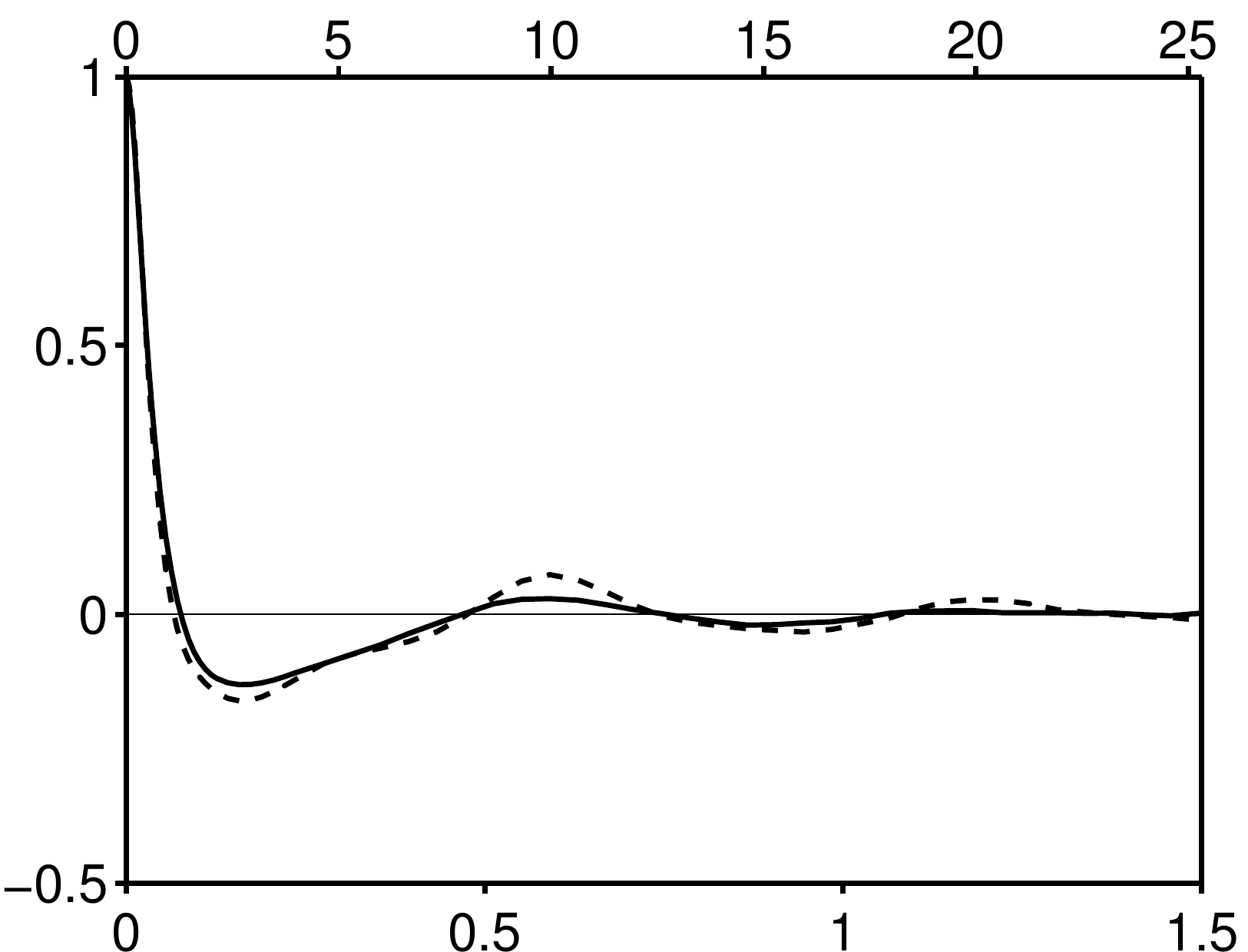}
      \\
      \centerline{$\tau/T_{\tilde{e}}$}
   \end{minipage}
   \caption{%
     Auto-correlation function of particle acceleration: 
     $(a)$ case A-G0; 
     $(b)$ case AL-G120.
     In $(b)$ the solid line corresponds to the vertical component,
     while the dashed line stands for the horizontal ones. 
     Note also that in $(b)$ the modified dissipation
     $\tilde{\varepsilon}$ (defined in
     \ref{equ-ek-box-avg-stat-avg-approx-1}) has been used to compute
     modified time-scales $\tau_{\tilde{\eta}}$ and $T_{\tilde{e}}$.   
   }
   \label{fig-two-phase-fluid-acc-auto-corr}
\end{figure}
Information on the time scales of forces acting on the particles is
provided in figure~\ref{fig-two-phase-fluid-acc-auto-corr} which shows
the auto-correlation of acceleration, $\rho_a(\tau)$. 
The shape of the auto-correlation function in case~A-G0 is fully
consistent with the data of Ref.\cite{YeoClimentMaxey:10,homann:10}, 
exhibiting first a rapid decay, a subsequent negative loop, and a
final slow approach to zero from below.  
The first zero-crossing in the present case~A-G0 occurs at
$\tau_{zero}/\tau_\eta=3.11$, while 
Yeo et al.\cite{YeoClimentMaxey:10} obtain $\tau_{zero}/\tau_\eta=3.2$ in their
case S2,  
and Homann \& Bec\cite{homann:10} report a value of
$\tau_{zero}/\tau_\eta=2.5$ for a single neutrally-buoyant particle
with $D/\eta=6$.  

In comparison the auto-correlation in case~AL-G120 exhibits a much
faster initial decay, with a zero-crossing at
$\tau_{zero}/\tau_{\tilde{\eta}}=1.26$ ($1.13$) for the vertical (horizontal)
component. This faster decorrelation can 
be attributed to the crossing-trajectory effect\citep{csanady:63} due
to settling particles spending a smaller amount of time under the
influence of a given flow structure as compared to a zero-gravity
case. 
It can also be observed that the auto-correlation in case~AL-G120
mildly oscillates  around the zero value.
Although the amplitude of the oscillation is very small ($0.03$ and
$0.07$ in the vertical and horizontal component, respectively), it is
instructive to investigate it further. 
The oscillation period measures $0.59T_e$, which 
corresponds roughly to one half of the value of the
particle return time based upon the time-averaged settling
velocity and the vertical extent of the periodic computational domain,
viz.\ $T_r={\cal L}_z/\langle w_{rel}\rangle_t=1.23T_e$. 
Therefore, this slightly oscillatory behavior of $\rho_a(\tau)$ is
an artifact of the finite vertical domain size. 
Since the secondary maximum of $\rho_a$ is located at a temporal
separation equal to one half of the return time, 
and because this time interval is not larger than the large-eddy
time-scale, the oscillations of $\rho_a$ are most likely caused by
particles having a certain probability to interact 
repeatedly 
with the same flow structure while settling through the vertically
periodic domain. 
Consequently, this feature can be linked to the two-point correlation
function of figure~\ref{fig-single-phase-isotropy-param-2p-corr}$(b)$
which was found to remain finite at the largest separations in the
vertical direction. 
It can therefore be concluded that the presently chosen vertical
elongation of the computational domain does not completely prevent the 
Lagrangian signals to be affected by the finite return time. 
It does, however, alleviate the restrictions stemming from the long
vertical extent of the wakes left behind by settling particles. 
\section{Summary and conclusion}
In the present study we have considered the motion of finite-size,
inertial, spherical particles in statistically stationary homogeneous
turbulence without mean velocity gradients. 
The Navier-Stokes equations for a constant density fluid are solved
with the aid of an immersed boundary technique for imposing the
no-slip condition at the fluid-particle interfaces while using a
fixed computational grid. 
The turbulent motion is generated by the large-scale random forcing
procedure of Eswaran and Pope\cite{eswaran:88}. 
Since only a limited number of Fourier modes are forced, full
transforms of the forcing fields can be avoided through a direct
evaluation of the Fourier series in physical space. This means that
the computation of the forcing term can be done efficiently in local
memory, avoiding any inter-processor communication. 

Since our objective is to 
understand the influence of gravity upon the fluid-particle
interaction, special care has been taken to address the requirements
due to particles which are upon average settling with respect to the
mean fluid-phase velocity. 
Therefore, we have considered the case of
isotropically-forced turbulence in vertically elongated computational
domains. 

We have carried out simulations in medium-sized systems at low to
moderate turbulent Reynolds numbers. An initial validation is
performed for single-phase flow. The comparison with reference
data\cite{jimenez:93} in cubical boxes shows an excellent agreement at
the corresponding Reynolds number. 
Concerning elongated boxes we obtain essentially the same flow
statistics (including a very good isotropy) when the domain is
elongated by a factor of two in one spatial direction while only every
second Fourier wavenumber in that direction is forced. 
Since the spectral energy content of the largest modes which are not
forced (i.e.\ those interspersed between the forced ones) does not
remain zero under the dynamics of the Navier-Stokes equations,
the simulated field in an elongated box is not a simple copy of a
corresponding field from a cubical simulation. 
At the same time two-point correlation data \revision{shows}{show} that the fields in the
cubical sub-volumes of the elongated simulation remain somewhat
correlated. 
However, we also show that once settling particles are added to the
simulation in a vertically-elongated box, the vertical two-point
correlation of the velocity field is reduced by the fact that the
particle positions only possess the fundamental periodicity.  

Particulate flow simulations involving a dilute suspension of inertial
particles at Galileo numbers $Ga=0$ and $120$ have been performed at
low turbulent Reynolds number. 
The first conclusion is that the random forcing procedure does not
pose any numerical stability problems, allowing us to obtain a
well-defined statistically stationary state and to collect statistics
over arbitrarily long temporal intervals. 
The zero-gravity case has been thoroughly validated with respect to
available experimental and numerical data. 

Concerning the particulate flow case with finite gravity, the present
simulations suggest that the average particle settling velocity is
reduced by 1.5\% with respect to the one of an isolated particle in
quiescent surroundings. 
An evaluation of the standard non-linear drag model\cite{homann:13}
with the present parameters yields the correct trend, 
and non-linear drag is also a likely candidate for the observed positive
skewness of the p.d.f.\ of vertical particle acceleration. 
On the other hand, the p.d.f.s of the vertical component of the fluid
and of the particle velocity in this case become negatively skewed,
presumably as a result of the presence of mean wakes, and similarly to
previous results in vertical plane channel flow\cite{uhlmann:08a}. 
Finally, the crossing-trajectories effect leads to a substantial
reduction of the auto-correlation of the particle acceleration in the
finite-gravity case as compared to zero gravity. 

The present study shows that the simulation of forced turbulent flow
seeded with fully-resolved, settling particles, satisfying the
requirements mentioned in the introduction, is feasible today. 
Our first results at relatively low Reynolds number obtained in
medium-sized systems  
already bear some interesting features. 
In the future we will consider larger systems while increasing 
the Reynolds number. This will allow us to investigate in more detail
the interaction between the finite-size particles and the turbulent
flow structures. 
\begin{acknowledgments}
  Fruitful discussions with T.\ Doychev are gratefully acknowledged. 
  Thanks is also due to M.\ Garc\'ia-Villalba who provided helpful
  comments on a draft of this manuscript. 
  This work was supported by the German Research Foundation (DFG) under
  project UH 242/1-2. 
  The simulations where partially performed at LRZ M\"unchen (grant
  pr83la) and at SCC Karlsruhe. 
  The computer resources, technical expertise as well as assistance
  provided by the staff are gratefully acknowledged. 
\end{acknowledgments}

\begin{appendix}
\section{Evaluating the turbulence forcing in physical space}
\label{sec-app-fou-trans}
The formula for evaluating the turbulence forcing term at a discrete 
physical space grid node $\mathbf{x}_{ijk}=(x_i,y_j,z_k)$ from given
discrete Fourier 
coefficients $\hat{\mathbf{f}}^{(t)}(\boldsymbol{\kappa}_{lmn})
=\hat{\mathbf{f}}^{(t)}(\kappa_{1,l},\kappa_{2,m},\kappa_{3,n})$ 
can be written as follows: 
\begin{equation}\label{equ-forcing-transform-fou-fis-naive}
  \mathbf{f}^{(t)}(\mathbf{x}_{ijk})
  =
  \sum_{l=-N_f}^{N_f}
  \sum_{m=-N_f}^{N_f}
  \sum_{n=-N_f}^{N_f}
  \left(
    \hat{\mathbf{f}}_{lmn}^{(t)}             
    \exp(I \kappa_{1,l} x_i)
    \exp(I \kappa_{2,m} y_j)
    \exp(I \kappa_{3,n} z_k)
  \right)
  \,,
  \quad
  \forall\,\,i,j,k
  \,,
\end{equation}
where for each direction $N_f$ is the number of positive Fourier
wavenumbers which receive a non-zero forcing contribution. 
The $i$th element of the wavenumber vector in the $\alpha$-direction
is defined as:
\begin{equation}\label{equ-def-wavevector-elt}
  \kappa_{\alpha,i}
  =
  \frac{2\pi i}{{\cal L}_\alpha}
  \,.
\end{equation}
Note that the coefficients for the negative wavenumbers 
in one of the coordinate directions 
are actually
reconstructed from the corresponding positive ones taking into account
that the result in physical space is real-valued.  
A direct implementation of (\ref{equ-forcing-transform-fou-fis-naive})
corresponds to a sextuple loop with 
${\cal O}(N_f^3N_{x,loc}N_{y,loc}N_{z,loc})$ 
operations, where 
$N_{x,loc}$, $N_{y,loc}$, $N_{z,loc}$ 
are the numbers of physical space grid points in each coordinate
direction which are held in local memory by a given process in
a three-dimensional Cartesian domain decomposition. 

In practice the summation in
(\ref{equ-forcing-transform-fou-fis-naive}) is split up into three
consecutive steps, as shown in the following:
\begin{subequations}\label{equ-forcing-transform-fou-fis-step}
\begin{eqnarray}\label{equ-forcing-transform-fou-fis-step-1}
  \mathbf{A}_{mni}
  &=&
  \sum_{l=-N_f}^{N_f}
  \left(
    \hat{\mathbf{f}}_{lmn}^{(t)}             
    \exp(I \kappa_{1,l} x_i)
  \right)
  \,,
  \quad
  \forall\,\,m,n,i
  \,,
  \\
  \mathbf{B}_{nij}
  &=&
  \sum_{m=-N_f}^{N_f}
  \left(
    \mathbf{A}_{mni}
    \exp(I \kappa_{2,m} y_j)
  \right)
  \,,
  \quad
  \forall\,\,n,i,j
  \,,
  \\
  \mathbf{f}^{(t)}(\mathbf{x}_{ijk})
  &=&
  \sum_{n=-N_f}^{N_f}
  \left(
    \mathbf{B}_{nij}
    \exp(I \kappa_{3,n} z_k)
  \right)
  \,,
  \quad
  \forall\,\,i,j,k
  \,.
\end{eqnarray}
\end{subequations}
The operation count for each of the steps in
(\ref{equ-forcing-transform-fou-fis-step}) is  
${\cal O}(N_f^3N_{x,loc})$, 
${\cal O}(N_f^2N_{x,loc}N_{y,loc})$, and  
${\cal O}(N_fN_{x,loc}N_{y,loc}N_{z,loc})$, 
respectively. 
\section{The time-splitting algorithm for particulate flow with
  turbulence forcing}
\label{sec-app-time-scheme}
In the present numerical method the flow equations are discretized in
time with the aid of a three-step Runge-Kutta scheme for the advection
terms in conjunction with a Crank-Nicolson scheme for the viscous
terms. In each Runge-Kutta sub-step (with super-script index ``$k$'') a 
fractional step procedure is employed, consisting of: 
(i) a first predictor step which serves to compute a preliminary flow
field $\tilde{\mathbf{u}}$ from which the necessary force density
$\mathbf{f}^{(ibm)}$ for imposing a no-slip condition at the
fluid-solid interface of each particle is deduced; 
(ii) a second predictor step yielding a solution $\mathbf{u}^\ast$ to
the momentum equation which does in general not verify the
divergence-free condition;  
(iii) a corrector step which yields a divergence-free velocity
field $\mathbf{u}^{k}$ through a pressure-projection operation.
Therefore, the time-discretized Navier-Stokes equations for the $k$th
Runge-Kutta step can be expressed as the following sequence of
operations: 
\begin{subequations}\label{equ-hybrid-discr-algo}
\begin{eqnarray}\nonumber
\tilde{\mathbf{u}}&=&
\mathbf{u}^{k-1}
+\Delta t\left(
2\alpha_k\nu\nabla^2\mathbf{u}^{k-1}
-2\alpha_k\nabla p^{k-1}\right.
\\\label{equ-hybrid-discr-algo-lag-rhs}
&&\left.
-\gamma_k\left((\mathbf{u}\cdot\nabla)\mathbf{u}\right)^{k-1}
-\zeta_k\left((\mathbf{u}\cdot\nabla)\mathbf{u}\right)^{k-2}
+2\alpha_k\mathbf{f}^{(t)}(t^n)
\right)
\,,
\\\label{equ-hybrid-discr-algo-lag-interpol}
\tilde{{U}}_\beta(\mathbf{X}_l^{(m)})&=&
\sum_{ijk}\tilde{{u}}_\beta(\mathbf{x}_{ijk}^{(\beta)})\,
\delta_h(\mathbf{x}_{ijk}^{(\beta)}-\mathbf{X}_l^{(m)})\,\Delta x^3
\,,
\quad\forall\,l;\,m;\,\beta
\\\label{equ-hybrid-discr-algo-lag-force}
\mathbf{F}(\mathbf{X}_l^{(m)})&=&
\frac{\mathbf{U}^{(d)}(\mathbf{X}_l^{(m)})
-\tilde{\mathbf{U}}(\mathbf{X}_l^{(m)})}{\Delta t}
\,,
\quad\qquad\qquad\forall\,l;\,m
\\\label{equ-hybrid-discr-algo-eul-force}
{f}_\beta^{(ibm)}(\mathbf{x}_{ijk})&=&
\sum_{m=1}^{N_p}
\sum_{l=1}^{N_L}{F}_\beta(\mathbf{X}_l^{(m)})\,
\delta_h(\mathbf{x}_{ijk}^{(\beta)}-\mathbf{X}_l^{(m)})\,\Delta V_l^{(m)}
\,,
\quad\forall\,\beta;\,i;\,j;\,k
\\\label{equ-hybrid-discr-algo-predict}
\nabla^2\mathbf{u}^\ast-\frac{\mathbf{u}^\ast}{\alpha_k\nu\Delta t}&=&
-\frac{1}{\nu\alpha_k}\left(\frac{\tilde{\mathbf{u}}}{\Delta
      t}+\mathbf{f}^{(ibm)}-\langle\mathbf{f}^{(ibm)}\rangle_{\Omega}\right) 
  +\nabla^2\mathbf{u}^{k-1}
\,,
\\\label{equ-hybrid-discr-algo-poisson}
\nabla^2\phi&=&\frac{\nabla\cdot\mathbf{u}^\ast}{2\alpha_k\Delta t}\,,
\\\label{equ-hybrid-discr-algo-update-vel}
\mathbf{u}^{k}&=&\mathbf{u}^\ast-2\alpha_k\Delta
t\nabla\phi\,,
\\\label{equ-hybrid-discr-algo-update-press}
p^{k}&=&p^{k-1}+\phi-\alpha_k\Delta t\,\nu\nabla^2\phi
\,,
\end{eqnarray}
\end{subequations}
where the set of coefficients $\alpha_k$, $\gamma_k$, $\zeta_k$ (with
$k=1,2,3$) is given in Ref.\citenum{rai:91}. The intermediate variable $\phi$ 
is the so-called ``pseudo-pressure'', which together with the
predictor velocity fields $\tilde{\mathbf{u}}$ and
$\mathbf{u}^\ast$ is discarded at the end of the Runge-Kutta
sub-step. 
The upper-case letters denote quantities evaluated at a set of 
Lagrangian force points located at positions $\mathbf{X}_l^{(m)}$ on
the surface of the $m$th particle (out of a total number of $N_p$
particles), the sub-script ranging as $l=1\ldots N_L$. 
The transfer between Eulerian coordinates (i.e.\ staggered grid nodes 
$\mathbf{x}_{ijk}^{(\beta)}$ associated with the $\beta$-component of
velocity) and Lagrangian force points is performed with the aid of
Peskin's regularized delta function $\delta_h$, with a forcing volume
$\Delta V_l^{(m)}$ associated to each Lagrangian force point. 
The quantity $\mathbf{U}^{(d)}(\mathbf{X}_l^{(m)})$ in
(\ref{equ-hybrid-discr-algo-lag-force}) represents the current local
velocity of the $m$th solid particle's force point with index $l$. It
is directly linked to the current values of the particle's
translational and angular velocity, which in turn, are obtained from
integrating the Newton equations for rigid body motion. 
Also note that the turbulence forcing field $\mathbf{f}^{(t)}$ is not
updated during the Runge-Kutta sub-steps; instead it is evaluated once
per full time step.  

In the case of triply-periodic boundary conditions, the 
spatial average of the force term
needs to be subtracted from the momentum equation in order to allow
the system to attain a statistically stationary
state~\cite{fogelson:88,hoefler:00}.  
More specifically, at each time step we compute the following average 
\begin{equation}\label{equ-app-time-splitting-periodic-driving-term}
  \langle\mathbf{f}^{(ibm)}\rangle_{\Omega}(t)
  =
  \frac{1}{||\Omega||}
  \int_\Omega\mathbf{f}^{(ibm)}(\mathbf{x},t)\mbox{d}\mathbf{x}
  \,,
\end{equation}
which is then subtracted from the momentum balance (cf.\ Helmholtz
equation~\ref{equ-hybrid-discr-algo-predict} during the predictor step).  
\section{Definition of averaging operators}
\label{sec-app-avg-ops}
Here we use the same averaging procedures as in Ref.\citenum{uhlmann:14a}
which will be briefly summarized in the following. 
We define a fluid-phase indicator function $\Phi_f(\mathbf{x},t)$, 
which specifies whether a given point $\mathbf{x}$ at time $t$ is
located inside the region $\Omega_f(t)$ occupied by the fluid at time
$t$, viz.
\begin{equation}
  \Phi_f(\mathbf{x},t) = \left\{
    \begin{array}{l l}
      1 \qquad \mbox{if} \quad \mathbf{x} \in \Omega_f(t) \\
      0 \qquad \mbox{else} \\
    \end{array} \right. 
  \,,
  \label{eq:Indicator_f}
\end{equation}
and the corresponding indicator function for the particle phase
$\Phi_p(\mathbf{x},t)$ is then simply given by $\Phi_p=1-\Phi_f$. 
The fluid-phase averaging for any fluid-related quantity
$\psi_f(\mathbf{x})$ is denoted by the operator $\langle
\cdot\rangle_{\Omega_f}$, 
while the indiscriminate average (over the entire computational
domain) is denoted by $\langle \cdot \rangle_{\Omega}$. 
These two averages are related by the following identity: 
\begin{equation}
  \label{equ-app-ident-fluid-only-comp-avg}
  \langle\psi_f  \rangle_{\Omega_f}
  =
  \frac{||\Omega||}{||\Omega_f||} \langle \psi_f \, \Phi_f
  \rangle_{\Omega}
  \,.
\end{equation}
The operator $\langle \cdot \rangle_p$ for a given quantity
$\psi_p^{(i)}$ related to the $i$th particle represents the
instantaneous average over the number of particles.  
Finally, the time average is indicated by the operator $\langle \cdot
\rangle_{t}$. 
\section{A priori estimation of turbulence forcing parameters}
\label{sec-single-phase-parametrization}
Eswaran and Pope\cite{eswaran:88} provide estimates of the dissipation
rate, the Kolmogorov length scale and the Reynolds number based upon
the values chosen for the free parameters of their forcing scheme
(i.e.\ $T_L$, $\kappa_f$, $\epsilon^\ast$).   
They first introduce a non-dimensional forcing time scale 
$T_L^{*}=T_L(\varepsilon^*)^{1/3}\kappa_0^{2/3}$ (where $\kappa_0$ is
the lowest wavenumber in Fourier space). 
It is then proposed that the quantity 
$\varepsilon_{T}^{*}=4\varepsilon^{*}N_f/(1+T_L^*N_f^{1/3}/\beta)$ 
should serve as an estimate for the dissipation rate ($N_f$
being the total number of forced wavenumbers, and $\beta=0.8$ an
adjusted constant). 
From this an estimate for the Kolmogorov length can be obtained from 
$\eta_T=(\nu^3/\varepsilon_T^*)^{1/4}$. 
Furthermore, Eswaran and Pope\cite{eswaran:88} suggest the following
empirical expression for estimating the Taylor-scale Reynolds number: 
$Re_\lambda^T=8.5/((\eta_T \kappa_0)^{5/6} N_f^{2/9})$. 

In our simulations we have indeed confirmed that the Kolmogorov length
$\eta$ is reasonably well predicted by the quantity $\eta_T$. 
However, the Reynolds number $Re_\lambda$ is overpredicted by the
quantity $Re_\lambda^T$ 
(by approximately 14\% in case~A and by 28\% in case~B). 
This discrepancy might be due to the fact that we are considering here 
Reynolds numbers larger than the ones used in the original work.

In what follows an alternative estimate is proposed. 
We define a macroscopic scale $\mathcal{L}_c=2\pi/\kappa_c$ 
from the central wavenumber of the forced interval 
$\kappa_c=(\kappa_0+\kappa_f)/2$, which should take values close to
the large-eddy length-scale $L$. 
We can then derive a large-scale Reynolds number 
$Re_c=\mathcal{L}_c u_{rms}/\nu\approx Re_L$. 
Therein $u_{rms}$ is estimated by assuming that $\varepsilon
\approx\varepsilon_{T}^{*}$ and that $T_e=u^2_{rms}/\varepsilon\approx
T_L$. 
Using the relation between large-scale Reynolds number and
Taylor-scale Reynolds number\cite{pope:00}, $Re_\lambda=(20
  Re_L/3)^{1/2}$, we arrive at the following estimate based upon the
  forcing parameters: 
$Re_\lambda^{T2}
=
(20{\cal L}_c(T_L\varepsilon_{T}^{*})^{1/2}/(3\nu))^{1/2}$. 
It is found that the resulting values of $Re_\lambda^{T2}$ 
provide a better match to our simulation data, 
with 8\% (3\%) discrepancy with respect to $Re_\lambda$ in case~A
(case~B).  

In the main text we use as a priori reference scales for normalizing
the dissipation rate and the kinetic energy the quantities 
$\varepsilon_{ref}=\varepsilon_{T}^{*}$ and
$k_{ref}=3\varepsilon_{T}^{*}T_L/2$, respectively, which are purely
based upon the imposed forcing parameters. 
\revision{}{%
  \section{Fluid velocity probabilities in the single-phase case}
  \label{sec-single-phase-fluid-vel-pdf}
  \begin{figure}%
    \begin{minipage}{1ex}
      \rotatebox{90}{pdf}
    \end{minipage}
    \begin{minipage}{0.46\linewidth}
     \centerline{$(a)$}
       \includegraphics[width=\linewidth]
       {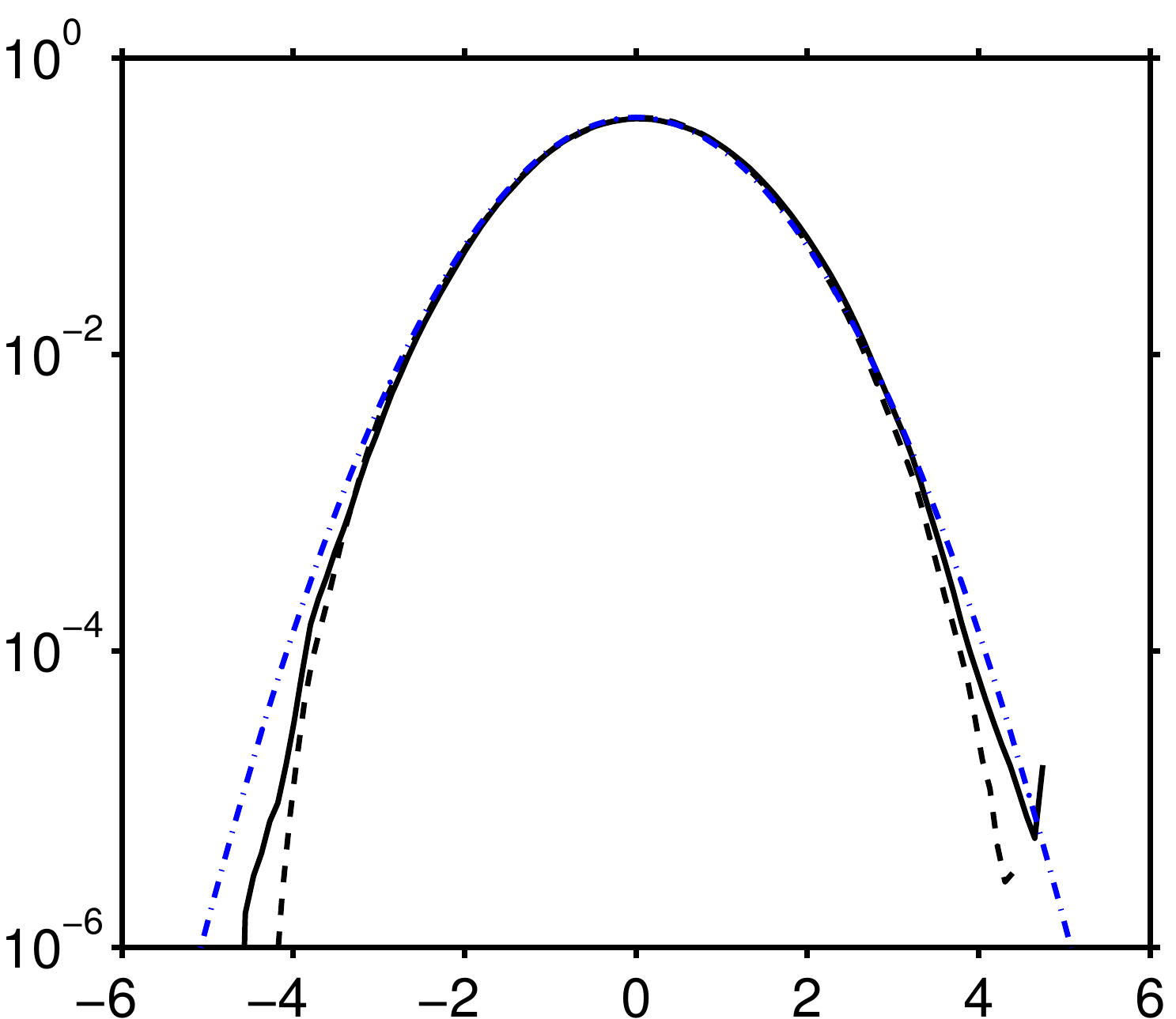}
     \centerline{$u^\prime/u_{rms}$}
    \end{minipage}
    \hfill
    \begin{minipage}{1ex}
      \rotatebox{90}{pdf}
    \end{minipage}
    \begin{minipage}{0.46\linewidth}
      \centerline{$(b)$}
       \includegraphics[width=\linewidth]
       {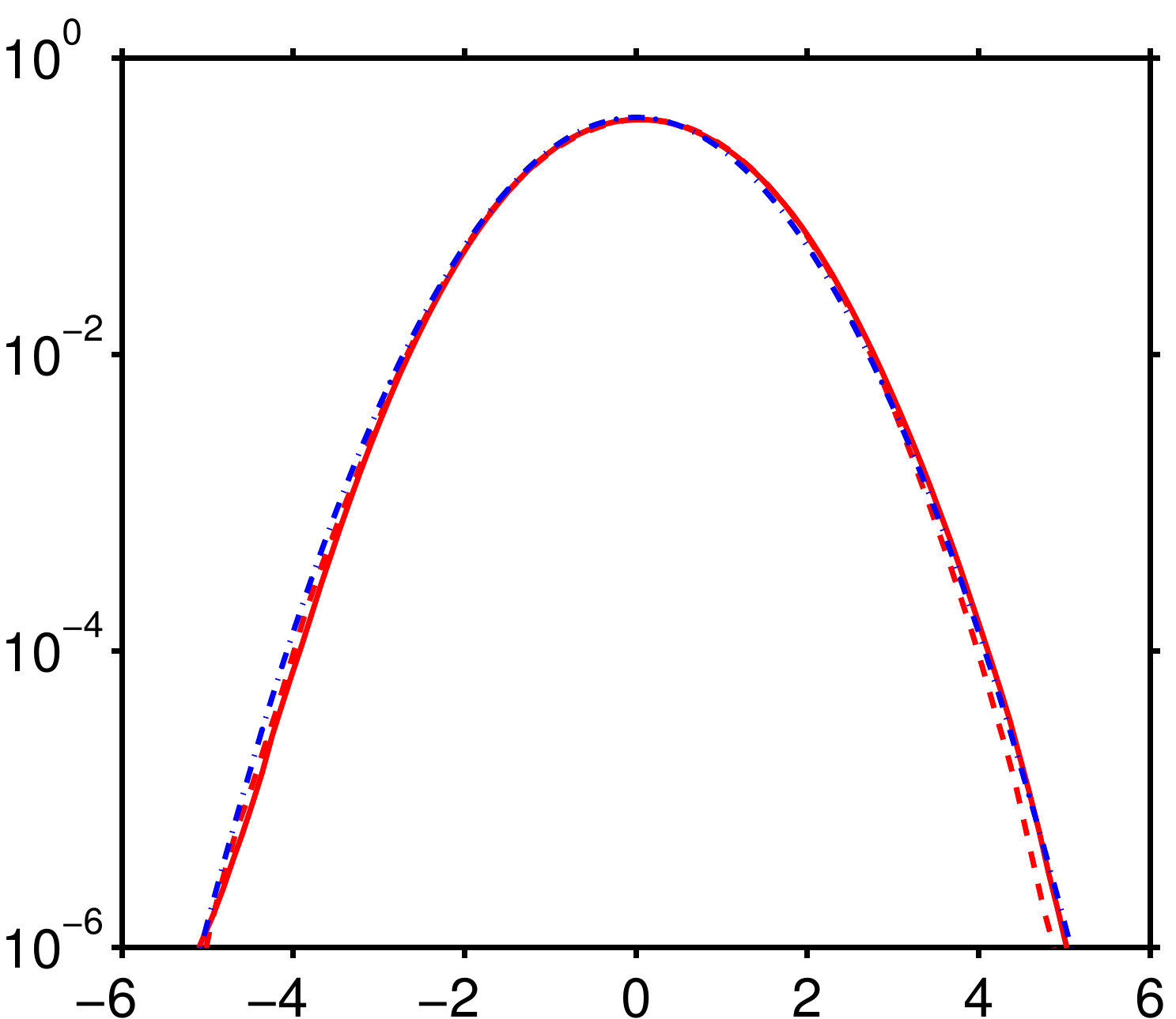}
     \centerline{$u^\prime/u_{rms}$}
    \end{minipage}
    \caption{%
      \revision{}{%
        Probability density functions of fluid velocity fluctuations
        in the single-phase case.  
        Data at lower Reynolds number is shown in $(a)$: 
        {\color{black}\solidthick},~case A, 
        {\color{black}\dashed},~case AL. 
        The graph in $(b)$ collects the data at higher Reynolds number: 
        {\color{red}\solidthick},~case B, 
        {\color{red}\dashed},~case BL. 
        The blue chain-dotted line shows a Gaussian reference curve. 
      }
    }
    \label{fig-single-phase-pdf-uf}
  \end{figure}
For the sake of completeness, the p.d.f.\ of the fluid velocity
flucutations in the single-phase cases of
section~\ref{sec-single-phase} are shown in
figure~\ref{fig-single-phase-pdf-uf}. 
It can be seen that while at lower Reynolds number there exists a
slight deviation from Gaussianity, the higher Reynolds data almost
perfectly collapse with the Gaussian reference curve. 
}
\section{Reformulation of the kinetic energy budget}
\label{sec-app-ekin}
The purpose of the present appendix is to bring the two-way
coupling term (\ref{equ-def-two-way-coupling-box-avg}) in the
box-averaged kinetic energy equation (\ref{equ-ek-box-avg}) into  
a more tractable form. 

Since the force which imposes the no-slip condition at the
fluid-particle interfaces, $\mathbf{f}^{(ibm)}$, is localized around
the particles, any volume integral involving $\mathbf{f}^{(ibm)}$ can
be expressed as a sum of integrals over each particle's surface, e.g.\
we can write: 
\begin{equation}\label{equ-integral-u-dot-force-ibm}
  \langle\mathbf{u}\cdot\mathbf{f}^{(ibm)}\rangle_\Omega
  \equiv
  \frac{1}{||\Omega||}
  \int_{\Omega}
    \left(
      \mathbf{u}\cdot\mathbf{f}^{(ibm)}
    \right)
    \mbox{d}\mathbf{x}
  =
  \frac{1}{||\Omega||}
  \sum_{i=1}^{N_p}
  \int_{\mathbf{x}\in{\cal S}^{(i)}}
    \left(
      \mathbf{u}\cdot\mathbf{f}^{(ibm)}
    \right)
    \mbox{d}\mathbf{x}
    \,,
\end{equation}
where $||\Omega||$ is the volume of the triply-periodic computational
domain and ${\cal S}^{(i)}$ is the surface of the $i$th particle.
Substituting the no-slip value at the surface of the particle, i.e.\ 
\begin{equation}\label{equ-def-particle-surface-vel}
  \mathbf{u}(\mathbf{x}\in{\cal S}^{(i)}(t),t)
  =
  \mathbf{u}_p^{(i)}(t)
  +
  \boldsymbol{\omega}_p^{(i)}(t)\times
  \mathbf{r}^{(i)}(\mathbf{x},t)
  \,,
\end{equation}
into (\ref{equ-integral-u-dot-force-ibm}) yields
\begin{eqnarray}\label{equ-integral-u-dot-force-ibm-2}
  \langle\mathbf{u}\cdot\mathbf{f}^{(ibm)}\rangle_\Omega
    &=&
  \frac{1}{||\Omega||}
  \sum_{i=1}^{N_p}
  \left[
    \mathbf{u}_p^{(i)}
    \cdot
    \int_{\mathbf{x}\in{\cal S}^{(i)}}
    \mathbf{f}^{(ibm)}
    \mbox{d}\mathbf{x}
    +
    \boldsymbol{\omega}_p^{(i)}
    \cdot
    \int_{\mathbf{x}\in{\cal S}^{(i)}}
    \left(
      \mathbf{r}^{(i)}
      \times
      \mathbf{f}^{(ibm)}
    \right)
    \mbox{d}\mathbf{x}
  \right]
  ,
\end{eqnarray}
where the vector $\mathbf{r}$ pointing from the $i$th
particle's center to a location $\mathbf{x}$ on its surface has been
defined as
$\mathbf{r}^{(i)}(\mathbf{x},t)=\mathbf{x}-\mathbf{x}_p^{(i)}(t)$, 
and the invariance with respect to circular shifts
of the operands of the triple scalar product 
$(\boldsymbol{\omega}_p
\times
\mathbf{r})
\cdot\mathbf{f}^{(ibm)}$ 
has been used. 

In order to replace the first integral on the r.h.s.\ of
(\ref{equ-integral-u-dot-force-ibm-2}) we resort to the Newton-Euler
equation for rigid body motion of a spherical particle which reads in
our present numerical framework\cite{uhlmann:08a}:
\begin{equation}\label{equ-newton-euler-b3-mu-phys-fluids-2008}
  V_p\left(\rho_p-\rho_f\right)\frac{\mbox{d}\mathbf{u}_p^{(i)}}{\mbox{d}t} 
  =
  -\rho_f\int_{\mathbf{x}\in{\cal
      S}^{(i)}}\mathbf{f}^{(ibm)}\,\mbox{d}\mathbf{x}
  +
  V_p\left(\rho_p-\rho_f\right)\mathbf{g}
  +
  \mathbf{F}_c^{(i)}
  \,,
\end{equation}
where 
$\mathbf{F}_c^{(i)}$ is
the sum of the forces due to solid contact of the $i$th particle with
any other particle. Note that $\mathbf{F}_c^{(i)}$ is a force (unit:
mass times acceleration), while $\mathbf{f}^{(ibm)}$ is a force
density (force per unit mass).  
The Newton-Euler equation for the angular
acceleration in the context of our immersed boundary approach reads
\citep{uhlmann:04}: 
\begin{equation}\label{equ-newton-euler-rotation-note-2013}
  I_p\frac{\mbox{d}\boldsymbol{\omega}_p^{(i)}}{\mbox{d}t}
  \left(
    1-\frac{\rho_f}{\rho_p}
  \right)
  =
  -\rho_f  
  \int_{\mathbf{x}\in{\cal S}^{(i)}}
  \left(
    \mathbf{r}^{(i)}
    \times
    \mathbf{f}^{(ibm)}
  \right)
  \mbox{d}\mathbf{x}
  \,,
\end{equation}
where the moment of inertia for a sphere is $I_p=2\rho_pV_pR^2/5$ (with
$R=D/2$ the sphere radius). Note that in writing 
(\ref{equ-newton-euler-rotation-note-2013}) we have assumed that the 
collision model does not generate torque on the particles (as is
indeed the case here) and, therefore, no collision-related term enters
this equation; the opposite is true if the collision model has a
tangential component.  

With the aid of relations
(\ref{equ-newton-euler-b3-mu-phys-fluids-2008}) and
(\ref{equ-newton-euler-rotation-note-2013}) the integrals 
appearing in (\ref{equ-integral-u-dot-force-ibm-2}) can be
eliminated. 
Similarly, relation (\ref{equ-newton-euler-b3-mu-phys-fluids-2008})
can be used to eliminate the integral
$\langle\mathbf{f}^{(ibm)}\rangle_\Omega$ appearing in the second term
on the right-hand side of relation
(\ref{equ-def-two-way-coupling-box-avg}).  
Finally, we obtain the following expression for the fluid-particle
coupling term: 
\begin{equation}
  \Psi^{(p)}(t)
  =
  \phi_s\left(\frac{\rho_p}{\rho_f}-1\right)
  \mathbf{u}_{rel,\Omega}
  \cdot\mathbf{g}
  +
  \Psi^{(p)}_{accel}(t)
  +
  \Psi^{(p)}_{coll}(t)
  \,,
\end{equation}
where we have defined an apparent slip velocity with respect to the
box-averaged velocity field: 
\begin{equation}\label{equ-def-apparnt-slip-composite}
  \mathbf{u}_{rel,\Omega}
  =
  \langle\mathbf{u}_p^{(i)}\rangle_p
  -
  \langle\mathbf{u}\rangle_\Omega
  \,.
\end{equation}
The two expressions regrouping all acceleration terms 
and the inter-particle collision term, respectively, 
read as follows:
\begin{subequations}
\begin{eqnarray}%
  \Psi^{(p)}_{accel}(t)&=&
  \phi_s\left(\frac{\rho_p}{\rho_f}-1\right)
    \left[
      \langle\mathbf{u}\rangle_\Omega
      \cdot
      \left\langle
        \frac{\mbox{d}\mathbf{u}_p}{\mbox{d}t} 
      \right\rangle_p
      -
      \left\langle
        \mathbf{u}_p
        \cdot
        \frac{\mbox{d}\mathbf{u}_p}{\mbox{d}t} 
      \right\rangle_p
      -
      \frac{D^2}{10}
      \left\langle
        \boldsymbol{\omega}_p
        \cdot
        \frac{\mbox{d}\boldsymbol{\omega}_p}{\mbox{d}t}
      \right\rangle_p
    \right]
    ,
    \\
  \Psi^{(p)}_{coll}(t)&=&
  \frac{1}{||\Omega||}
    \sum_{i=1}^{N_p}
    \mathbf{u}_p^{(i)}
    \cdot
    \frac{\mathbf{F}_c^{(i)}}{\rho_f}
  \,.
\end{eqnarray}
\end{subequations}
\revision{}{%
  \section{Vorono\"i tesselation analysis}
  \label{sec-two-phase-voronoi-cell-vol-stddev}
\begin{figure}%
   \centering
   \begin{minipage}{2ex}
     \rotatebox{90}{$\sigma(V^{(i)}/\langle V^{(i)}\rangle_{p,t})$}
   \end{minipage}
   \begin{minipage}{0.45\linewidth}
      \includegraphics[width=\linewidth]
      {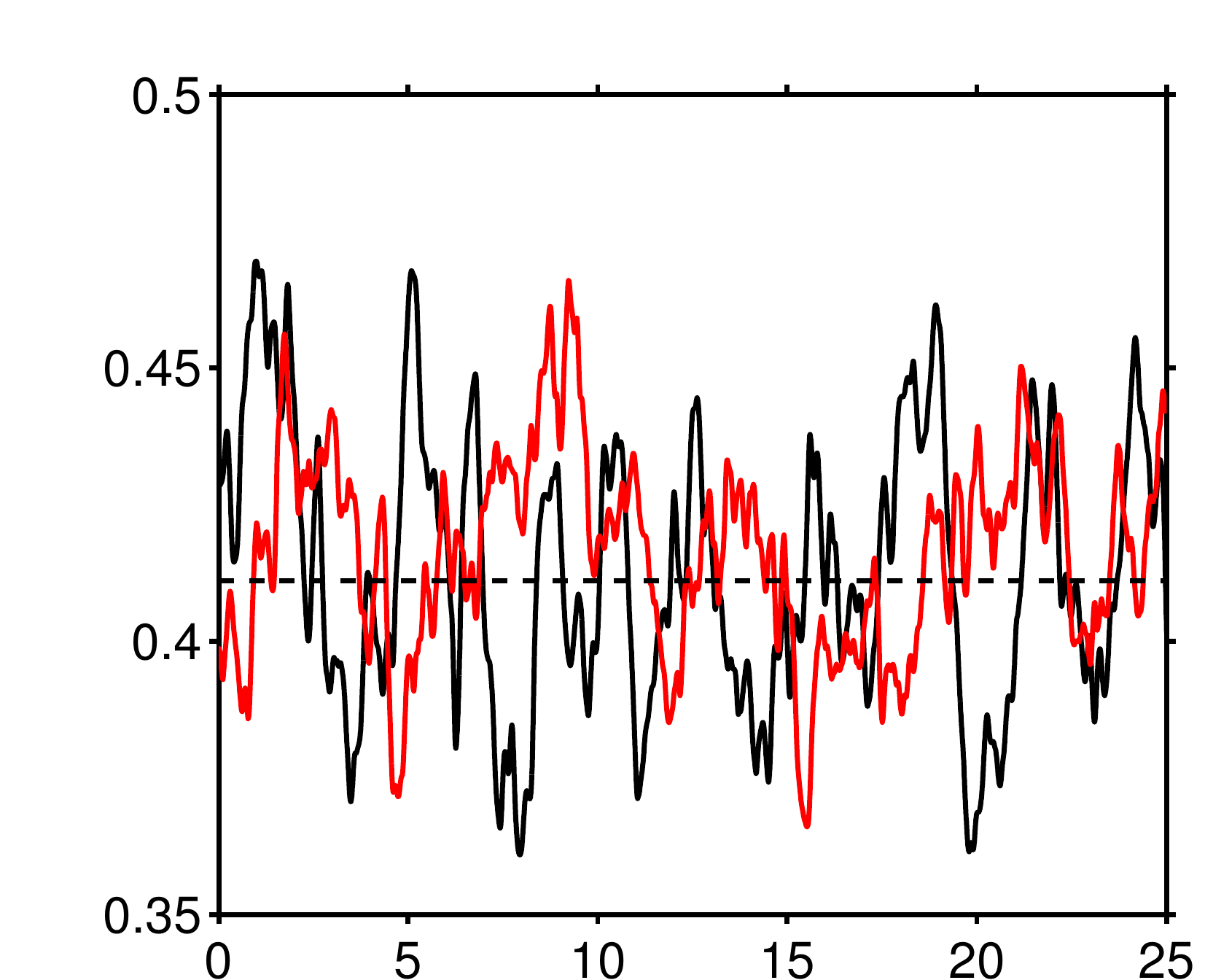}
      \\
      \centerline{$t/T_e$}
   \end{minipage}
   \caption{%
     \revision{}{%
       Time evolution of the standard deviation of the normalized volume
       of the cells of the three-dimensional Vorono\"i tesselation with
       the particle centers as ``sites''. 
     Line styles correspond to: 
     {\color{black}\solidthick},~case~A-G0 (time-evaraged value $0.418$); 
     {\color{red}\solidthick},~case~AL-G120 (time-evaraged value
     $0.416$),  
     {\color{black}\dashed},~value of 
     $0.411$, 
     corresponding to particles with the same solid volume fraction,
     distributed non-overlappingly by means of a random Poisson
     process \citep{uhlmann:14a}. 
     }
   }
   \label{fig-two-phase-voronoi-cell-vols}
\end{figure}
An analysis of the spatial particle distribution by means of Vorono\"i
tesselation\cite{monchaux:10} has been performed in the same manner as in
Ref.\citenum{uhlmann:14a}. The Vorono\"i cells' volume is an indicator
for the inverse of a concentration; the p.d.f.\ of the normalized cell
volumes being typically gamma-distributed\cite{ferenc:07}, with the
standard-deviation being a good scalar indicator of clustering. 
According to the data shown in
figure~\ref{fig-two-phase-voronoi-cell-vols} in both cases A-G0 and
AL-G120 the standard deviations of the Vorono\"i cell volumes
fluctuate in time around the value which corresponds to a particle
distribution through a random Poisson process. Therefore, we conclude
that significant particle clustering does not occur. 
}
\end{appendix}
\addcontentsline{toc}{section}{References}
\end{document}